\documentclass[aps,prd,nofootinbib,twocolumn,superscriptaddress,floatfix,notitlepage]{revtex4-1}
\usepackage[export]{adjustbox} 
\usepackage[utf8]{inputenc}
\usepackage[dvipsnames]{xcolor}
\usepackage{graphicx}
\usepackage{amsmath,amsfonts,amssymb}
\usepackage[breaklinks,colorlinks,urlcolor=Plum,citecolor=MidnightBlue,linkcolor=WildStrawberry]{hyperref}

\usepackage{bm}
\usepackage{acronym}
\usepackage{xspace}
\usepackage{multirow}
\usepackage{tabularx}
\usepackage{orcidlink}
\usepackage{ragged2e}

\usepackage{setspace}
\usepackage{caption}
\usepackage{subcaption}

\usepackage{comment}

\usepackage{listings}
\usepackage[ruled,vlined]{algorithm2e}

\SetCommentSty{mycommfont}

\newacro{S/N}{signal-to-noise ratio}

\newacro{PN}{post-Newtonian}

\newacro{O3}{the third observing run}

\newacro{O2}{the second observing run}

\newacro{CW}{\emph{Continuous Wave}}

\newcommand{\bea}{\begin{equation}\begin{aligned}} 
\newcommand{\eea}{\end{aligned}\end{equation}}
\newcommand{\be}{\begin{equation}}
\newcommand{\ee}{\end{equation}}

\newcommand{\msun}{M_{\odot}}

\newcommand{\fpbh}{f_{\rm PBH}}
\newcommand{\meanm}{\langle m \rangle }
\newcommand{\varm}{\langle m^2 \rangle - \langle m \rangle^2 }
\newcommand{\td}{{\rm d}}
\newcommand{\Ra}{R_{\mathrm{ABH}}}
\newcommand{\mmin}{m_{\mathrm{min}}}
\newcommand{\mmax}{m_{\mathrm{max}}}

\newcommand{\Ndet}{N_{\mathrm{det}}}
\newcommand{\Nobs}{N_{\mathrm{obs}}}

\definecolor{rossocorsa}{rgb}{0.83, 0.0, 0.0}

\newcommand{\revComments}[1]{{\color{black}{#1}}} 

\begin{document}

\title{Constraints on primordial black holes from LIGO-Virgo-KAGRA O3 events}

\author{M.~Andr\'es-Carcasona\orcidlink{0000-0002-8738-1672}}
\email{mandres@ifae.es (corresponding author)}
\affiliation{Institut de Física d’Altes Energies (IFAE), Barcelona Institute of Science and Technology, E-08193 Barcelona, Spain}
\author{A.J.~Iovino\orcidlink{0000-0002-8531-5962}}
\email{antoniojunior.iovino@uniroma1.it}
\affiliation{Dipartimento di Fisica, ``Sapienza'' Universit\`a di Roma, Piazzale Aldo Moro 5, 00185, Roma, Italy}
\affiliation{Istituto Nazionale di Fisica Nucleare, sezione di Roma, Piazzale Aldo Moro 5, 00185, Roma, Italy}
\affiliation{Department of Theoretical Physics and Gravitational Wave Science Center, 24 quai E. Ansermet, CH-1211 Geneva 4, Switzerland}
\author{V.~Vaskonen\orcidlink{0000-0003-0003-2259}}
\email{ville.vaskonen@pd.infn.it}
\affiliation{Keemilise ja Bioloogilise F\"u\"usika Instituut, R\"avala pst. 10, 10143 Tallinn, Estonia}
\affiliation{Dipartimento di Fisica e Astronomia, Universit\`a degli Studi di Padova, Via Marzolo 8, 35131 Padova, Italy}
\affiliation{Istituto Nazionale di Fisica Nucleare, Sezione di Padova, Via Marzolo 8, 35131 Padova, Italy}
\author{H.~Veerm\"ae\orcidlink{0000-0003-1845-1355}}
\email{hardi.veermae@cern.ch}
\affiliation{Keemilise ja Bioloogilise F\"u\"usika Instituut, R\"avala pst. 10, 10143 Tallinn, Estonia}
\author{M.~Mart\'inez\orcidlink{0000-0002-3135-945X}}
\affiliation{Institut de Física d’Altes Energies (IFAE), Barcelona Institute of Science and Technology, E-08193 Barcelona, Spain}
\affiliation{Catalan Institution for Research and Advanced Studies (ICREA), E-08010 Barcelona, Spain}
\author{O.~Pujol\`as\orcidlink{0000-0003-4263-1110}}
\affiliation{Institut de Física d’Altes Energies (IFAE), Barcelona Institute of Science and Technology, E-08193 Barcelona, Spain}
\author{Ll.M.~Mir\orcidlink{0000-0002-4276-715X}}
\affiliation{Institut de Física d’Altes Energies (IFAE), Barcelona Institute of Science and Technology, E-08193 Barcelona, Spain}

\date{\today}

\begin{abstract}
Primordial black holes (PBH) can efficiently form black hole binaries in the early universe. We update the resulting constraints on PBH abundance using data from the third observational run (O3) of LIGO-Virgo-KAGRA. To capture a wide range of PBH scenarios, we consider a variety of mass functions, including critical collapse in the QCD epoch \revComments{and primordial non-Gaussianities}. Applying hierarchical Bayesian analysis to \revComments{a population of binaries} consisting of primordial and astrophysical black holes, we find that, in every scenario, the PBHs can make up at most $f_{\rm PBH} \lesssim 10^{-3}$ of dark matter in the mass range $1-200\msun$. The shape and strength of the constraints are not significantly affected by the type of non-Gaussianities, the modifications to the mass function during the QCD epoch, or the modelling of the astrophysical PBH population.
\end{abstract}

\maketitle

\section{Introduction}
The first detection of gravitational waves (GWs) by LIGO in 2015~\cite{LIGOScientific:2016aoc} has allowed for a new independent way to study black holes (BHs) and the growing number of confirmed events by the LIGO-Virgo-KAGRA collaboration (LVK)~\cite{LIGOScientific:2018mvr, LIGOScientific:2020ibl, KAGRA:2021vkt} is being used to infer their population~\cite{KAGRA:2021duu}. To explain the origin of the observed events, two conceptually different scenarios can be considered: astrophysical and primordial. Although the existence of astrophysical black holes (ABHs) is beyond doubt, the characteristics of their population are not well understood. ABHs form binaries either in isolation or by dynamical capture in dense stellar environments~\cite{Mandel:2018hfr, Woosley:2002zz} and, depending on their formation channel, different ABH binary populations can be expected.

Primordial black holes (PBHs) can have formed in the early universe via a wide range of different mechanisms. The most studied and best-understood formation scenario is the gravitational collapse of primordial overdensities after they enter the cosmological horizon~\cite{Carr:1974nx, Carr:1975qj, Ivanov:1994pa, Garcia-Bellido:1996mdl, Ivanov:1997ia}. Other possibilities involve the collapse of cosmic string loops~\cite{Hawking:1987bn,Polnarev:1988dh,Garriga:1993gj,Caldwell:1995fu,MacGibbon:1997pu}, false vacuum bubbles~\cite{Hawking:1982ga,Kodama:1982sf,Garriga:2015fdk,Maeso:2021xvl,Lewicki:2023ioy}, domain wall networks~\cite{Liu:2019lul,Ferrer:2018uiu,Gouttenoire:2023gbn,Ferreira:2024eru}, or non-linear field dynamics during preheating~\cite{Green:2000he,Bassett:2000ha,Suyama:2004mz,Suyama:2006sr,Martin:2019nuw,Auclair:2020csm,Martin:2020fgl}.

PBHs can explain the totality of dark matter but this possibility is limited only in the mass range $10^{-16}-10^{-12}~\msun$~\cite{Carr:2020gox}. Lighter PBHs would be evaporating today and can constitute only a small portion of the DM~\cite{Auffinger:2022khh}. In contrast, heavier PBHs are constrained by microlensing observations\footnote{The bounds arising from microlensing are subject to uncertainties in the Milky Way dark matter halo profile~\cite{Garcia-Bellido:2024yaz}.}~\cite{EROS-2:2006ryy, Niikura:2017zjd, Oguri:2017ock, Niikura:2019kqi} and for PBHs heavier than around $10^{3}~\msun$ the strongest constraint arises from modification of the cosmic microwave background (CMB) spectrum due to accreting PBHs~\cite{Serpico:2020ehh, Piga:2022ysp}. In this paper, we focus on the range $0.1 - 10^3 \,M_{\odot}$ that can be directly probed by LVK GW observations.

Previous works have explored the use of GW data to find direct or indirect evidence of PBHs. Specifically targeted searches of subsolar mass compact objects, which would provide a smoking gun signal of the existence of PBHs, have so far been unsuccessful~\cite{LIGOScientific:2018glc, LIGOScientific:2019kan, Nitz:2020bdb, Phukon:2021cus, Nitz:2021mzz, Nitz:2021vqh, Nitz:2022ltl, Miller:2020kmv, Miller:2021knj, Morras:2023jvb, Mukherjee:2021ags, Mukherjee:2021itf, Andres-Carcasona:2022prl, Miller:2024fpo}.\footnote{Candidate sub-solar events have been claimed in the literature~\cite{Prunier:2023cyv, Morras:2023jvb}, although without sufficient statistical evidence.} Additionally, studies of the LVK GW data including both PBH and ABH populations have not found enough statistical significance to claim the existence of PBHs~\cite{Hutsi:2020sol, Hall:2020daa, Wong:2020yig, Franciolini:2021tla, DeLuca:2021wjr}. However, some of the component masses, in particular, in GW190521~\cite{LIGOScientific:2020iuh} and GW230529$\_$181500~\cite{LIGOScientific:2024elc}, fall in regions where astrophysical models do not predict them, potentially suggesting for a PBH population. Another method of studying the possible PBH population uses the stochastic GW background and the lack of its detection has also been recast as an upper limit to the PBH abundance~\cite{Raidal:2017mfl, Raidal:2018bbj, Vaskonen:2019jpv, Hutsi:2020sol, Romero-Rodriguez:2021aws, Franciolini:2022tfm}. 

In this paper, we update the constraints on the PBH abundance using the GW data from LVK up to the third observational run (O3) studying the population of the observed events. One of our aims is to derive constraints which do not depend significantly on the underlying formation scenario. Thus, we consider a variety of different PBH mass functions. We take a closer look at mass functions arising in scenarios in which PBHs are produced via critical collapse of large overdensities~\cite{Musco:2008hv, Niemeyer:1999ak, Yokoyama:1998xd}. In such scenarios, PBH formation is affected by the cosmological background. In particular, the quantum chromodynamics (QCD) phase transition, during which the cosmological equation of state is softened, catalyzes the formation of PBHs in the stellar mass range~\cite{Jedamzik:1996mr, Byrnes:2018clq, Musco:2023dak, Carr:2019kxo, Clesse:2020ghq, Bagui:2021dqi, Escriva:2022bwe, Carr:2023tpt}. We also account for the sizeable effect primordial non-Gaussianities can have on the mass function and show that their effect on the constraints is, however, quite marginal.

This paper is organized as follows: In Sec.~\ref{sec:methods}, we explain the methodology used to set up the constraints. In Sec.~\ref{sec:models}, we present the models employed for the BH binary populations. The results are presented in Sec.~\ref{sec:results} and we conclude in Sec.~\ref{sec:conc}. Natural units $\hbar=c=G_{\rm N}=1$ are used throughout this paper.

\section{Methods} \label{sec:methods}

In this paper, we mainly adopt a hierarchical Bayesian approach. We define the population parameters that describe the production rate of binary mergers as $\Lambda$ and the data measured from $\Nobs$ observed events as $\{d\}$. The associated likelihood is~\cite{LIGOScientific:2020kqk, Thrane:2018qnx,Mandel:2018mve, Mastrogiovanni:2023zbw}
\be \label{eq:hierarchicalBayesian}
    \mathcal{L}(\{d\}|\Lambda) \propto e^{-N(\Lambda)}\prod_{i=1}^{\Nobs}\int \mathcal{L}(d_i|\theta)\pi(\theta|\Lambda)\td \theta\, ,
\ee
where $N$ is the expected number of observed events. On the other hand, $\mathcal{L}(d_i|\theta)$ is the event likelihood, given some parameters $\theta$, and $\pi(\theta|\Lambda)$ is called the hyperprior and defines the distribution of mass, spin, redshift and PBH merger rate.

The single event likelihood is usually not available. Instead, posterior samples are provided. Using Bayes' theorem, the $N_p$ posterior samples and the likelihood are related as $p(\theta|d_i,\Lambda)\propto\mathcal{L}(d_i|\theta)\pi_{\varnothing}(\theta|\Lambda)$, where $\pi_{\varnothing}(\theta|\Lambda)$ is the prior used for the initial parameter estimation. Therefore, the integral appearing in Eq.~\eqref{eq:hierarchicalBayesian} for each event can be approximated using Monte-Carlo integration as~\cite{Mastrogiovanni:2023zbw,LIGOScientific:2020kqk}
\be
    \int\!\mathcal{L}(d_i|\theta)\pi(\theta|\Lambda)\td \theta 
    \!\approx\!\frac{1}{N_p}\!\sum_{j=1}^{N_p}\!\frac{\pi(\theta_{ij}|\Lambda)}{\pi_{\varnothing}(\theta_{ij}|\Lambda)}
    \!\equiv\!\frac{1}{N_p}\!\sum_{j=1}^{N_p}w_{ij} .
\ee
To ensure numerical stability during this integration, the effective number of posterior samples per each event, defined as~\cite{Mastrogiovanni:2023zbw,Talbot:2023pex,LIGOScientific:2021aug}
\be
    N_{\mathrm{eff},i} = \bigg[  \displaystyle\sum_j^{N_p} w_{ij} \bigg]^2 \!\bigg/\, \displaystyle\sum_j^{N_p} w_{ij}^2 \,,
\ee
is set to $20$. 

The number of expected events is~\cite{Mastrogiovanni:2023zbw,LIGOScientific:2021aug,LIGOScientific:2020kqk,Vaskonen:2019jpv,Hutsi:2020sol}
\be \label{eq:Nexpected}
    N(\Lambda) = T\int p_{\mathrm{det}}(\theta)\pi(\theta|\Lambda)\td \theta\, ,
\ee
where $p_{\mathrm{det}}(\theta)$ denotes the detection probability. It depends primarily on the masses and redshift of the system~\cite{LIGOScientific:2020kqk}. For this reason, the effect of the spin on the detection probability will be ignored in this work. We use a semianalytical estimate of $p_{\mathrm{det}}$, following the approach of Ref.~\cite{Vaskonen:2019jpv},
\be
    p_{\mathrm{det}}(\theta)=\int_{\frac{\mathrm{SNR}_c}{\mathrm{SNR}(\theta)}}^1 p(\omega)\td \omega \,,
\ee
where SNR denotes the signal-to-noise ratio and $p(\omega)$ the probability density function of the projection parameter $\omega\in[0,1]$ defined as (see e.g.~\cite{Hutsi:2020sol})
\be
    \omega^2 = \frac{(1+\iota^2)^2}{4}F_+(\alpha,\delta,\psi)^2+\iota^2F_\times(\alpha,\delta,\psi)^2 ~,
\ee
where $F_+(\alpha,\delta,\psi)$ and $F_\times(\alpha,\delta,\psi)$ represent the antenna patterns of an L-shaped detector. Assuming a uniform distribution of the inclination $\iota\in(-1,1)$, right ascension $\alpha\in(0,2\pi)$, cosine of the declination $\cos(\delta)\in(-1,1)$ and polarization $\psi \in (0,2\pi)$; the PDF of the projection parameter can be obtained and integrated to give the detection probability. For this work, the critical SNR for detection is set to $\mathrm{SNR_c}=8$, and the SNR of an injection is estimated using the methodology described in Appendix~B of Ref.~\cite{Hutsi:2020sol} and the fitting coefficients from Ref.~\cite{Ajith:2007kx}. A comparison of this method with a set of injections \revComments{can be found in Refs.~\cite{LIGOScientific:2016kwr,LIGOScientific:2018jsj}}. The hyperprior used is
\be
    \pi(\theta|\Lambda) = \frac{1}{1+z}\frac{\td V_c}{\td z} \frac{\td R}{\td m_1 \td m_2}(\theta|\Lambda) \, ,
\ee
where $V_c$ is the comoving volume and $R$ the merger rate. 

Following the same prescription as for the posterior samples, we evaluate Eq.~\eqref{eq:Nexpected} using Monte-Carlo integration over $N_g$ generated samples as~\cite{Mastrogiovanni:2023zbw}
\be
    N(\Lambda)\approx \frac{T}{N_g}\sum_{j= 1}^{N_{\mathrm{det}}}\frac{\pi(\theta_j|\Lambda)}{\pi_{\mathrm{inj}}(\theta_j)}\equiv \frac{T}{N_g}\sum_{j=1}^{N_{\mathrm{det}}}s_j\, ,
\ee
where $\pi_{\mathrm{inj}}(\theta_j)$ is the prior probability of the $j$th event. Similarly, a numerical stability estimator can be defined for the injections as \cite{Farr:2019rap,Mastrogiovanni:2023zbw}
\be
    N_{\mathrm{eff,inj}} = \frac{\bigg[\displaystyle \sum_j^{N_{\mathrm{det}}} s_j \bigg]^2}{\displaystyle\sum_j^{N_{\mathrm{det}}} s_j^2-\displaystyle\frac{1}{N_{g}}\bigg[\displaystyle\sum_j^{N_{\mathrm{det}}} s_j\bigg]^2}\, ,
\ee
which we set to $N_{\mathrm{eff,inj}}>4\Nobs$. Finally, the log-likelihood is evaluated as~\cite{Mastrogiovanni:2023zbw}
\be
    \ln\!\mathcal{L}(\{d\}|\Lambda) 
    \!\approx\!-\frac{T_\mathrm{obs}}{N_g}\sum_{j}^{\Ndet}s_j +\!\sum_{i}^{\Nobs}\ln\!\left[ \frac{T_\mathrm{obs}}{N_{p}}\sum_j^{N_p}w_{ij} \right]\!.
\ee
We implement this computation on top of the ICAROGW code, as it already includes these functionalities~\cite{Mastrogiovanni:2023zbw}.

In the case without any detected events, the likelihood~\eqref{eq:hierarchicalBayesian} reduces to~\cite{Vaskonen:2019jpv,Hutsi:2020sol}
\be
    \mathcal{L}(\{d\}|\Lambda) \propto e^{-N(\Lambda)} \, ,
\ee
corresponding to a simple Poisson process. Therefore, in the case where all the events detected by LVK are considered to be of astrophysical origin, the 2$\sigma$ confidence level upper limits can be placed by discarding the region of the parameter space where $N>3$~\cite{Vaskonen:2020lbd,Hutsi:2020sol}.

To analyze the scenarios with multiple BH binary populations, two approaches will be followed: an agnostic and a model-dependent one. \revComments{The agnostic analysis tests how well a PBH model fits subsets of the observed events. The Bayesian inference using the likelihood~\eqref{eq:hierarchicalBayesian} is performed with various subsets and the combined posterior is then obtained by taking the maximum of the individual posteriors. Testing all possible subsets is computationally unfeasible, and therefore the strategy used in Ref.~\cite{Hutsi:2020sol} is followed. The mass range is divided into bands; for each one, all the events with the primary mass falling inside it are taken to be primordial. In this way, the least restrictive case is considered and the most conservative constraints can be obtained. As shown in Ref.~\cite{Hutsi:2020sol}, the results converge fast with a relatively small set of subsamples, but following this strategy, the procedure is further optimized.} 

For the model-dependent case, the merger rate and population of ABHs is assumed to follow a given distribution and the fit using the likelihood of Eq.~\eqref{eq:hierarchicalBayesian} with the full set of parameters of both the PBH and ABH models is performed. In any case, all limits arising from this method should lie between the constraints using the agnostic approach and the one assuming that all events are astrophysical. As we will show in section~\ref{sec:results}, this is indeed the case.

\revComments{To compare between models the Bayes factor is used. The log-Bayes factor between two models, $\mathcal{M}_1$ and $\mathcal{M}_2$, is defined as 
$
    \ln \mathcal{B}_{\mathcal{M}_1}^{\mathcal{M}_2}=\ln \mathcal{Z}_{\mathcal{M}_2} - \ln \mathcal{Z}_{\mathcal{M}_1}
$,
where $\mathcal{Z}$ denotes the Bayesian evidence for a model and is computed as
$
    \mathcal{Z}_{\mathcal{M}} = \int \mathcal{L}(\{d\} | \theta, \mathcal{M}) \pi(\theta | \mathcal{M}) \, \td \theta
$.
Typically, a Bayes factor of $\ln \mathcal{B}_{\mathcal{M}_1}^{\mathcal{M}_2}>5$ is needed to claim a strong evidence of model $\mathcal{M}_2$ compared to model $\mathcal{M}_1$.
}

\section{Modelling BH binary populations}
\label{sec:models}

\subsection{Primordial black hole population}

PBH binaries can arise from various channels. Characterized by the time of formation, one can distinguish binaries formed in the early universe before matter domination as the first gravitationally bound non-linear structures in the universe, and binaries formed later in DM haloes (for a review see~\cite{Raidal:2024bmm}). Both processes can be divided by the number of progenitors of the binary, i.e., whether the binary formation involves two or more PBHs. 

The early universe formation channels include a two-body channel in which close PBHs decouple from expansion and form a highly eccentric binary, with the initial angular momentum provided by tidal forces from surrounding PBHs and matter fluctuations~\cite{Nakamura:1997sm, Ioka:1998nz, Ali-Haimoud:2017rtz, Raidal:2018bbj, Vaskonen:2019jpv, Hutsi:2020sol}, and a three-body channel in which a compact configuration of three PBHs forms a three-body system after decoupling from expansion and a binary is formed after one of the PBHs is ejected from this system~\cite{Vaskonen:2019jpv}. The latter process produces generally much harder and less eccentric binaries than the former. However, as compact three-body systems are less likely than two-body ones, the three-body channel is subdominant to the two-body one, unless PBHs make up a considerable fraction, above $\mathcal{O}(10\%)$, of DM. In the late universe, PBHs can form binaries during close encounters in the DM haloes if they emit enough of their energy into GWs to become bound~\cite{Mouri:2002mc, Bird:2016dcv}, and, if the PBH abundance is sufficiently large, also three-body encounters in DM haloes can become relevant~\cite{Franciolini:2022ewd}. However, both late universe formation channels are subdominant~\cite{Raidal:2017mfl, DeLuca:2020jug, Franciolini:2022ewd, Raidal:2024bmm} to the early universe ones, and we will not consider their contributions in this study.

The merger rate from the early universe two-body channel is~\cite{Raidal:2018bbj, Vaskonen:2019jpv, Hutsi:2020sol}:
\begin{multline}\label{eq:R2}
    \frac{\td R_{\mathrm{PBH},2}}{\td m_1 \td m_2}
    = \frac{1.6\times 10^6}{\text{Gpc}^3\text{yr}^1}
    \fpbh^{\frac{53}{37}}\left[ \frac{t}{t_0}\right]^{-\frac{34}{37}}\left[ \frac{M}{\msun}\right]^{-\frac{32}{37}}  
    \\  \times 
    \eta^{-\frac{34}{37}} S  \left[ \psi,\fpbh,M \right]\frac{\psi(m_1)\psi(m_2)}{m_1 m_2}\, ,
\end{multline}
where $\fpbh\equiv \rho_{\mathrm{PBH}}/\rho_{\mathrm{DM}}$ denotes the total fraction of DM in PBHs, $t_0$ the current age of the universe, $M=m_1+m_2$ the total mass of the binary, $\eta=m_1m_2/M^2$ its symmetric mass ratio, $S$ the suppression factor, $\psi(m) \equiv \rho_{\rm PBH}^{-1}\td \rho_{\rm PBH}/\td \ln m$ the PBH mass function and $\meanm$ its mean. Note that $\int \td \ln m\, \psi(m) = 1$ and the average over any mass-dependent quantity $X$ is defined via the number density as\footnote{We remark that different conventions for the mass function have been used in Refs.~\cite{Raidal:2018bbj,Vaskonen:2019jpv,Hutsi:2020sol}.}
\be
    \langle X \rangle  
    \equiv \int \frac{\td n_{\rm PBH}}{n_{\rm PBH}} X
    = \frac{\int \td \ln m \, m^{-1}\, \psi(m) X }{\int \td \ln m \, m^{-1} \,\psi(m)}  \,,
\ee
where $\td n_{\rm PBH} = m^{-1} \td \rho_{\rm PBH}$ is the differential number density of PBHs.

We compute the suppression factor $S$ as the product of two contributions $S=S_1S_2(z)$. The first factor accounts for the initial configuration and perturbations in the surrounding matter and discards all those initial configurations that contain a third PBH within a distance smaller than $y$. It can be approximated as~\cite{Hutsi:2020sol}
\be\label{eq:S1}
    S_1\approx 1.42\left[ \frac{\langle m^2\rangle/\meanm }{\tilde{N}(y)+C}+\frac{\sigma_M^2}{\fpbh^2} \right]^{-\frac{21}{74}}e^{-\tilde{N}(y)}~,
\ee
where $\sigma_M\simeq 0.004$ is the rescaled variance of matter density perturbations,  $\tilde{N}(y)$ is the number of expected PBH to be formed inside a sphere of comoving radius $y$ and can be approximated as~\cite{Raidal:2018bbj}
\be
    \tilde{N}(y)\approx \frac{M}{\meanm}\frac{\fpbh}{\fpbh+\sigma_M}~,
\ee
and $C$ is a fitting factor
\begin{multline}
    C= \fpbh^2\frac{\langle m^2\rangle/\meanm^2}{\sigma_M^2}
    \\ \times 
    \left\{\left[ \frac{\Gamma(29/37)}{\sqrt{\pi}}U\left(\frac{21}{74},\frac{1}{2},\frac{5}{6}\frac{\fpbh^2}{\sigma_M^2}\right)\right]^{-\frac{74}{21}}-1\right\}^{-1}\, ,
\end{multline}
$U$ denotes the confluent hypergeometric function and $\Gamma$ is the gamma function. The second term of the suppression factor, $S_2(z)$, accounts for the disruption of PBH binaries due to encounters with other PBHs in the late universe~\cite{Vaskonen:2019jpv}. It can be approximated by~\cite{Hutsi:2020sol}
\be\label{eq:S2}
    S_2(z)\approx\min\left[ 1,\,0.01\chi^{-0.65}e^{-0.03\ln^2\chi}\right]~,
\ee
where $\chi = (t(z)/t_0)^{0.44}\fpbh$. This approximation is accurate if $z \lesssim 100$, which is within our region of interest.

The merger rate of binaries formed from compact three-body configurations in the early universe is~\cite{Vaskonen:2019jpv, Raidal:2024bmm}
\bea\label{eq:R3}
    \frac{\td R_{\rm PBH,3}}{\td m_1 \td m_2}
    &\approx \frac{7.9 \times 10^4}{\rm Gpc^{3}\,yr } 
    \left[ \frac{t}{t_0}\right]^{\frac{\gamma}{7} - 1}
    f_{\rm PBH}^{\frac{144 \gamma}{259}+\frac{47}{37}}
    \\ 
    \times&
    \left[ \frac{\langle m \rangle}{\msun}\right]^{\frac{5 \gamma -32}{37}}\!
    \left(\frac{M}{2\langle m \rangle}\right)^{\frac{179 \gamma }{259}-\frac{2122}{333}}\!\!\!
    (4\eta)^{-\frac{3 \gamma }{7}-1}  
    \\ 
    \times&
    \mathcal{K} \,
    \frac{e^{-3.2 (\gamma - 1)}\gamma}{28/9-\gamma}
    \bar{\mathcal{F}}(m_1,m_2)
    \frac{\psi(m_1)\psi(m_2)}{m_1 m_2}\, ,
\eea
where $\gamma \in [1,2]$ characterizes the dimensionless distribution of angular momenta $j$, which is assumed to take the form $P(j) = \gamma j^{\gamma-1}$ after the initial three-body system has ejected one of the PBHs. The value $\gamma=2$ corresponds to the equilibrium distribution while studies suggest a super-thermal distribution with $\gamma = 1$~\cite{Raidal:2018bbj,2019Natur.576..406S}. The factor
\bea\label{eq:barF_R3}
    \bar{\mathcal{F}}(m_1,m_2) 
    &\equiv \int_{m\leq m_1,m_2} 
    \!\!\!\!\td\ln m \, \psi(m) \frac{\langle m\rangle}{m} \\
    &\times \bigg[ 2 \mathcal{F}(m_1,m_2,m) + \mathcal{F}(m,m_1,m_2) \bigg]\,,
\eea
with
\bea
    \mathcal{F} (m_1,m_2,m_3)
    &= m_1^{\frac{5}{3}} m_2^{\frac{5}{3}} m_3^{\frac{7}{9}}\left(\frac{m_1+m_2}{2}\right)^{\frac{4}{9}} \\
    &\times \left(\frac{m_1+m_2+m_3}{3}\right)^{\frac{2}{9}}\langle m\rangle^{-\frac{43}{9}} \,
\eea
accounts for the composition of masses in the initial 3-body system and assumes that the lightest PBH gets ejected. Thus, the 3-body channel tends to generate binaries from the heavier tail of the PBH mass spectrum. The factor $\mathcal{K}$ accounts for the hardening of the early binary in encounters with other PBHs. We will use $\gamma=1$ and $\mathcal{K} = 4$ as suggested by numerical simulations~\cite{Raidal:2018bbj}.\footnote{The most conservative choice corresponds to $\gamma=2$ and $\mathcal{K} = 1$.} The rate shown in Eq.~\eqref{eq:R3} assumes a monochromatic mass function. 

The binaries formed from three-body systems are significantly less eccentric and much harder than those contributing to the merger rate of Eq.~\eqref{eq:R2}. This means that, unlike for the rate~\eqref{eq:R2}, the binary-single PBH encounters in DM (sub)structures will approximately preserve the angular momentum distribution and harden the binaries. As a result, these encounters will not diminish the merger rate.\footnote{We note that Eq.~\eqref{eq:R3} does not account for the potential enhancement of the merger rate due to the hardening of the binaries by binary-PBH encounters. However, the effect of this process was estimated to be relatively weak for binaries formed in PBH DM haloes~\cite{Franciolini:2022ewd} and we expect this estimate to be valid also for the early binaries.} Since the suppression factors of Eq.~\eqref{eq:S1} and Eq.~\eqref{eq:S2} only apply to Eq.~\eqref{eq:R2}, the three-body channel can become dominant for relatively large PBH abundances, when $f_{\rm PBH} \gtrsim 0.1$. 

Unlike the rate of Eq.\eqref{eq:R2}, the rate of Eq.~\eqref{eq:R3} is expected to be enhanced when PBHs are formed in clusters since the density of very compact three-PBH configurations would be larger. Thus, more binaries would be formed as such configurations decouple from expansion. For this reason, the constraints derived from Eq.~\eqref{eq:R3} are more robust in ruling out $f_{\rm PBH} \approx 1$ as well as initially clustered PBH scenarios. On a more speculative note, as a potential caveat for avoiding the merger rate constraints, one can imagine extreme cases of clustering, in which case the mean PBH separation is so small that all the early binaries would merge within much less than a Hubble time even if they were circular -- thus, they would not contribute to the current merger rate. However, extreme clustering of PBHs in the stellar mass range can introduce noticeable isocurvature perturbations at relatively large scales and will likely clash with the Lyman-$\alpha$ observations~\cite{DeLuca:2022uvz}.

\revComments{Finally, the merger rate estimates in both the two- and three-body formation channels can receive sizable corrections for very large mass ratio binaries made possible in scenarios in which the PBH mass function spans several orders of magnitude. This is because, in these scenarios, the heaviest PBHs are likely to be initially surrounded by multiple light PBHs instead of forming relatively isolated two or three-body configurations. The impact of such configurations on binary formation and the PBH merger rate has not been estimated in the literature. Nevertheless, as the typical mass function widths and PBH binary mass ratios are unlikely to exceed an order of magnitude in the scenarios considered here, we do not expect such effects to significantly alter our results.}

\begin{figure*}   
    \includegraphics[width=0.95\columnwidth]{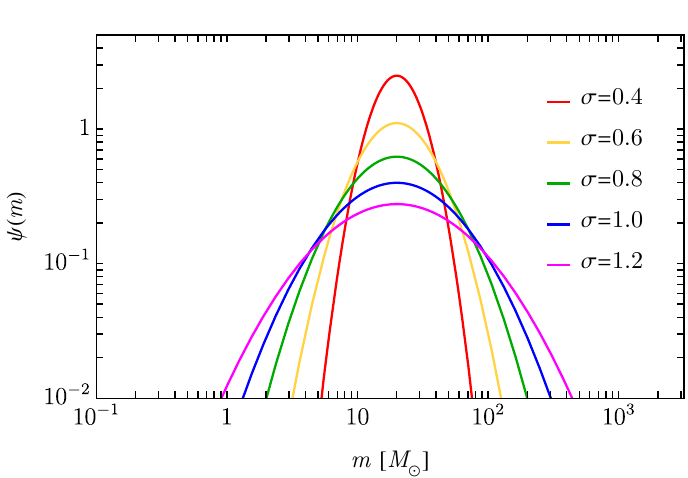}
    \quad
    \includegraphics[width=0.95\columnwidth]{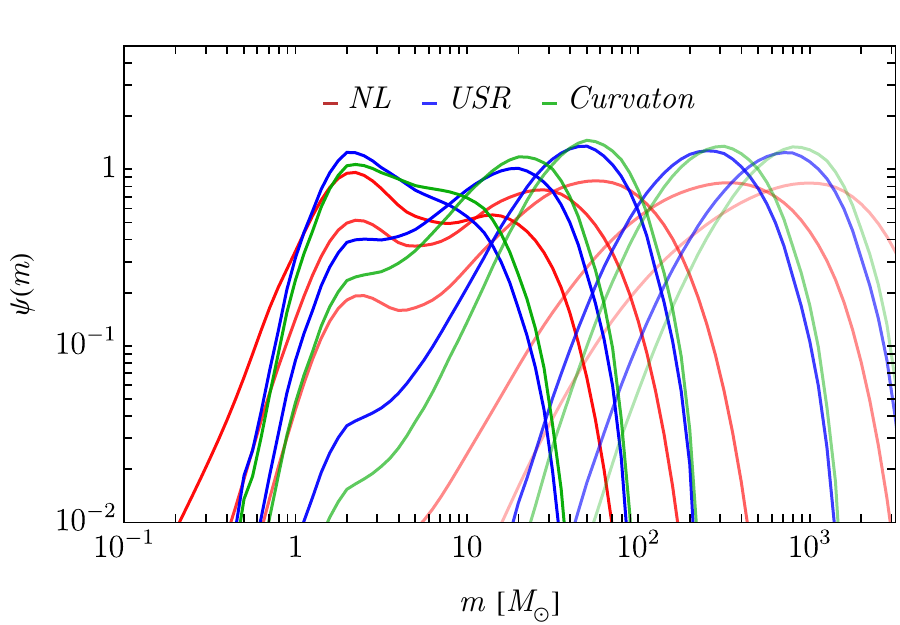}
    \caption{ \justifying Examples of extended mass functions considered in this work. \emph{Left panel:}  Log-normal mass function with $m_c=20 \msun$. \emph{Right panel:} Critical collapse mass function arising from a broken power-law curvature power spectrum with $\alpha=4$, $\beta=0.5$ different values of $k_*$, including only non-linearities (red), primordial non-Gaussianities from USR (blue), and curvaton (green) models. For the primordial non-Gaussianities, in the case of USR models, the choice of $\beta$ also determines the amount of primordial non-Gaussianities. For the curvaton model, we fix $r_{\rm dec}=0.1$.}
    \label{fig:mass_functions}
\end{figure*}

In this article, we consider the following PBH mass functions:

\subsubsection{Monochromatic}

The monochromatic mass function 
\be
    \psi(m) = m_c\delta(m-m_c)\,,
\ee
where $\delta$ is the Dirac delta function and $m_c$ is the mass where the distribution has the peak. It is likely the simplest possible and approximates models that predict a sharp peak at a particular mass. By construction, the mean of this distribution is $\meanm=m_c$ and the variance vanishes, $\varm=0$.

If PBHs of different masses contribute independently to an observable that is constrained, the constraints obtained for a monochromatic mass function can be easily recast to other mass functions~\cite{Carr:2017jsz}. This is, however, not the case for the constraints arising from the PBH mergers and the constraint for each mass function must be derived separately.

\subsubsection{Log-normal}

The log-normal mass function
\be
    \psi(m)=\frac{1}{\sqrt{2\pi}\sigma}\exp\left\{ -\frac{\ln^2(m/m_c)}{2\sigma^2} \right\}\,,
\ee
where $m_c$ and $\sigma^2$ denote the mode and the width of the distribution. It is a good approximation of wider peaks in the mass spectrum and it is predicted by several PBH formation mechanisms. The mean of this distribution is $\meanm = m_c e^{-\sigma^2/2}$ and the variance $\varm= \meanm^2 (e^{\sigma^2}-1)$. The log-normal mass function is depicted in Fig.~\ref{fig:mass_functions} for different widths.

\subsubsection{Critical collapse mass function}

The masses of PBHs formed from large curvature fluctuations follow the critical scaling law~\cite{Choptuik:1992jv, Niemeyer:1997mt, Niemeyer:1999ak} that gives rise to a PBH mass function with a power-law low-mass tail with an exponential high-mass cut-off~\cite{Vaskonen:2020lbd}. The exact shape of the mass function depends on the shape of the curvature power spectrum and the amount of non-Gaussianities~\cite{Bugaev:2013vba, Nakama:2016gzw, Byrnes:2012yx, Young:2013oia, Yoo:2018kvb, Kawasaki:2019mbl, Yoo:2019pma,
Riccardi:2021rlf, Taoso:2021uvl, Meng:2022ixx, Escriva:2022pnz, Young:2014ana, Shibata:1999zs, Musco:2018rwt, Young:2022phe}. Therefore, instead of using an ansatz directly on the mass function, we compute it numerically starting from the curvature power spectrum accounting for different types of non-Gaussianities. In this work, we assume a \emph{broken power-law} (BPL) shape for the curvature power spectrum:
\be \label{eq:PPL}
    \mathcal{P}_{\zeta}(k)
    = A \,\frac{\left(\alpha+\beta\right)^{\lambda}}{\left[\beta\left(k / k_*\right)^{-\alpha/\lambda}+\alpha\left(k / k_*\right)^{\beta/\lambda}\right]^{\lambda}},
\ee
where $\alpha, \beta>0$ describe, respectively, the growth and decay of the spectrum around the peak and $\lambda$ characterizes the width of the peak. Typically, the $\alpha \approx 4$~\cite{Byrnes:2018txb,Karam:2022nym}.

To properly compute the abundance of PBHs, two kinds of non-Gaussianities must be considered. First, even for Gaussian curvature perturbations, the density fluctuations will inevitably inherit non-Gaussianities from  the nonlinear relation between curvature and density perturbations in the long-wavelength approximation~\cite{Harada:2015yda, Musco:2018rwt, DeLuca:2019qsy, Young:2019yug, Germani:2019zez}
\be\label{eq:NL}
    \delta(\vec{x}, t) = -\frac{2}{3} \Phi\left(\frac{1}{a H}\right)^2 e^{-2 \zeta} \left[\nabla^2 \zeta + \frac{1}{2} \partial_i \zeta \partial_i \zeta\right] \,,
\ee
where $a$ denotes the scale factor and $H$ the Hubble rate, $\Phi=2/3$.\footnote{The $\Phi$ parameter is given by $\Phi= 3(1+w)/(5+3w)$ for a constant equation of state parameter $w$~\cite{Polnarev:2006aa}, and we have dropped the explicit $\vec{x}$ and $t$ dependence for the sake of brevity. We assume that PBHs form during radiation domination: $w=1/3$.} We refer to this type of non-Gaussianities as non-linearities (NL).

Second, $\zeta$ itself can be a non-Gaussian field -- we refer to such cases as primordial non-Gaussianities. We consider two specific cases in which the primordial non-Gaussianities can be worked out explicitly.

The first case is the quasi-inflection-point models of single field inflation, also called ultra-slow-roll (USR) models, where the peak in $\mathcal{P}_{\zeta}$ arises from a brief ultra-slow-roll like phase which is typically followed by constant-roll inflation~\cite{Garcia-Bellido:2017mdw,Pi:2017gih, Kannike:2017bxn,Ballesteros:2020qam,Inomata:2016rbd,Iacconi:2021ltm,Kawai:2021edk,Bhaumik:2019tvl,Cheong:2019vzl,Inomata:2018cht,Dalianis:2018frf,Motohashi:2019rhu,Hertzberg:2017dkh,Ballesteros:2017fsr,Karam:2022nym,Rasanen:2018fom,Balaji:2022rsy,Frolovsky:2023hqd,Dimopoulos:2017ged,Germani:2017bcs,Choudhury:2013woa,Ragavendra:2023ret,Cheng:2021lif,Franciolini:2023lgy,Karam:2023haj,Mishra:2023lhe,Cole:2023wyx,Karam:2023haj,Frosina:2023nxu,Franciolini:2022pav,Choudhury:2024one,Wang:2024vfv,Stamou:2021qdk,Stamou:2024lqf,Heydari:2021gea,Heydari:2021qsr,Heydari:2023xts}. In this case, the non-Gaussianities can be related to the large $k$ spectral slope as~\cite{Atal:2019cdz, Tomberg:2023kli}\footnote{This relation is derived assuming a constant-roll phase following the USR-like phase and was found to hold also in the stochastic formalism~\cite{Tomberg:2023kli}. Nevertheless, as quantum diffusion generally depends on the shape of the inflationary potential~\cite{Biagetti:2018pjj, Ezquiaga:2018gbw, Ezquiaga:2019ftu}, this relation may be modified if one accounts for the transition into the USR phase.}
\be\label{eq:zeta_IP}
    \zeta = -\frac{2}{\beta}\ln\left(1-\frac{\beta}{2}\zeta_{\rm G}\right) \,.
\ee
Scenarios for stellar mass PBHs typically have $\beta\lesssim0.5$.

Second, in curvaton models~\cite{Enqvist:2001zp, Lyth:2001nq, Sloth:2002xn, Lyth:2002my, Dimopoulos:2003ii, Kohri:2012yw, Kawasaki:2012wr, Kawasaki:2013xsa, Carr:2017edp, Ando:2017veq, Ando:2018nge, Chen:2019zza, Liu:2020zzv, Pi:2021dft, Cai:2021wzd, Liu:2021rgq, Chen:2023lou, Torrado:2017qtr, Chen:2023lou, Wilkins:2023asp, Ferrante:2023bgz, Inomata:2023drn} the small-scale angular perturbations are determined by curvaton fluctuations, while scales associated with CMB observations are dominated by the inflaton contribution. The non-Gaussianities are described by~\cite{Sasaki:2006kq,Pi:2022ysn}
\be\label{eq:zeta_cur}
    \zeta = \ln\big[X(r_{\rm dec},\zeta_{\rm G})\big] \,,
\ee
where $X(r_{\rm dec})$ is a function of the weighted fraction of the curvaton energy density to the total energy density at the time of curvaton decay $r_{\rm dec}$. Since the shape of the power spectrum and the parameter $r_{\rm dec}$ are non-trivially related, we leave $\beta$ and $r_{\rm dec}$ as free parameters in our analysis. For simplicity, we omit the contribution of the curvaton self-interactions that may modify the non-Gaussianities (for instance, see~\cite{Enqvist:2008gk, Enqvist:2009zf, Huang:2008zj, Fonseca:2011aa, Byrnes:2011gh, Kobayashi:2012ba, Liu:2020zlr, Hooper:2023nnl}). 

To estimate the mass function we follow the prescription presented in~\cite{Ferrante:2022mui} (see also~\cite{Gow:2022jfb}) based on threshold statistics on the compaction function $\mathcal{C}$, including NLs and primordial non-Gaussianities for several benchmark cases. \revComments{The mass of the PBHs formed when the horizon mass was 
\be
    M_H = 4.8\times 10^{-2} M_\odot \left[\frac{106.75}{g_*} \right]^{\frac12} \left[\frac{\rm GeV}{T} \right]^{2} \,, 
\ee
is
\be
    m(\mathcal{C}) = \mathcal{K} M_H\left[\mathcal{C}-\mathcal{C}_{\mathrm{th}}\right]^{\kappa} \,,
\ee
where the parameters $\mathcal{K}$ and $\kappa$ and the threshold $\mathcal{C}_{\mathrm{th}}$ depend on the shape of the curvature power spectrum~\cite{Musco:2020jjb,Musco:2023dak,Ianniccari:2024ltb}. The PBH mass function is then
\be
    \psi(m) \!\propto\!
    \int \! \td \ln M_{H} \left[\frac{M_\odot}{M_{H}}\right]^{\frac{1}{2}}\!
    \frac{\beta(m, M_{H})}{7.9\!\times\!10^{-10}} 
    \left[\frac{106.75}{g_{*s}^{4}/g_*^{3}} \right]^{\frac{1}{4}} \!,
\ee
where $\beta(m, M_{H})$ is the differential fraction of the energy density collapsing into PBHs of mass $m$ when the horizon mass is $M_{H}$. We estimate it by integrating the joint probability distribution function $P_{\mathrm{G}}$ of the Gaussian components,
\bea\label{eq:beta}
    \beta(m, M_{H}) = 
    \int_{\mathcal{D}} &\td\mathcal{C}_{\rm G} \td\zeta_{\rm G} P_{\rm G}(\mathcal{C}_{\rm G},\zeta_{\rm G}) \\
    &\times \frac{m}{M_H} \delta\left[\ln \frac{m}{m(\mathcal{C}_{\rm G},\zeta_{\rm G})}\right]
    \,,
\eea
where} the domain of integration is 
$\mathcal{D} =
\left\{
    \mathcal{C}(\mathcal{C}_{\rm G},\zeta_{\rm G}) > \mathcal{C}_{\rm th}  
    ~\land~\mathcal{C}_1(\mathcal{C}_{\rm G},\zeta_{\rm G}) < 2\Phi
\right\}$. 
The compaction function $\mathcal{C} = \mathcal{C}_1 - \mathcal{C}_1^2/(4\Phi)$ can be built from the linear component $\mathcal{C}_1 = \mathcal{C}_{\rm G} \, \td F/\td\zeta_{\rm G}$, where $\mathcal{C}_{\rm G} = -2\Phi\,r\,\zeta_{\rm G}^{\prime}$. The Gaussian components are distributed as
\be
    P_{\mathrm{G}}
    = \frac{\exp{\left[-\displaystyle\frac{1}{2\left(1-\gamma_{c r}^2\right)}\left(\frac{\mathcal{C}_{\mathrm{G}}}{\sigma_c}-\frac{\gamma_{c r} \zeta_{\mathrm{G}}}{\sigma_r}\right)^2 \!-\! \frac{\zeta_{\mathrm{G}}^2}{2 \sigma_r^2}\right]}}{2 \pi \sigma_c \sigma_\tau \sqrt{1-\gamma_{c r}^2}}.
\ee
with the correlators given by
\begin{subequations}\label{eq:Variances}
\begin{align}
    & \sigma_c^2=\frac{4 \Phi^2}{9} \int_0^{\infty} \frac{\td k}{k}\left(k r_m\right)^4 W^2\left(k, r_m\right)P^{T}_\zeta
    \,, \\
    &\sigma_{c r}^2=\frac{2 \Phi}{3} \! \int_0^{\infty} \!\! \frac{\td k}{k} \!\left(k r_m\right)^2 \!W\!\!\left(k, r_m\right) \!W_s\!\left(k, r_m\right) \!P^{T}_\zeta\! 
    \,, \\
    &\sigma_r^2=\int_0^{\infty} \frac{\td k}{k} W_s^2\left(k, r_m\right) P^{T}_\zeta \,,
\end{align}
\end{subequations}
with $P^{T}_\zeta=T^2\left(k, r_m\right) P_\zeta(k)$, and  $\gamma_{c r} \equiv \sigma_{c r}^2 / \sigma_c \sigma_\tau$.
We have defined $W\left(k, r_m\right), $ $W_s\left(k, r_m\right)$ and $T\left(k, r_m\right)$ 
as the top-hat window function, the spherical-shell window function, and the radiation transfer function, computed assuming radiation domination~\cite{Young:2022phe}.\footnote{The softening of the equation of state near the QCD transitions is expected to slightly affect the evolution of sub-horizon modes. Since this is mitigated by the window function that also smooths out sub-horizon modes, we neglect this effect here.} We follow the prescription given in~\cite{Musco:2020jjb} to compute the values of the threshold $\mathcal{C}_{\rm th}$ and the position of the maximum of the compaction function $r_m$, which depend on the shape of the power spectrum.  We note that corrections to the horizon crossing and from the non-linear radiation transfer function can affect the PBH abundance when compared to the current prescription, which relies on average compaction profiles~\cite{Ianniccari:2024bkh}. However, since the effect of the radiation transfer function involves uniform rescaling of all variances in Eqs.~\eqref{eq:Variances}, we expect minimal alteration to the shape of the mass functions and thus also to our results.

The softening of the equation of state during the thermal evolution of the universe on PBH formation, in particular, the QCD phase transition, can enhance the formation of PBHs~\cite{Jedamzik:1996mr, Byrnes:2018clq}. It is accounted for by considering that $\gamma\left(M_H\right), \mathcal{K}\left(M_H\right), \mathcal{C}_{\rm th}\left(M_H\right)$ and $\Phi\left(M_H\right)$ are functions of the horizon mass around $m=\mathcal{O}\left(M_\odot\right)$~\cite{Franciolini:2022tfm, Musco:2023dak}.

Examples of critical mass functions arising from curvature power spectra with $\beta = 0.5$ and $\alpha = 4$ are shown in the right panel of Fig.~\ref{fig:mass_functions} for different primordial non-Gaussianities. The enhancement around $1\,\msun$ arises due to the QCD phase transition. As a result, the QCD phase transition can generate multimodal mass distributions. However, as seen from Fig.~\ref{fig:mass_functions}, peaks around $1\,\msun$ are not universal. In particular, the effect of the QCD phase transition is \revComments{strongly reduced} for mass functions peaked much above $1\,\msun$. 

For mass functions peaking close to $1\,\msun$, the QCD effect tends to be more pronounced in the non-linearities only (NL) scenario, indicating that primordial non-Gaussianities tend to curtail the impact of the QCD phase transition. In general, we find that primordial non-Gaussianities tend to reduce the width of the mass function in both the USR and curvaton models.

\subsection{Astrophysical black hole population}
\label{sec:ABHs}

To describe the ABH population, we use the phenomenological POWER-LAW + PEAK model~\cite{KAGRA:2021duu}. This general parametrization allows us to cover a wide range of potential ABH scenarios that could serve as a foreground for PBH mergers. \revComments{We stress, however, that our ABH ansatz is not derived from a precise computation of astrophysical BH merger rate, but is rather intended to provide a generic shape that can fit the GW data. Nevertheless, the choice of priors can encode astrophysical considerations, which would not be present in a PBH binary population. This includes imposing cut-offs based on limitations on BH formation due to the TOV limit or physics of pair-instability supernovae. More importantly, the model ABH considered here should be general enough to encode a wide range of possibilities for the ABH population in order to suppress selection effects that would generate artificial preferences for PBHs.} The differential merger rate for this model is 
\be\label{eq:R_ABH}
    \frac{\td \Ra}{\td m_1 \td m_2}
    \!=\! R_{\mathrm{ABH}}^0 p_{\mathrm{ABH}}^z(z)p_{\mathrm{ABH}}^{m_1}(m_1)p_{\mathrm{ABH}}^{m_2}(m_2|m_1) .
\ee
The probability density function of the primary mass is modeled as a combination of a power law and a Gaussian peak 
\bea
    p_{\mathrm{ABH}}^{m_1}(m_1) = &\left[ (1-\lambda)P_{\mathrm{ABH}}(m_1)+\lambda G_{\mathrm{ABH}}(m_1)\right] \\ &\times S(m_1|\delta_m,\mmin) \, ,
\eea
where
\bea
    P_{\mathrm{ABH}}(&m_1|\alpha,\mmin,\mmax) \\
    &\propto \Theta(m\!-\!\mmin) \Theta(\mmax\!-\!m) m_1^{-\alpha} \,,\\
    G_{\mathrm{ABH}}(&m_1|\mu_G,\sigma_G,\mmin,\mmax) \\ 
    &\propto \Theta(m\!-\!\mmin) \Theta(\mmax\!-\!m) e^{-\frac{(m_1-\mu_G)^2}{2\sigma_G^2}}\,
\eea
are restricted to masses between $\mmin$ and $\mmax$ and normalized. The term $S(m_1|\delta_m,\mmin)$ is a smoothing function, which rises from 0 to 1 over the interval $(\mmin, \mmin + \delta_m)$:
\bea 
&S(m_1|\delta_m,\mmin) = \\ &\begin{cases} 
      0 & m< \mmin \\
      [f(m-\mmin,\delta_m)+1]^{-1} & \mmin\leq m< \mmin+\delta_m \\
      1 & m\geq \mmin+\delta_m\, .
   \end{cases}
\eea
with
\be
f(m',\delta_m) = \exp\left( \frac{\delta_m}{m'}+\frac{\delta_m}{m'-\delta_m}\right)
\ee

The distribution of the secondary mass is modelled as a power law,
\be
    p_{\mathrm{ABH}}^{m_2}(m_2|m_1,\beta,\mmin) \propto \left(\frac{m_2}{m_1} \right)^\beta \,,
\ee
where the normalization ensures that the secondary mass is bounded by $\mmin \leq m_2 \leq m_1$. Finally, the redshift evolution considered is 
\be
    p_{\mathrm{ABH}}^z(z|\kappa) \propto (1+z)^\kappa\,,
\ee
where we leave $\kappa$ as a free parameter during inference, unlike in Ref.~\cite{Franciolini:2022tfm} where it was fixed to $\kappa = 2.9$ corresponding to the best fit value in~\cite{KAGRA:2021duu}.

\section{Results}\label{sec:results}

\begin{figure}
    \centering
    \includegraphics[width=0.98\columnwidth]{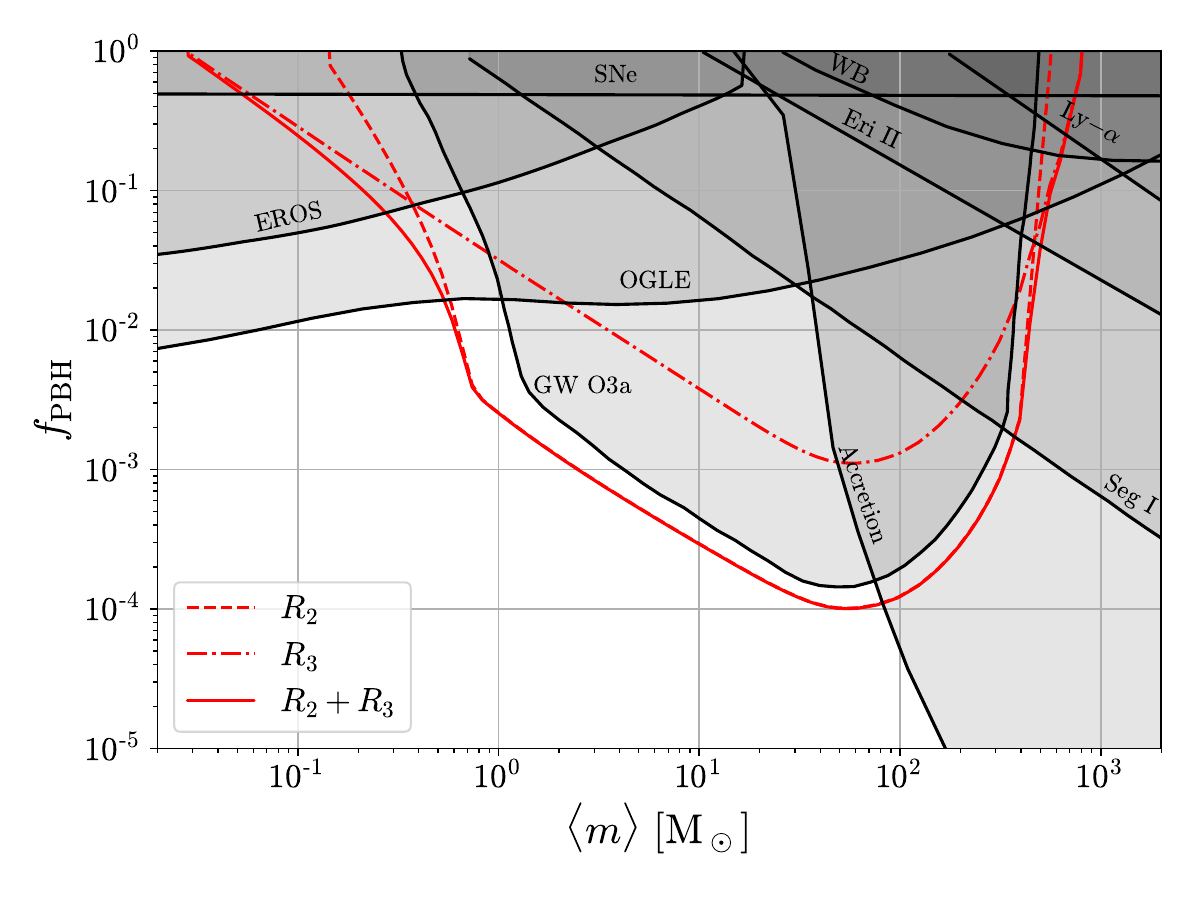}
    \caption{ \justifying Constraints for the monochromatic mass function detailing the contribution from $R_{\rm PBH,2}$ and $R_{\rm PBH,3}$ assuming that none of the observed events has a primordial origin. The constraints shown in gray are described in the main text and assume a monochromatic mass function.}
    \label{fig:monofpbh}
\end{figure}

In the following, we will consider pessimistic and optimistic scenarios. In the pessimistic case, none of the observed events has a primordial origin. This scenario gives the most stringent constraints. In the optimistic case, we consider the possibility that a subset of the observed GW events were due to PBH mergers. We will approach this possibility in two ways: 1) fitting the data to a model that contains a mixed population of ABHs and PBHs and 2) fitting all different subsets of events with a PBH population. Both approaches have their advantages. The first approach makes it possible to impose specific characteristics of ABH binaries, which are, however, highly uncertain at present. The second method is completely agnostic about the ABH population and will thus result in the most conservative constraints.

\begin{figure*}
    \centering
    \includegraphics[width=0.95\columnwidth]{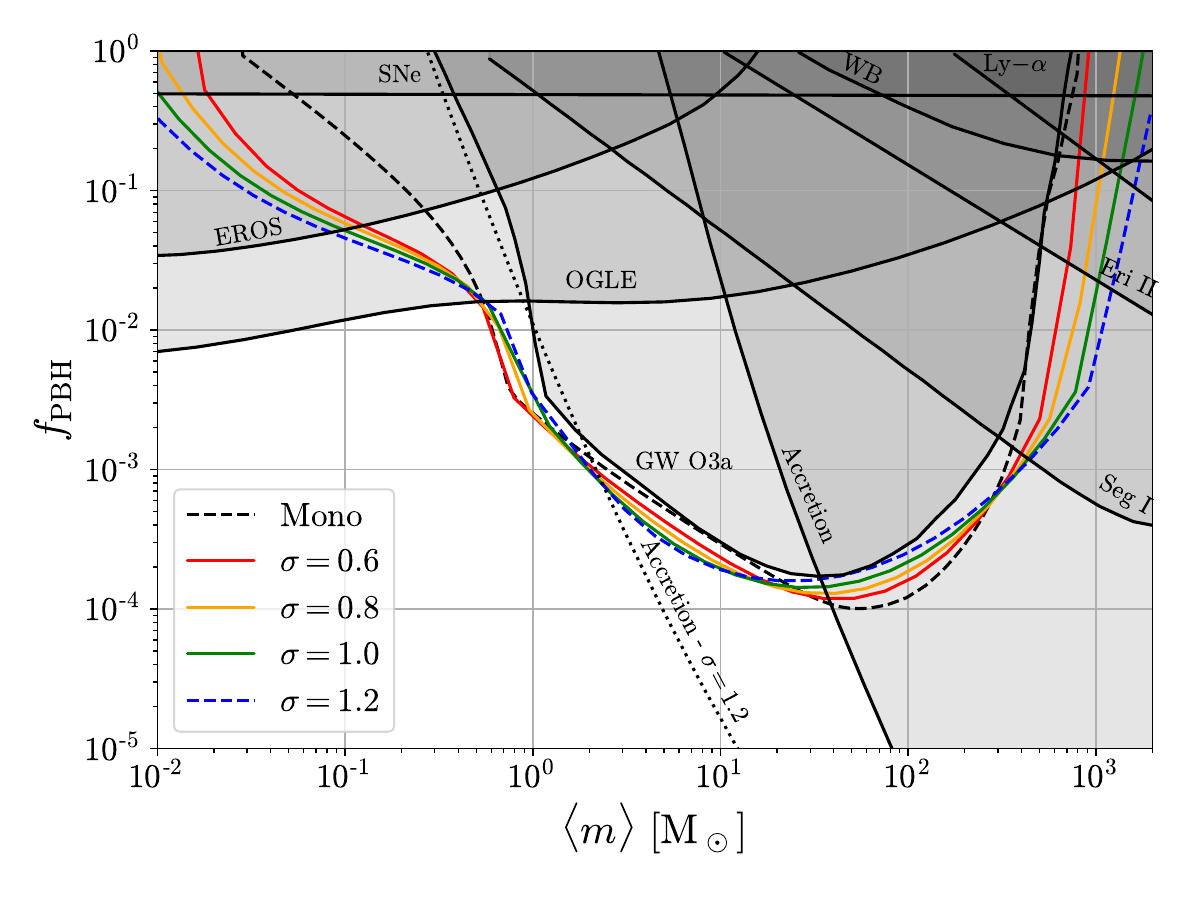}\quad
    \includegraphics[width=0.95\columnwidth]{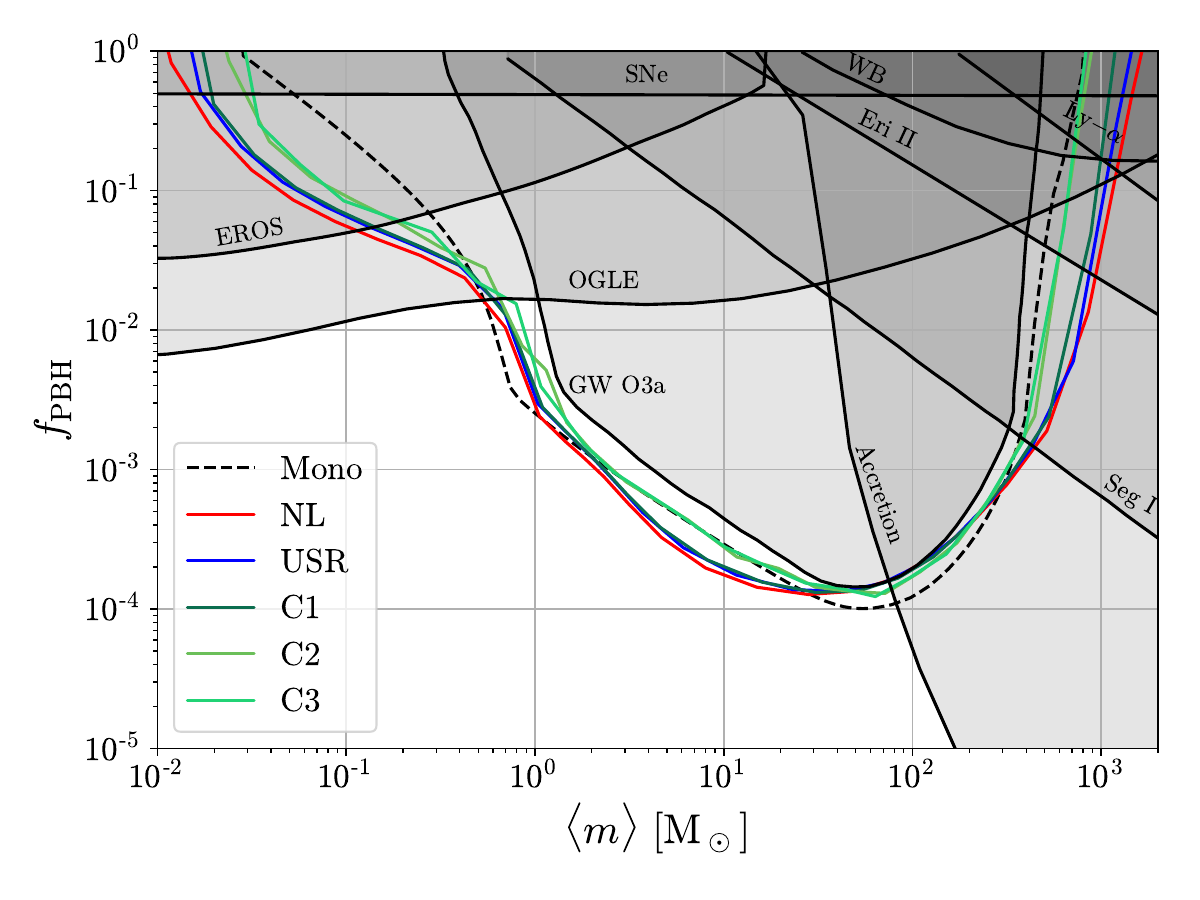}
    \caption{\justifying Same as Fig.~\eqref{fig:monofpbh}, but for log-normal mass function \emph{(left panel)} and critical collapse mass functions \emph{(right panel)}. Parameters for the latter are reported in Table~\ref{tab:par}. The black dashed curve shows the constraint for a monochromatic mass function for comparison. The grey constraints are shown for a log-normal power spectrum with $\sigma=0.6$. The dotted line on the left panel shows, the accretion constraint for $\sigma=1.2$.}
    \label{fig:extfpbh}
\end{figure*}

\subsection{Scenarios without primordial events}

Consider first the pessimistic case in which all events observed by LVK have an astrophysical origin. Although there is presently no evidence to indicate otherwise~\cite{KAGRA:2021duu,KAGRA:2021vkt}, this analysis assesses the current sensitivity of LVK to PBH populations in various PBH scenarios instead of constraining their abundance.

The limits that can be established in this way are shown in Figs.~\ref{fig:monofpbh} and~\ref{fig:extfpbh}. As expected, more observing time and a lower strain noise improve the existing limits obtained from the O3a data in the same region of the parameter space. In particular, for all of the considered mass functions, the new analysis reduces the window next to $\mathcal{O}(1)$ $\msun$ compared to O3a results~\cite{Hutsi:2020sol}. In Fig.~\ref{fig:monofpbh}, considering a monochromatic mass function, we show the contributions from the two- and the three-body channels. The latter begins to dominate when $f_{\rm PBH} > 0.1$ and strengthens the bounds at the edges of the constrained region. 

The other constraints reported here are GW O3a~\cite{Hutsi:2020sol}, EROS~\cite{EROS-2:2006ryy}, OGLE~\cite{Mroz:2024mse,Mroz:2024wag}, Seg1~\cite{Koushiappas:2017chw}, Planck~\cite{Serpico:2020ehh}~\footnote{The bounds arising from CMB observations may be relaxed under more conservative assumptions about accretion physics~\cite{Agius:2024ecw, Facchinetti:2022kbg}.}, Eri II~\cite{Brandt:2016aco}, WB~\cite{Monroy-Rodriguez:2014ula}, Ly$-\alpha$~\cite{Murgia:2019duy} and SNe~\cite{Zumalacarregui:2017qqd}. We report the more stringent constraints. For a more comprehensive overview, see e.g. the review~\cite{Carr:2020gox}.

The limits for extended mass functions are shown in Fig.~\ref{fig:extfpbh}, assuming that all events have an astrophysical origin. The right panel shows the limits for the log-normal mass function with various widths. For small $\sigma$ the limit agrees with the monochromatic case. As illustrated in Fig.~\ref{fig:monofpbh}, the constrained region in the sub-solar mass range arises mostly from the three-body channel. In Fig.~\ref{fig:extfpbh}, we see that this tail is more sensitive to the width of the mass function than the constraint curves above the solar mass, which arise mostly from the two-body channel. 

The left panel of Fig.~\ref{fig:extfpbh}, shows the constraints obtained for the critical collapse mass functions in different models of non-Gaussianities listed in Table.~\ref{tab:par}. We have considered three scenarios: non-Gaussianities arising only due to the non-linear relation~\eqref{eq:NL} between the curvature perturbation and the density contrast, and models with additional primordial non-Gaussianities in the curvature perturbation arising in quasi-inflection-point models~\eqref{eq:zeta_IP} and curvaton models~\eqref{eq:zeta_cur}. Since the mass function depends mildly on the PBH abundance, we consider benchmark cases with $f_{\rm PBH}\simeq 10^{-3}$. 

Crucially, we find that the constraints are nearly identical in all cases considered and match approximately the constraints obtained in the monochromatic case. In particular, when comparing the constraints for log-normal and critical collapse mass functions, we do not find that the enhancement of mass spectra around $1 \msun$ due to the QCD phase transition has a noticeable effect on the constraints. We also find the impact of non-Gaussianities to be quite mild.

We remark that the mass functions considered here are relatively narrow as the span at most about 2 orders of magnitude. For much wider mass functions spanning several orders of magnitude, the bulk of the events would not fall into the LVK mass range and thus the constraints on $f_{\rm PBH}$ would be less stringent. On top of that, the available merger rate estimates have been derived assuming that the PBHs have masses of a similar order, thus the theoretical uncertainties are expected to increase for extremely wide mass functions. On the other hand, one must also consider that all constraints are affected by the shape of the mass function~\cite{Carr:2017jsz}. Compare, for example, the gray constraints in Figs.~\ref{fig:monofpbh} and~\ref{fig:extfpbh} corresponding to monochromatic and the relatively narrow log-normal mass function with $\sigma=0.6$. One can see that the constraint from accretion has moved towards lower mean masses. As shown by the dashed line in Fig.~\ref{fig:extfpbh}, for $\sigma=1.2$, the constraint from accretion will exclude most of the parameter space accessible by LVK. Thus, PBH scenarios with wider mass functions (for instance, with $\sigma > 1.2$) are less relevant for LVK as other observables such as accretion exclude the accessible PBH mass range.

\begin{table}
    \centering
    \begin{tabular}{p{3.cm}p{2.4cm}p{1.2cm}p{1.4cm}}
         \hline \hline
         &  $\log_{10} A$ &  $\beta$& $r_{\rm dec}$\\
         \hline
         only NL&  $[-1.9,-1.8]$& 3.0 & - \\
         USR&  $[-2.3,-2.2]$&  3.0& - \\
         curvaton (C1)&  $[-2.8,-2.7]$&  0.5& $0.1$\\
         curvaton (C2)&  $[-2.5,-2.4]$&  3.0& $0.1$\\
         curvaton (C3)&  $[-1.8,-1.7]$&  3.0& $0.9$\\
         \hline \hline
    \end{tabular}
    \caption{\justifying Parameters used in the right panel of Fig.~\ref{fig:extfpbh}. We fix $\alpha=4$ and choose the amplitude $A$ (with one digit precision) in each case so that $f_{\rm PBH}\approx 10^{-3}$. The amplitude depends mildly on $k_{*}$, so we list the range of $A$ corresponding to $f_{\rm PBH}\approx 10^{-3}$ in the range $k_{*} \in [10^4,10^8] {\rm Mpc}^{-1}$. }
    \label{tab:par}
\end{table}

Enhanced primordial power spectra, responsible for the PBH scenarios considered in the left panel of Fig.~\ref{fig:extfpbh}, will also source a scalar-induced gravitational wave background (SIGW) at higher orders of perturbation theory~\cite{Tomita:1975kj, Matarrese:1993zf, Acquaviva:2002ud, Mollerach:2003nq, Ananda:2006af, Baumann:2007zm} (for a recent review see~\cite{Domenech:2021ztg}). In particular, stellar mass PBHs produced from critical collapse are related to a SIGW close to the frequency range probed by pulsar timing array (PTA) experiments. The recent evidence of GWs from PTA observations~\cite{NANOGrav:2023gor, EPTA:2023fyk, Reardon:2023gzh, Xu:2023wog} might be related to sub-solar mass PBH scenarios~\cite{Franciolini:2023pbf,Ellis:2023oxs}, to which LVK is less sensitive. Moreover, the connection between the PBH abundance and the related SIGW depends on spectral shape and the non-Gaussianities~\cite{Franciolini:2023pbf} as well as non-linear contributions to PBH formation~\cite{DeLuca:2023tun}. For this reason, Fig.~\ref{fig:extfpbh} does not include constraints that may potentially arise from PTA observations~\cite{Iovino:2024uxp}.

\subsection{Scenarios combining primordial and astrophysical events}

\begin{figure*}
    \centering
    \includegraphics[width=\textwidth]{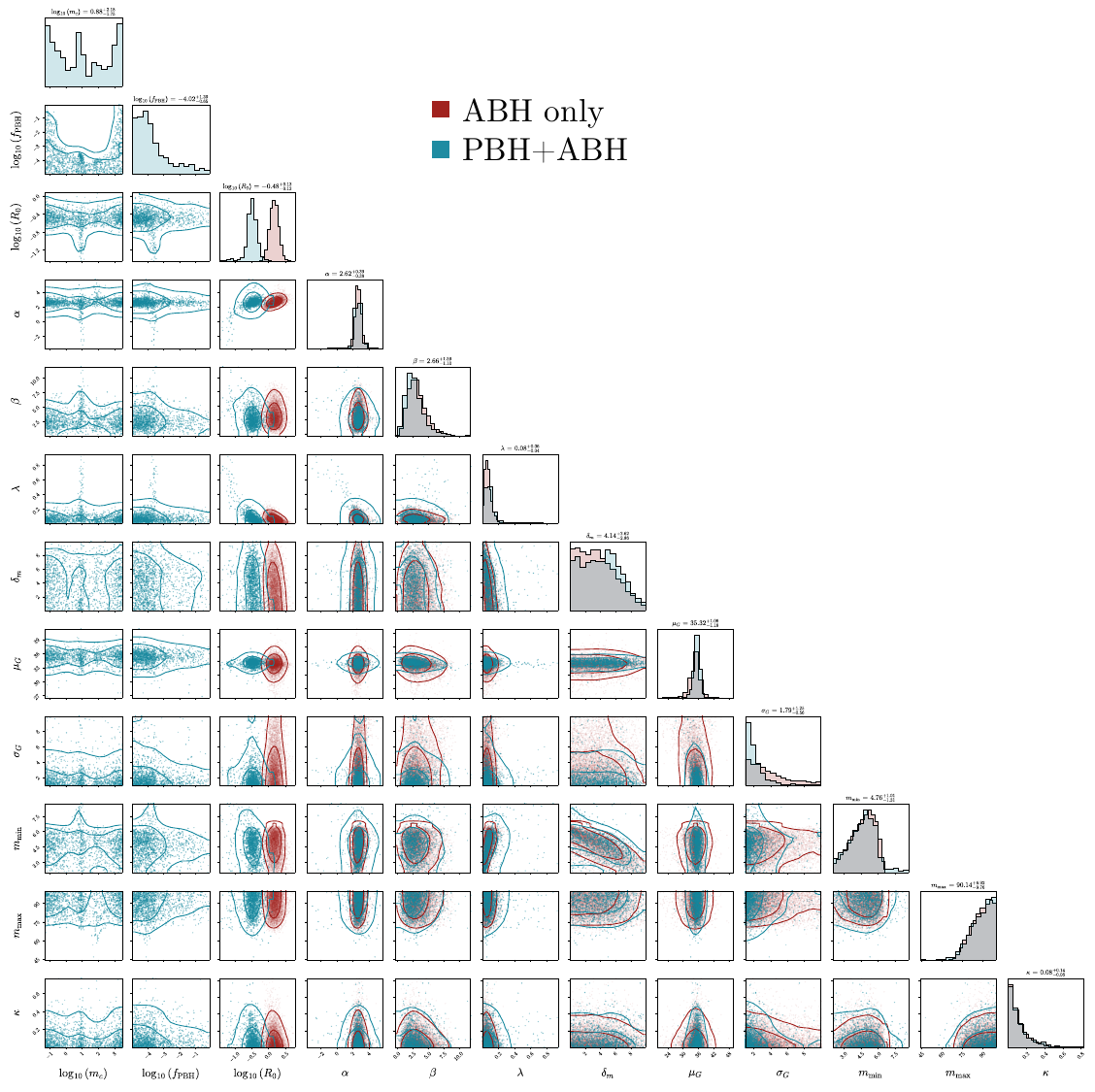}
    \caption{\justifying Posteriors for the BH binary merger rate model combining PBH and ABH binaries (blue) and for the model containing only ABH binaries (red). The PBH binaries are assumed to have a log-normal mass function with width $\sigma=0.6$ and the ABH binaries are described by a phenomenological POWER-LAW + PEAK model given in section~\ref{sec:ABHs}. All masses are given in units of $\msun$ and $R_0$ in $\rm Gpc^{-3}yr^{-1}$. }
    \label{fig:PBHplusABH_LN_PP_sig06}
\end{figure*}

Allowing for the possibility, that some of the observed GW events have a primordial origin, does generally soften the constraints. In this case, two approaches are followed: a model-dependent and an agnostic one. For this work, we consider all the events that have a signal-to-noise ratio larger than $11$ and an inverse false alarm rate more than $4$ years. This ensures the usage of highly confident detections only.

\begin{figure}[tbp]
    \centering
    \includegraphics[width=\columnwidth]{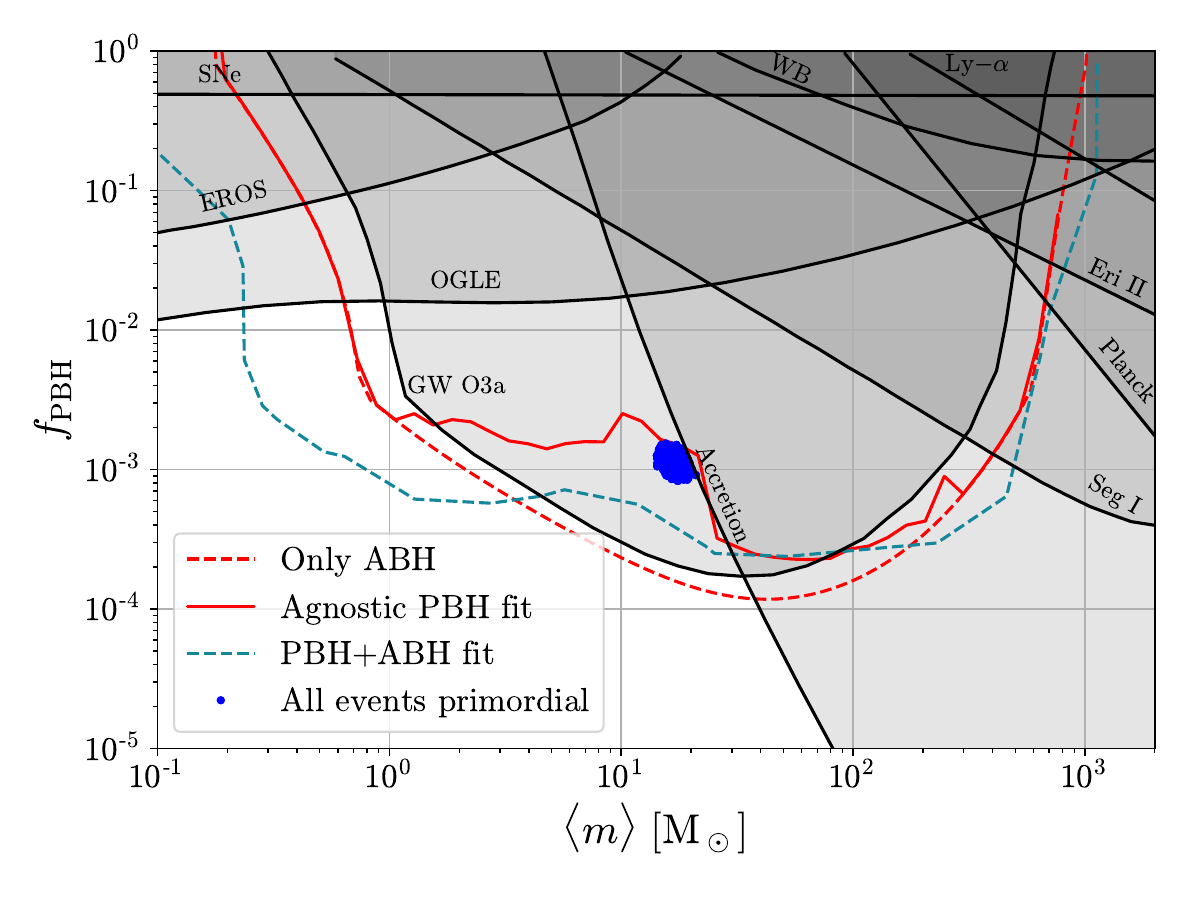}
    \caption{\justifying Constraints for a log-normal mass function with $\sigma = 0.6$ assuming that an arbitrary fraction of events can be primordial (solid red line) and that only ABH binaries have been observed (dashed red line). The dotted blue line shows the constraints from the fit of the PBH+ABH model (from Fig.~\ref{fig:PBHplusABH_LN_PP_sig06}). The blue shaded area indicates the $2\sigma$ confidence region in the scenario where all events are primordial. The contribution from the three-body channel is omitted here.}
    \label{fig:LN06}
\end{figure}

The results are displayed in Fig.~\ref{fig:LN06} for a log-normal mass function with a fixed width of $\sigma=0.6$. As expected, the constraints are weakened compared to scenarios assuming no primordial events. 

As shown in the previous section, the constraints are only mildly dependent on the shape of the mass function and, analogously to Figs.~\ref {fig:extfpbh} and~\ref{fig:monofpbh}, we expect that very similar constraints will hold for critical collapse mass functions. We neglected the contribution of the three-body formation channel~\eqref{eq:R3} in this analysis since, as was established in the previous subsection, it is mildly relevant only in a region of parameter space constrained by other observations.

In the case where an ABH population is modelled explicitly on top of a PBH one, the parameters defining both ABH and PBH binary populations must be fitted simultaneously. Such a fit is shown in Fig.~\ref{fig:PBHplusABH_LN_PP_sig06} using a nested sampling algorithm to scan the parameter space and the priors specified in Table~. It used the phenomenological POWER LAW+PEAK model for the ABH population and the log-normal mass function with $\sigma=0.6$ for the primordial one. As a consistency check, we also performed the ABH population fit without PBHs (see Fig.~\ref{fig:PBHplusABH_LN_PP_sig06}) and found that it agrees with existing LVK results~\cite{KAGRA:2021duu}. The parameters of the ABH population in the combined ABH+PBH fit were relatively similar to the ABH-only fit. One relevant difference was that the base rate of the ABH merger rate (see Eq.~\eqref{eq:R_ABH}) in the fit that included PBHs, $\log_{10} R_0 = -0.48(13)$, was reduced when compared to the ABH only fit $\log_{10} R_0 = -0.16(13)$, indicating that the combined fit prefers scenarios containing PBH subpopulations.

\begin{table}[htbp]
	\centering
	\begin{tabular}{wc{2cm}  wc{4.5cm} wc{1.5cm}} 
		\hline \hline
		\textbf{Parameter} & \textbf{Prior} & \textbf{Units}\\
		\hline \hline
            $R_0$ & LogU$(10^{-2},10^3)$ & Gpc$^{-3}$yr$^{-1}$\\ 
		$\alpha$ & U$(-4,12)$ & - \\ 
            $\beta$ & U$(-4,12)$ & - \\ 
            $\mmin$ & U$(2,10)$ & $\msun$ \\ 
            $\mmax$ & U$(40,100)$ & $\msun$ \\ 
            $\delta_m$ & U$(0,10)$ & $\msun$ \\ 
            $\mu_G$ & U$(20,50)$ & $\msun$ \\ 
            $\sigma_G$ & U$(1,10)$ & $\msun$ \\ 
            $\lambda$ & U$(0,1)$ & - \\
            $\kappa$ & U$(0,5)$ & - \\
            $\fpbh$ & LogU$(10^{-5},1)$ & - \\
            $m_c$ & LogU$(10^{-1},10^3)$ & $\msun$ \\
            \hline\hline
	\end{tabular}
      \caption{Priors used for the PBH+ABH fit.}
      \label{tab:priors}
\end{table}

The $f_{\rm PBH}$ - $m_c$ projection of the posteriors on the top of in Fig.~\ref{fig:PBHplusABH_LN_PP_sig06} implies a constraint on the PBH abundance assuming an ABH binary population obeying the POWER LAW+PEAK model. It is consistent with the agnostic ones shown in Fig.~\ref{fig:LN06} except in the subsolar mass range, where it implies a stronger constraint. This is likely a result of insufficient resolution in the scan, because the most stringent constant is given by the red dashed line in Fig.~\ref{fig:PBHplusABH_LN_PP_sig06} corresponding to an expected $N=3$ primordial events. Below the dashed line, the expected number of observable PBH events is too small to yield a statistically significant constraint on the PBH abundance. These uncertainties in the fit to the ABH+PBH model illustrate a noticeable numerical drawback when compared to the agnostic approach, which was computationally much faster and more accurate than the scan of the ABH+PBH model -- we were not able to resolve the subsolar mass region with sufficient accuracy with the available computational resources.

\revComments{The (log-)Bayes factor for the various fits can be compared to see the most favoured scenario. These are summarized in Tab.~\ref{tab:logBayes}  for the log-normal mass function with a fixed width of $\sigma = 0.6$. The most favoured model is, in this case, the ABH-only scenario, while the least favoured one is the PBH-only. We stress that the aim of this analysis is not to look for PBHs in the data but to constrain them. The preference for the ABH model indicates that the chosen parametrization for the ABH foreground is general enough -- the implied constraints on the PBH abundance would arise for a wide range of potential ABH foregrounds. As a byproduct, we found that PBH-only scenarios are strongly disfavored compared to the scenarios that include ABH populations, which aligns with earlier studies~\cite{Hutsi:2020sol, Hall:2020daa, Wong:2020yig, Franciolini:2021tla, DeLuca:2021wjr}. 
}
 \begin{table}[tbp]
	\centering
        \begin{tabular}{ wl{3.cm}  wc{1.7cm} wc{1.7cm} wc{1.7cm}  }
		\hline \hline
		\textbf{Scenario} & $\ln\mathcal{B}$ & $f_{\rm PBH}$ & $\langle m \rangle$ \\
		\hline \hline
            ABH only    & $0$       & --    & --\\
            PBH and ABH & $-0.94$   & $9.5\times 10^{-5}$ & $6.34 \msun$ \\
            PBH only    & $-38.4$   & $1.1\times 10^{-3}$ & $16.8 \msun$ \\
            \hline\hline
	\end{tabular}
    \caption{\justifying Bayes factors and the best fit PBH parameters for the various fits compared to the ABH-only fit. A log-normal PBH mass function with $\sigma = 0.6$ is assumed.}
    \label{tab:logBayes}
\end{table}

\section{Conclusions}
\label{sec:conc}

We have updated the constraints on PBHs using data from the third observational run (O3) of the LVK experiment by adapting a hierarchical Bayesian approach. Overall, we have found that in the mass range $1-300\msun$, the PBH abundance is constrained $f_{\rm PBH} \lesssim 10^{-3}$ in all scenarios considered.

We have shown that the constraints are insensitive to the detailed shape of the mass function and the order of magnitude of these constraints can be captured already by the monochromatic mass function. As a specific and theoretically well-motivated example, we have considered mass functions generated by the critical collapse of primordial inhomogeneities. We have included the effect of the QCD phase transition, which enhances the mass function around $1 \msun$, as well as the effect of non-Gaussianities. Although both phenomena have a non-negligible impact on the shape of the mass function, we have found them to have a minor impact on the constraints.

To account for the foreground of ABH binaries, we have followed two approaches. First, in the agnostic approach, which does not require an explicit model for ABH binaries, we have allowed for the possibility that any subset of the observed BH-BH merger events could have a primordial origin. Second, we have performed a fit in a mixed population of primordial and astrophysical BH binaries, with the latter described by the phenomenological POWER-LAW + PEAK model. Both approaches yielded a similar constraint on $f_{\rm PBH}$ indicating that the constraints on PBHs are relatively insensitive to the modelling of ABH binaries.

The analysis included the contribution from PBH binaries formed from three-body systems in the early universe. Behind the early universe two-body channel, it gives generally the second strongest contribution and can dominate when $f_{\rm PBH} \gtrsim 0.1$. Consistently with this expectation, we have found that including this channel mildly widens the constrained region. Overall, including the three-body PBH binary formation channel affects the LVK constraints in parameter regions already excluded by other observations, and does thus not have a significant impact on the current PBH constraints.

We have considered PBH scenarios with mass functions spanning at most a few orders of magnitude in masses and without significant initial clustering. However, we have argued that even with these assumptions relaxed, combining LVK mergers with other observables, the abundance of stellar mass PBHs remains strongly constrained. In particular, we demonstrated that CMB observations, \revComments{without the most conservative assumptions about the accretion physics}, exclude stellar mass PBHs with very wide mass functions. \revComments{Similarly, while the merger rate estimates account for the inherent Poisson clustering of PBHs, they do not include the effect of potential initial clustering. However, strong initial clustering is in tension with Lyman-$\alpha$ observations~\cite{DeLuca:2022uvz} and, while mild initial clustering can modify the constraints by weakening the two-body merger rate and strengthening the three-body merger rate, it can not remove them. Further research is needed to quantify more precisely how very wide mass functions and initial clustering can alter the PBH merger rate and the resulting constraints.}

\vspace{5mm}
\begin{acknowledgments}
We thank Gabriele Franciolini and Juan Urrutia for useful comments. M.A.C. is supported by the 2022 FI-00335 grant by the Generalitat de Catalunya. A.J.I. acknowledges the financial support provided under the ``Progetti per Avvio alla Ricerca Tipo 1", protocol number AR1231886850F568. V.V. and H.V. are supported by the Estonian Research Council grants PRG803, PSG869, RVTT3 and RVTT7 and the Center of Excellence program TK202. V.V. is also supported by the European Union's Horizon Europe research and innovation program under the Marie Sk\l{}odowska-Curie grant agreement No. 101065736. The corner plots use the \textit{corner.py} library \cite{corner}. This research has made use of data, software, and/or web tools obtained from the Gravitational Wave Open Science Center (\url{https://www.gw-openscience.org/}), a service of the LIGO Laboratory, the LIGO Scientific Collaboration, and the Virgo Collaboration. This material is based upon work supported by NSF's LIGO Laboratory which is a major facility fully funded by the National Science Foundation. We acknowledge the support from the Departament de Recerca i Universitats from Generalitat de Catalunya to the Grup de Recerca ‘Grup de Fisica Teorica UAB/IFAE’ (Code: 2021 SGR 00649) and the Spanish Ministry of Science and Innovation (PID2020-115845GB- I00/AEI/10.13039/501100011033). IFAE is partially funded by the CERCA program of the Generalitat de Catalunya. 
\end{acknowledgments}

\bibliography{ref}

\begin{thebibliography}{227}%
\makeatletter
\providecommand \@ifxundefined [1]{%
 \@ifx{#1\undefined}
}%
\providecommand \@ifnum [1]{%
 \ifnum #1\expandafter \@firstoftwo
 \else \expandafter \@secondoftwo
 \fi
}%
\providecommand \@ifx [1]{%
 \ifx #1\expandafter \@firstoftwo
 \else \expandafter \@secondoftwo
 \fi
}%
\providecommand \natexlab [1]{#1}%
\providecommand \enquote  [1]{``#1''}%
\providecommand \bibnamefont  [1]{#1}%
\providecommand \bibfnamefont [1]{#1}%
\providecommand \citenamefont [1]{#1}%
\providecommand \href@noop [0]{\@secondoftwo}%
\providecommand \href [0]{\begingroup \@sanitize@url \@href}%
\providecommand \@href[1]{\@@startlink{#1}\@@href}%
\providecommand \@@href[1]{\endgroup#1\@@endlink}%
\providecommand \@sanitize@url [0]{\catcode `\\12\catcode `\$12\catcode `\&12\catcode `\#12\catcode `\^12\catcode `\_12\catcode `\%12\relax}%
\providecommand \@@startlink[1]{}%
\providecommand \@@endlink[0]{}%
\providecommand \url  [0]{\begingroup\@sanitize@url \@url }%
\providecommand \@url [1]{\endgroup\@href {#1}{\urlprefix }}%
\providecommand \urlprefix  [0]{URL }%
\providecommand \Eprint [0]{\href }%
\providecommand \doibase [0]{http://dx.doi.org/}%
\providecommand \selectlanguage [0]{\@gobble}%
\providecommand \bibinfo  [0]{\@secondoftwo}%
\providecommand \bibfield  [0]{\@secondoftwo}%
\providecommand \translation [1]{[#1]}%
\providecommand \BibitemOpen [0]{}%
\providecommand \bibitemStop [0]{}%
\providecommand \bibitemNoStop [0]{.\EOS\space}%
\providecommand \EOS [0]{\spacefactor3000\relax}%
\providecommand \BibitemShut  [1]{\csname bibitem#1\endcsname}%
\let\auto@bib@innerbib\@empty
\bibitem [{\citenamefont {Abbott}\ \emph {et~al.}(2016{\natexlab{a}})\citenamefont {Abbott} \emph {et~al.}}]{LIGOScientific:2016aoc}%
  \BibitemOpen
  \bibfield  {author} {\bibinfo {author} {\bibfnamefont {B.~P.}\ \bibnamefont {Abbott}} \emph {et~al.} (\bibinfo {collaboration} {LIGO Scientific, Virgo}),\ }\href {\doibase 10.1103/PhysRevLett.116.061102} {\bibfield  {journal} {\bibinfo  {journal} {Phys. Rev. Lett.}\ }\textbf {\bibinfo {volume} {116}},\ \bibinfo {pages} {061102} (\bibinfo {year} {2016}{\natexlab{a}})},\ \Eprint {http://arxiv.org/abs/1602.03837} {arXiv:1602.03837 [gr-qc]} \BibitemShut {NoStop}%
\bibitem [{\citenamefont {Abbott}\ \emph {et~al.}(2019{\natexlab{a}})\citenamefont {Abbott} \emph {et~al.}}]{LIGOScientific:2018mvr}%
  \BibitemOpen
  \bibfield  {author} {\bibinfo {author} {\bibfnamefont {B.~P.}\ \bibnamefont {Abbott}} \emph {et~al.} (\bibinfo {collaboration} {LIGO Scientific, Virgo}),\ }\href {\doibase 10.1103/PhysRevX.9.031040} {\bibfield  {journal} {\bibinfo  {journal} {Phys. Rev. X}\ }\textbf {\bibinfo {volume} {9}},\ \bibinfo {pages} {031040} (\bibinfo {year} {2019}{\natexlab{a}})},\ \Eprint {http://arxiv.org/abs/1811.12907} {arXiv:1811.12907 [astro-ph.HE]} \BibitemShut {NoStop}%
\bibitem [{\citenamefont {Abbott}\ \emph {et~al.}(2021{\natexlab{a}})\citenamefont {Abbott} \emph {et~al.}}]{LIGOScientific:2020ibl}%
  \BibitemOpen
  \bibfield  {author} {\bibinfo {author} {\bibfnamefont {R.}~\bibnamefont {Abbott}} \emph {et~al.} (\bibinfo {collaboration} {LIGO Scientific, Virgo}),\ }\href {\doibase 10.1103/PhysRevX.11.021053} {\bibfield  {journal} {\bibinfo  {journal} {Phys. Rev. X}\ }\textbf {\bibinfo {volume} {11}},\ \bibinfo {pages} {021053} (\bibinfo {year} {2021}{\natexlab{a}})},\ \Eprint {http://arxiv.org/abs/2010.14527} {arXiv:2010.14527 [gr-qc]} \BibitemShut {NoStop}%
\bibitem [{\citenamefont {Abbott}\ \emph {et~al.}(2023{\natexlab{a}})\citenamefont {Abbott} \emph {et~al.}}]{KAGRA:2021vkt}%
  \BibitemOpen
  \bibfield  {author} {\bibinfo {author} {\bibfnamefont {R.}~\bibnamefont {Abbott}} \emph {et~al.} (\bibinfo {collaboration} {KAGRA, VIRGO, LIGO Scientific}),\ }\href {\doibase 10.1103/PhysRevX.13.041039} {\bibfield  {journal} {\bibinfo  {journal} {Phys. Rev. X}\ }\textbf {\bibinfo {volume} {13}},\ \bibinfo {pages} {041039} (\bibinfo {year} {2023}{\natexlab{a}})},\ \Eprint {http://arxiv.org/abs/2111.03606} {arXiv:2111.03606 [gr-qc]} \BibitemShut {NoStop}%
\bibitem [{\citenamefont {Abbott}\ \emph {et~al.}(2023{\natexlab{b}})\citenamefont {Abbott} \emph {et~al.}}]{KAGRA:2021duu}%
  \BibitemOpen
  \bibfield  {author} {\bibinfo {author} {\bibfnamefont {R.}~\bibnamefont {Abbott}} \emph {et~al.} (\bibinfo {collaboration} {KAGRA, VIRGO, LIGO Scientific}),\ }\href {\doibase 10.1103/PhysRevX.13.011048} {\bibfield  {journal} {\bibinfo  {journal} {Phys. Rev. X}\ }\textbf {\bibinfo {volume} {13}},\ \bibinfo {pages} {011048} (\bibinfo {year} {2023}{\natexlab{b}})},\ \Eprint {http://arxiv.org/abs/2111.03634} {arXiv:2111.03634 [astro-ph.HE]} \BibitemShut {NoStop}%
\bibitem [{\citenamefont {Mandel}\ and\ \citenamefont {Farmer}(2022)}]{Mandel:2018hfr}%
  \BibitemOpen
  \bibfield  {author} {\bibinfo {author} {\bibfnamefont {I.}~\bibnamefont {Mandel}}\ and\ \bibinfo {author} {\bibfnamefont {A.}~\bibnamefont {Farmer}},\ }\href {\doibase 10.1016/j.physrep.2022.01.003} {\bibfield  {journal} {\bibinfo  {journal} {Phys. Rept.}\ }\textbf {\bibinfo {volume} {955}},\ \bibinfo {pages} {1} (\bibinfo {year} {2022})},\ \Eprint {http://arxiv.org/abs/1806.05820} {arXiv:1806.05820 [astro-ph.HE]} \BibitemShut {NoStop}%
\bibitem [{\citenamefont {Woosley}\ \emph {et~al.}(2002)\citenamefont {Woosley}, \citenamefont {Heger},\ and\ \citenamefont {Weaver}}]{Woosley:2002zz}%
  \BibitemOpen
  \bibfield  {author} {\bibinfo {author} {\bibfnamefont {S.~E.}\ \bibnamefont {Woosley}}, \bibinfo {author} {\bibfnamefont {A.}~\bibnamefont {Heger}}, \ and\ \bibinfo {author} {\bibfnamefont {T.~A.}\ \bibnamefont {Weaver}},\ }\href {\doibase 10.1103/RevModPhys.74.1015} {\bibfield  {journal} {\bibinfo  {journal} {Rev. Mod. Phys.}\ }\textbf {\bibinfo {volume} {74}},\ \bibinfo {pages} {1015} (\bibinfo {year} {2002})}\BibitemShut {NoStop}%
\bibitem [{\citenamefont {Carr}\ and\ \citenamefont {Hawking}(1974)}]{Carr:1974nx}%
  \BibitemOpen
  \bibfield  {author} {\bibinfo {author} {\bibfnamefont {B.~J.}\ \bibnamefont {Carr}}\ and\ \bibinfo {author} {\bibfnamefont {S.~W.}\ \bibnamefont {Hawking}},\ }\href {\doibase 10.1093/mnras/168.2.399} {\bibfield  {journal} {\bibinfo  {journal} {Mon. Not. Roy. Astron. Soc.}\ }\textbf {\bibinfo {volume} {168}},\ \bibinfo {pages} {399} (\bibinfo {year} {1974})}\BibitemShut {NoStop}%
\bibitem [{\citenamefont {Carr}(1975)}]{Carr:1975qj}%
  \BibitemOpen
  \bibfield  {author} {\bibinfo {author} {\bibfnamefont {B.~J.}\ \bibnamefont {Carr}},\ }\href {\doibase 10.1086/153853} {\bibfield  {journal} {\bibinfo  {journal} {Astrophys. J.}\ }\textbf {\bibinfo {volume} {201}},\ \bibinfo {pages} {1} (\bibinfo {year} {1975})}\BibitemShut {NoStop}%
\bibitem [{\citenamefont {Ivanov}\ \emph {et~al.}(1994)\citenamefont {Ivanov}, \citenamefont {Naselsky},\ and\ \citenamefont {Novikov}}]{Ivanov:1994pa}%
  \BibitemOpen
  \bibfield  {author} {\bibinfo {author} {\bibfnamefont {P.}~\bibnamefont {Ivanov}}, \bibinfo {author} {\bibfnamefont {P.}~\bibnamefont {Naselsky}}, \ and\ \bibinfo {author} {\bibfnamefont {I.}~\bibnamefont {Novikov}},\ }\href {\doibase 10.1103/PhysRevD.50.7173} {\bibfield  {journal} {\bibinfo  {journal} {Phys. Rev. D}\ }\textbf {\bibinfo {volume} {50}},\ \bibinfo {pages} {7173} (\bibinfo {year} {1994})}\BibitemShut {NoStop}%
\bibitem [{\citenamefont {Garcia-Bellido}\ \emph {et~al.}(1996)\citenamefont {Garcia-Bellido}, \citenamefont {Linde},\ and\ \citenamefont {Wands}}]{Garcia-Bellido:1996mdl}%
  \BibitemOpen
  \bibfield  {author} {\bibinfo {author} {\bibfnamefont {J.}~\bibnamefont {Garcia-Bellido}}, \bibinfo {author} {\bibfnamefont {A.~D.}\ \bibnamefont {Linde}}, \ and\ \bibinfo {author} {\bibfnamefont {D.}~\bibnamefont {Wands}},\ }\href {\doibase 10.1103/PhysRevD.54.6040} {\bibfield  {journal} {\bibinfo  {journal} {Phys. Rev. D}\ }\textbf {\bibinfo {volume} {54}},\ \bibinfo {pages} {6040} (\bibinfo {year} {1996})},\ \Eprint {http://arxiv.org/abs/astro-ph/9605094} {arXiv:astro-ph/9605094} \BibitemShut {NoStop}%
\bibitem [{\citenamefont {Ivanov}(1998)}]{Ivanov:1997ia}%
  \BibitemOpen
  \bibfield  {author} {\bibinfo {author} {\bibfnamefont {P.}~\bibnamefont {Ivanov}},\ }\href {\doibase 10.1103/PhysRevD.57.7145} {\bibfield  {journal} {\bibinfo  {journal} {Phys. Rev. D}\ }\textbf {\bibinfo {volume} {57}},\ \bibinfo {pages} {7145} (\bibinfo {year} {1998})},\ \Eprint {http://arxiv.org/abs/astro-ph/9708224} {arXiv:astro-ph/9708224} \BibitemShut {NoStop}%
\bibitem [{\citenamefont {Hawking}(1989)}]{Hawking:1987bn}%
  \BibitemOpen
  \bibfield  {author} {\bibinfo {author} {\bibfnamefont {S.~W.}\ \bibnamefont {Hawking}},\ }\href {\doibase 10.1016/0370-2693(89)90206-2} {\bibfield  {journal} {\bibinfo  {journal} {Phys. Lett. B}\ }\textbf {\bibinfo {volume} {231}},\ \bibinfo {pages} {237} (\bibinfo {year} {1989})}\BibitemShut {NoStop}%
\bibitem [{\citenamefont {Polnarev}\ and\ \citenamefont {Zembowicz}(1991)}]{Polnarev:1988dh}%
  \BibitemOpen
  \bibfield  {author} {\bibinfo {author} {\bibfnamefont {A.}~\bibnamefont {Polnarev}}\ and\ \bibinfo {author} {\bibfnamefont {R.}~\bibnamefont {Zembowicz}},\ }\href {\doibase 10.1103/PhysRevD.43.1106} {\bibfield  {journal} {\bibinfo  {journal} {Phys. Rev. D}\ }\textbf {\bibinfo {volume} {43}},\ \bibinfo {pages} {1106} (\bibinfo {year} {1991})}\BibitemShut {NoStop}%
\bibitem [{\citenamefont {Garriga}\ and\ \citenamefont {Sakellariadou}(1993)}]{Garriga:1993gj}%
  \BibitemOpen
  \bibfield  {author} {\bibinfo {author} {\bibfnamefont {J.}~\bibnamefont {Garriga}}\ and\ \bibinfo {author} {\bibfnamefont {M.}~\bibnamefont {Sakellariadou}},\ }\href {\doibase 10.1103/PhysRevD.48.2502} {\bibfield  {journal} {\bibinfo  {journal} {Phys. Rev. D}\ }\textbf {\bibinfo {volume} {48}},\ \bibinfo {pages} {2502} (\bibinfo {year} {1993})},\ \Eprint {http://arxiv.org/abs/hep-th/9303024} {arXiv:hep-th/9303024} \BibitemShut {NoStop}%
\bibitem [{\citenamefont {Caldwell}\ and\ \citenamefont {Casper}(1996)}]{Caldwell:1995fu}%
  \BibitemOpen
  \bibfield  {author} {\bibinfo {author} {\bibfnamefont {R.~R.}\ \bibnamefont {Caldwell}}\ and\ \bibinfo {author} {\bibfnamefont {P.}~\bibnamefont {Casper}},\ }\href {\doibase 10.1103/PhysRevD.53.3002} {\bibfield  {journal} {\bibinfo  {journal} {Phys. Rev. D}\ }\textbf {\bibinfo {volume} {53}},\ \bibinfo {pages} {3002} (\bibinfo {year} {1996})},\ \Eprint {http://arxiv.org/abs/gr-qc/9509012} {arXiv:gr-qc/9509012} \BibitemShut {NoStop}%
\bibitem [{\citenamefont {MacGibbon}\ \emph {et~al.}(1998)\citenamefont {MacGibbon}, \citenamefont {Brandenberger},\ and\ \citenamefont {Wichoski}}]{MacGibbon:1997pu}%
  \BibitemOpen
  \bibfield  {author} {\bibinfo {author} {\bibfnamefont {J.~H.}\ \bibnamefont {MacGibbon}}, \bibinfo {author} {\bibfnamefont {R.~H.}\ \bibnamefont {Brandenberger}}, \ and\ \bibinfo {author} {\bibfnamefont {U.~F.}\ \bibnamefont {Wichoski}},\ }\href {\doibase 10.1103/PhysRevD.57.2158} {\bibfield  {journal} {\bibinfo  {journal} {Phys. Rev. D}\ }\textbf {\bibinfo {volume} {57}},\ \bibinfo {pages} {2158} (\bibinfo {year} {1998})},\ \Eprint {http://arxiv.org/abs/astro-ph/9707146} {arXiv:astro-ph/9707146} \BibitemShut {NoStop}%
\bibitem [{\citenamefont {Hawking}\ \emph {et~al.}(1982)\citenamefont {Hawking}, \citenamefont {Moss},\ and\ \citenamefont {Stewart}}]{Hawking:1982ga}%
  \BibitemOpen
  \bibfield  {author} {\bibinfo {author} {\bibfnamefont {S.~W.}\ \bibnamefont {Hawking}}, \bibinfo {author} {\bibfnamefont {I.~G.}\ \bibnamefont {Moss}}, \ and\ \bibinfo {author} {\bibfnamefont {J.~M.}\ \bibnamefont {Stewart}},\ }\href {\doibase 10.1103/PhysRevD.26.2681} {\bibfield  {journal} {\bibinfo  {journal} {Phys. Rev. D}\ }\textbf {\bibinfo {volume} {26}},\ \bibinfo {pages} {2681} (\bibinfo {year} {1982})}\BibitemShut {NoStop}%
\bibitem [{\citenamefont {Kodama}\ \emph {et~al.}(1982)\citenamefont {Kodama}, \citenamefont {Sasaki},\ and\ \citenamefont {Sato}}]{Kodama:1982sf}%
  \BibitemOpen
  \bibfield  {author} {\bibinfo {author} {\bibfnamefont {H.}~\bibnamefont {Kodama}}, \bibinfo {author} {\bibfnamefont {M.}~\bibnamefont {Sasaki}}, \ and\ \bibinfo {author} {\bibfnamefont {K.}~\bibnamefont {Sato}},\ }\href {\doibase 10.1143/PTP.68.1979} {\bibfield  {journal} {\bibinfo  {journal} {Prog. Theor. Phys.}\ }\textbf {\bibinfo {volume} {68}},\ \bibinfo {pages} {1979} (\bibinfo {year} {1982})}\BibitemShut {NoStop}%
\bibitem [{\citenamefont {Garriga}\ \emph {et~al.}(2016)\citenamefont {Garriga}, \citenamefont {Vilenkin},\ and\ \citenamefont {Zhang}}]{Garriga:2015fdk}%
  \BibitemOpen
  \bibfield  {author} {\bibinfo {author} {\bibfnamefont {J.}~\bibnamefont {Garriga}}, \bibinfo {author} {\bibfnamefont {A.}~\bibnamefont {Vilenkin}}, \ and\ \bibinfo {author} {\bibfnamefont {J.}~\bibnamefont {Zhang}},\ }\href {\doibase 10.1088/1475-7516/2016/02/064} {\bibfield  {journal} {\bibinfo  {journal} {JCAP}\ }\textbf {\bibinfo {volume} {02}},\ \bibinfo {pages} {064} (\bibinfo {year} {2016})},\ \Eprint {http://arxiv.org/abs/1512.01819} {arXiv:1512.01819 [hep-th]} \BibitemShut {NoStop}%
\bibitem [{\citenamefont {Maeso}\ \emph {et~al.}(2022)\citenamefont {Maeso}, \citenamefont {Marzola}, \citenamefont {Raidal}, \citenamefont {Vaskonen},\ and\ \citenamefont {Veerm\"ae}}]{Maeso:2021xvl}%
  \BibitemOpen
  \bibfield  {author} {\bibinfo {author} {\bibfnamefont {D.~N.}\ \bibnamefont {Maeso}}, \bibinfo {author} {\bibfnamefont {L.}~\bibnamefont {Marzola}}, \bibinfo {author} {\bibfnamefont {M.}~\bibnamefont {Raidal}}, \bibinfo {author} {\bibfnamefont {V.}~\bibnamefont {Vaskonen}}, \ and\ \bibinfo {author} {\bibfnamefont {H.}~\bibnamefont {Veerm\"ae}},\ }\href {\doibase 10.1088/1475-7516/2022/02/017} {\bibfield  {journal} {\bibinfo  {journal} {JCAP}\ }\textbf {\bibinfo {volume} {02}},\ \bibinfo {pages} {017} (\bibinfo {year} {2022})},\ \Eprint {http://arxiv.org/abs/2112.01505} {arXiv:2112.01505 [astro-ph.CO]} \BibitemShut {NoStop}%
\bibitem [{\citenamefont {Lewicki}\ \emph {et~al.}(2023)\citenamefont {Lewicki}, \citenamefont {Toczek},\ and\ \citenamefont {Vaskonen}}]{Lewicki:2023ioy}%
  \BibitemOpen
  \bibfield  {author} {\bibinfo {author} {\bibfnamefont {M.}~\bibnamefont {Lewicki}}, \bibinfo {author} {\bibfnamefont {P.}~\bibnamefont {Toczek}}, \ and\ \bibinfo {author} {\bibfnamefont {V.}~\bibnamefont {Vaskonen}},\ }\href {\doibase 10.1007/JHEP09(2023)092} {\bibfield  {journal} {\bibinfo  {journal} {JHEP}\ }\textbf {\bibinfo {volume} {09}},\ \bibinfo {pages} {092} (\bibinfo {year} {2023})},\ \Eprint {http://arxiv.org/abs/2305.04924} {arXiv:2305.04924 [astro-ph.CO]} \BibitemShut {NoStop}%
\bibitem [{\citenamefont {Liu}\ \emph {et~al.}(2020)\citenamefont {Liu}, \citenamefont {Guo},\ and\ \citenamefont {Cai}}]{Liu:2019lul}%
  \BibitemOpen
  \bibfield  {author} {\bibinfo {author} {\bibfnamefont {J.}~\bibnamefont {Liu}}, \bibinfo {author} {\bibfnamefont {Z.-K.}\ \bibnamefont {Guo}}, \ and\ \bibinfo {author} {\bibfnamefont {R.-G.}\ \bibnamefont {Cai}},\ }\href {\doibase 10.1103/PhysRevD.101.023513} {\bibfield  {journal} {\bibinfo  {journal} {Phys. Rev. D}\ }\textbf {\bibinfo {volume} {101}},\ \bibinfo {pages} {023513} (\bibinfo {year} {2020})},\ \Eprint {http://arxiv.org/abs/1908.02662} {arXiv:1908.02662 [astro-ph.CO]} \BibitemShut {NoStop}%
\bibitem [{\citenamefont {Ferrer}\ \emph {et~al.}(2019)\citenamefont {Ferrer}, \citenamefont {Masso}, \citenamefont {Panico}, \citenamefont {Pujolas},\ and\ \citenamefont {Rompineve}}]{Ferrer:2018uiu}%
  \BibitemOpen
  \bibfield  {author} {\bibinfo {author} {\bibfnamefont {F.}~\bibnamefont {Ferrer}}, \bibinfo {author} {\bibfnamefont {E.}~\bibnamefont {Masso}}, \bibinfo {author} {\bibfnamefont {G.}~\bibnamefont {Panico}}, \bibinfo {author} {\bibfnamefont {O.}~\bibnamefont {Pujolas}}, \ and\ \bibinfo {author} {\bibfnamefont {F.}~\bibnamefont {Rompineve}},\ }\href {\doibase 10.1103/PhysRevLett.122.101301} {\bibfield  {journal} {\bibinfo  {journal} {Phys. Rev. Lett.}\ }\textbf {\bibinfo {volume} {122}},\ \bibinfo {pages} {101301} (\bibinfo {year} {2019})},\ \Eprint {http://arxiv.org/abs/1807.01707} {arXiv:1807.01707 [hep-ph]} \BibitemShut {NoStop}%
\bibitem [{\citenamefont {Gouttenoire}\ and\ \citenamefont {Vitagliano}(2023)}]{Gouttenoire:2023gbn}%
  \BibitemOpen
  \bibfield  {author} {\bibinfo {author} {\bibfnamefont {Y.}~\bibnamefont {Gouttenoire}}\ and\ \bibinfo {author} {\bibfnamefont {E.}~\bibnamefont {Vitagliano}},\ }\href@noop {} {\  (\bibinfo {year} {2023})},\ \Eprint {http://arxiv.org/abs/2311.07670} {arXiv:2311.07670 [hep-ph]} \BibitemShut {NoStop}%
\bibitem [{\citenamefont {Ferreira}\ \emph {et~al.}(2024)\citenamefont {Ferreira}, \citenamefont {Notari}, \citenamefont {Pujol\`as},\ and\ \citenamefont {Rompineve}}]{Ferreira:2024eru}%
  \BibitemOpen
  \bibfield  {author} {\bibinfo {author} {\bibfnamefont {R.~Z.}\ \bibnamefont {Ferreira}}, \bibinfo {author} {\bibfnamefont {A.}~\bibnamefont {Notari}}, \bibinfo {author} {\bibfnamefont {O.}~\bibnamefont {Pujol\`as}}, \ and\ \bibinfo {author} {\bibfnamefont {F.}~\bibnamefont {Rompineve}},\ }\href@noop {} {\  (\bibinfo {year} {2024})},\ \Eprint {http://arxiv.org/abs/2401.14331} {arXiv:2401.14331 [astro-ph.CO]} \BibitemShut {NoStop}%
\bibitem [{\citenamefont {Green}\ and\ \citenamefont {Malik}(2001)}]{Green:2000he}%
  \BibitemOpen
  \bibfield  {author} {\bibinfo {author} {\bibfnamefont {A.~M.}\ \bibnamefont {Green}}\ and\ \bibinfo {author} {\bibfnamefont {K.~A.}\ \bibnamefont {Malik}},\ }\href {\doibase 10.1103/PhysRevD.64.021301} {\bibfield  {journal} {\bibinfo  {journal} {Phys. Rev. D}\ }\textbf {\bibinfo {volume} {64}},\ \bibinfo {pages} {021301} (\bibinfo {year} {2001})},\ \Eprint {http://arxiv.org/abs/hep-ph/0008113} {arXiv:hep-ph/0008113} \BibitemShut {NoStop}%
\bibitem [{\citenamefont {Bassett}\ and\ \citenamefont {Tsujikawa}(2001)}]{Bassett:2000ha}%
  \BibitemOpen
  \bibfield  {author} {\bibinfo {author} {\bibfnamefont {B.~A.}\ \bibnamefont {Bassett}}\ and\ \bibinfo {author} {\bibfnamefont {S.}~\bibnamefont {Tsujikawa}},\ }\href {\doibase 10.1103/PhysRevD.63.123503} {\bibfield  {journal} {\bibinfo  {journal} {Phys. Rev. D}\ }\textbf {\bibinfo {volume} {63}},\ \bibinfo {pages} {123503} (\bibinfo {year} {2001})},\ \Eprint {http://arxiv.org/abs/hep-ph/0008328} {arXiv:hep-ph/0008328} \BibitemShut {NoStop}%
\bibitem [{\citenamefont {Suyama}\ \emph {et~al.}(2005)\citenamefont {Suyama}, \citenamefont {Tanaka}, \citenamefont {Bassett},\ and\ \citenamefont {Kudoh}}]{Suyama:2004mz}%
  \BibitemOpen
  \bibfield  {author} {\bibinfo {author} {\bibfnamefont {T.}~\bibnamefont {Suyama}}, \bibinfo {author} {\bibfnamefont {T.}~\bibnamefont {Tanaka}}, \bibinfo {author} {\bibfnamefont {B.}~\bibnamefont {Bassett}}, \ and\ \bibinfo {author} {\bibfnamefont {H.}~\bibnamefont {Kudoh}},\ }\href {\doibase 10.1103/PhysRevD.71.063507} {\bibfield  {journal} {\bibinfo  {journal} {Phys. Rev. D}\ }\textbf {\bibinfo {volume} {71}},\ \bibinfo {pages} {063507} (\bibinfo {year} {2005})},\ \Eprint {http://arxiv.org/abs/hep-ph/0410247} {arXiv:hep-ph/0410247} \BibitemShut {NoStop}%
\bibitem [{\citenamefont {Suyama}\ \emph {et~al.}(2006)\citenamefont {Suyama}, \citenamefont {Tanaka}, \citenamefont {Bassett},\ and\ \citenamefont {Kudoh}}]{Suyama:2006sr}%
  \BibitemOpen
  \bibfield  {author} {\bibinfo {author} {\bibfnamefont {T.}~\bibnamefont {Suyama}}, \bibinfo {author} {\bibfnamefont {T.}~\bibnamefont {Tanaka}}, \bibinfo {author} {\bibfnamefont {B.}~\bibnamefont {Bassett}}, \ and\ \bibinfo {author} {\bibfnamefont {H.}~\bibnamefont {Kudoh}},\ }\href {\doibase 10.1088/1475-7516/2006/04/001} {\bibfield  {journal} {\bibinfo  {journal} {JCAP}\ }\textbf {\bibinfo {volume} {04}},\ \bibinfo {pages} {001} (\bibinfo {year} {2006})},\ \Eprint {http://arxiv.org/abs/hep-ph/0601108} {arXiv:hep-ph/0601108} \BibitemShut {NoStop}%
\bibitem [{\citenamefont {Martin}\ \emph {et~al.}(2020{\natexlab{a}})\citenamefont {Martin}, \citenamefont {Papanikolaou},\ and\ \citenamefont {Vennin}}]{Martin:2019nuw}%
  \BibitemOpen
  \bibfield  {author} {\bibinfo {author} {\bibfnamefont {J.}~\bibnamefont {Martin}}, \bibinfo {author} {\bibfnamefont {T.}~\bibnamefont {Papanikolaou}}, \ and\ \bibinfo {author} {\bibfnamefont {V.}~\bibnamefont {Vennin}},\ }\href {\doibase 10.1088/1475-7516/2020/01/024} {\bibfield  {journal} {\bibinfo  {journal} {JCAP}\ }\textbf {\bibinfo {volume} {01}},\ \bibinfo {pages} {024} (\bibinfo {year} {2020}{\natexlab{a}})},\ \Eprint {http://arxiv.org/abs/1907.04236} {arXiv:1907.04236 [astro-ph.CO]} \BibitemShut {NoStop}%
\bibitem [{\citenamefont {Auclair}\ and\ \citenamefont {Vennin}(2021)}]{Auclair:2020csm}%
  \BibitemOpen
  \bibfield  {author} {\bibinfo {author} {\bibfnamefont {P.}~\bibnamefont {Auclair}}\ and\ \bibinfo {author} {\bibfnamefont {V.}~\bibnamefont {Vennin}},\ }\href {\doibase 10.1088/1475-7516/2021/02/038} {\bibfield  {journal} {\bibinfo  {journal} {JCAP}\ }\textbf {\bibinfo {volume} {02}},\ \bibinfo {pages} {038} (\bibinfo {year} {2021})},\ \Eprint {http://arxiv.org/abs/2011.05633} {arXiv:2011.05633 [astro-ph.CO]} \BibitemShut {NoStop}%
\bibitem [{\citenamefont {Martin}\ \emph {et~al.}(2020{\natexlab{b}})\citenamefont {Martin}, \citenamefont {Papanikolaou}, \citenamefont {Pinol},\ and\ \citenamefont {Vennin}}]{Martin:2020fgl}%
  \BibitemOpen
  \bibfield  {author} {\bibinfo {author} {\bibfnamefont {J.}~\bibnamefont {Martin}}, \bibinfo {author} {\bibfnamefont {T.}~\bibnamefont {Papanikolaou}}, \bibinfo {author} {\bibfnamefont {L.}~\bibnamefont {Pinol}}, \ and\ \bibinfo {author} {\bibfnamefont {V.}~\bibnamefont {Vennin}},\ }\href {\doibase 10.1088/1475-7516/2020/05/003} {\bibfield  {journal} {\bibinfo  {journal} {JCAP}\ }\textbf {\bibinfo {volume} {05}},\ \bibinfo {pages} {003} (\bibinfo {year} {2020}{\natexlab{b}})},\ \Eprint {http://arxiv.org/abs/2002.01820} {arXiv:2002.01820 [astro-ph.CO]} \BibitemShut {NoStop}%
\bibitem [{\citenamefont {Carr}\ \emph {et~al.}(2021{\natexlab{a}})\citenamefont {Carr}, \citenamefont {Kohri}, \citenamefont {Sendouda},\ and\ \citenamefont {Yokoyama}}]{Carr:2020gox}%
  \BibitemOpen
  \bibfield  {author} {\bibinfo {author} {\bibfnamefont {B.}~\bibnamefont {Carr}}, \bibinfo {author} {\bibfnamefont {K.}~\bibnamefont {Kohri}}, \bibinfo {author} {\bibfnamefont {Y.}~\bibnamefont {Sendouda}}, \ and\ \bibinfo {author} {\bibfnamefont {J.}~\bibnamefont {Yokoyama}},\ }\href {\doibase 10.1088/1361-6633/ac1e31} {\bibfield  {journal} {\bibinfo  {journal} {Rept. Prog. Phys.}\ }\textbf {\bibinfo {volume} {84}},\ \bibinfo {pages} {116902} (\bibinfo {year} {2021}{\natexlab{a}})},\ \Eprint {http://arxiv.org/abs/2002.12778} {arXiv:2002.12778 [astro-ph.CO]} \BibitemShut {NoStop}%
\bibitem [{\citenamefont {Auffinger}(2023)}]{Auffinger:2022khh}%
  \BibitemOpen
  \bibfield  {author} {\bibinfo {author} {\bibfnamefont {J.}~\bibnamefont {Auffinger}},\ }\href {\doibase 10.1016/j.ppnp.2023.104040} {\bibfield  {journal} {\bibinfo  {journal} {Prog. Part. Nucl. Phys.}\ }\textbf {\bibinfo {volume} {131}},\ \bibinfo {pages} {104040} (\bibinfo {year} {2023})},\ \Eprint {http://arxiv.org/abs/2206.02672} {arXiv:2206.02672 [astro-ph.CO]} \BibitemShut {NoStop}%
\bibitem [{\citenamefont {Garcia-Bellido}\ and\ \citenamefont {Hawkins}(2024)}]{Garcia-Bellido:2024yaz}%
  \BibitemOpen
  \bibfield  {author} {\bibinfo {author} {\bibfnamefont {J.}~\bibnamefont {Garcia-Bellido}}\ and\ \bibinfo {author} {\bibfnamefont {M.}~\bibnamefont {Hawkins}},\ }\href@noop {} {\  (\bibinfo {year} {2024})},\ \Eprint {http://arxiv.org/abs/2402.00212} {arXiv:2402.00212 [astro-ph.GA]} \BibitemShut {NoStop}%
\bibitem [{\citenamefont {Tisserand}\ \emph {et~al.}(2007)\citenamefont {Tisserand} \emph {et~al.}}]{EROS-2:2006ryy}%
  \BibitemOpen
  \bibfield  {author} {\bibinfo {author} {\bibfnamefont {P.}~\bibnamefont {Tisserand}} \emph {et~al.} (\bibinfo {collaboration} {EROS-2}),\ }\href {\doibase 10.1051/0004-6361:20066017} {\bibfield  {journal} {\bibinfo  {journal} {Astron. Astrophys.}\ }\textbf {\bibinfo {volume} {469}},\ \bibinfo {pages} {387} (\bibinfo {year} {2007})},\ \Eprint {http://arxiv.org/abs/astro-ph/0607207} {arXiv:astro-ph/0607207} \BibitemShut {NoStop}%
\bibitem [{\citenamefont {Niikura}\ \emph {et~al.}(2019{\natexlab{a}})\citenamefont {Niikura} \emph {et~al.}}]{Niikura:2017zjd}%
  \BibitemOpen
  \bibfield  {author} {\bibinfo {author} {\bibfnamefont {H.}~\bibnamefont {Niikura}} \emph {et~al.},\ }\href {\doibase 10.1038/s41550-019-0723-1} {\bibfield  {journal} {\bibinfo  {journal} {Nature Astron.}\ }\textbf {\bibinfo {volume} {3}},\ \bibinfo {pages} {524} (\bibinfo {year} {2019}{\natexlab{a}})},\ \Eprint {http://arxiv.org/abs/1701.02151} {arXiv:1701.02151 [astro-ph.CO]} \BibitemShut {NoStop}%
\bibitem [{\citenamefont {Oguri}\ \emph {et~al.}(2018)\citenamefont {Oguri}, \citenamefont {Diego}, \citenamefont {Kaiser}, \citenamefont {Kelly},\ and\ \citenamefont {Broadhurst}}]{Oguri:2017ock}%
  \BibitemOpen
  \bibfield  {author} {\bibinfo {author} {\bibfnamefont {M.}~\bibnamefont {Oguri}}, \bibinfo {author} {\bibfnamefont {J.~M.}\ \bibnamefont {Diego}}, \bibinfo {author} {\bibfnamefont {N.}~\bibnamefont {Kaiser}}, \bibinfo {author} {\bibfnamefont {P.~L.}\ \bibnamefont {Kelly}}, \ and\ \bibinfo {author} {\bibfnamefont {T.}~\bibnamefont {Broadhurst}},\ }\href {\doibase 10.1103/PhysRevD.97.023518} {\bibfield  {journal} {\bibinfo  {journal} {Phys. Rev. D}\ }\textbf {\bibinfo {volume} {97}},\ \bibinfo {pages} {023518} (\bibinfo {year} {2018})},\ \Eprint {http://arxiv.org/abs/1710.00148} {arXiv:1710.00148 [astro-ph.CO]} \BibitemShut {NoStop}%
\bibitem [{\citenamefont {Niikura}\ \emph {et~al.}(2019{\natexlab{b}})\citenamefont {Niikura}, \citenamefont {Takada}, \citenamefont {Yokoyama}, \citenamefont {Sumi},\ and\ \citenamefont {Masaki}}]{Niikura:2019kqi}%
  \BibitemOpen
  \bibfield  {author} {\bibinfo {author} {\bibfnamefont {H.}~\bibnamefont {Niikura}}, \bibinfo {author} {\bibfnamefont {M.}~\bibnamefont {Takada}}, \bibinfo {author} {\bibfnamefont {S.}~\bibnamefont {Yokoyama}}, \bibinfo {author} {\bibfnamefont {T.}~\bibnamefont {Sumi}}, \ and\ \bibinfo {author} {\bibfnamefont {S.}~\bibnamefont {Masaki}},\ }\href {\doibase 10.1103/PhysRevD.99.083503} {\bibfield  {journal} {\bibinfo  {journal} {Phys. Rev. D}\ }\textbf {\bibinfo {volume} {99}},\ \bibinfo {pages} {083503} (\bibinfo {year} {2019}{\natexlab{b}})},\ \Eprint {http://arxiv.org/abs/1901.07120} {arXiv:1901.07120 [astro-ph.CO]} \BibitemShut {NoStop}%
\bibitem [{\citenamefont {Serpico}\ \emph {et~al.}(2020)\citenamefont {Serpico}, \citenamefont {Poulin}, \citenamefont {Inman},\ and\ \citenamefont {Kohri}}]{Serpico:2020ehh}%
  \BibitemOpen
  \bibfield  {author} {\bibinfo {author} {\bibfnamefont {P.~D.}\ \bibnamefont {Serpico}}, \bibinfo {author} {\bibfnamefont {V.}~\bibnamefont {Poulin}}, \bibinfo {author} {\bibfnamefont {D.}~\bibnamefont {Inman}}, \ and\ \bibinfo {author} {\bibfnamefont {K.}~\bibnamefont {Kohri}},\ }\href {\doibase 10.1103/PhysRevResearch.2.023204} {\bibfield  {journal} {\bibinfo  {journal} {Phys. Rev. Res.}\ }\textbf {\bibinfo {volume} {2}},\ \bibinfo {pages} {023204} (\bibinfo {year} {2020})},\ \Eprint {http://arxiv.org/abs/2002.10771} {arXiv:2002.10771 [astro-ph.CO]} \BibitemShut {NoStop}%
\bibitem [{\citenamefont {Piga}\ \emph {et~al.}(2022)\citenamefont {Piga}, \citenamefont {Lucca}, \citenamefont {Bellomo}, \citenamefont {Bosch-Ramon}, \citenamefont {Matarrese}, \citenamefont {Raccanelli},\ and\ \citenamefont {Verde}}]{Piga:2022ysp}%
  \BibitemOpen
  \bibfield  {author} {\bibinfo {author} {\bibfnamefont {L.}~\bibnamefont {Piga}}, \bibinfo {author} {\bibfnamefont {M.}~\bibnamefont {Lucca}}, \bibinfo {author} {\bibfnamefont {N.}~\bibnamefont {Bellomo}}, \bibinfo {author} {\bibfnamefont {V.}~\bibnamefont {Bosch-Ramon}}, \bibinfo {author} {\bibfnamefont {S.}~\bibnamefont {Matarrese}}, \bibinfo {author} {\bibfnamefont {A.}~\bibnamefont {Raccanelli}}, \ and\ \bibinfo {author} {\bibfnamefont {L.}~\bibnamefont {Verde}},\ }\href {\doibase 10.1088/1475-7516/2022/12/016} {\bibfield  {journal} {\bibinfo  {journal} {JCAP}\ }\textbf {\bibinfo {volume} {12}},\ \bibinfo {pages} {016} (\bibinfo {year} {2022})},\ \Eprint {http://arxiv.org/abs/2210.14934} {arXiv:2210.14934 [astro-ph.CO]} \BibitemShut {NoStop}%
\bibitem [{\citenamefont {Abbott}\ \emph {et~al.}(2018)\citenamefont {Abbott} \emph {et~al.}}]{LIGOScientific:2018glc}%
  \BibitemOpen
  \bibfield  {author} {\bibinfo {author} {\bibfnamefont {B.~P.}\ \bibnamefont {Abbott}} \emph {et~al.} (\bibinfo {collaboration} {LIGO Scientific, Virgo}),\ }\href {\doibase 10.1103/PhysRevLett.121.231103} {\bibfield  {journal} {\bibinfo  {journal} {Phys. Rev. Lett.}\ }\textbf {\bibinfo {volume} {121}},\ \bibinfo {pages} {231103} (\bibinfo {year} {2018})},\ \Eprint {http://arxiv.org/abs/1808.04771} {arXiv:1808.04771 [astro-ph.CO]} \BibitemShut {NoStop}%
\bibitem [{\citenamefont {Abbott}\ \emph {et~al.}(2019{\natexlab{b}})\citenamefont {Abbott} \emph {et~al.}}]{LIGOScientific:2019kan}%
  \BibitemOpen
  \bibfield  {author} {\bibinfo {author} {\bibfnamefont {B.~P.}\ \bibnamefont {Abbott}} \emph {et~al.} (\bibinfo {collaboration} {LIGO Scientific, Virgo}),\ }\href {\doibase 10.1103/PhysRevLett.123.161102} {\bibfield  {journal} {\bibinfo  {journal} {Phys. Rev. Lett.}\ }\textbf {\bibinfo {volume} {123}},\ \bibinfo {pages} {161102} (\bibinfo {year} {2019}{\natexlab{b}})},\ \Eprint {http://arxiv.org/abs/1904.08976} {arXiv:1904.08976 [astro-ph.CO]} \BibitemShut {NoStop}%
\bibitem [{\citenamefont {Nitz}\ and\ \citenamefont {Wang}(2021{\natexlab{a}})}]{Nitz:2020bdb}%
  \BibitemOpen
  \bibfield  {author} {\bibinfo {author} {\bibfnamefont {A.~H.}\ \bibnamefont {Nitz}}\ and\ \bibinfo {author} {\bibfnamefont {Y.-F.}\ \bibnamefont {Wang}},\ }\href {\doibase 10.1103/PhysRevLett.126.021103} {\bibfield  {journal} {\bibinfo  {journal} {Phys. Rev. Lett.}\ }\textbf {\bibinfo {volume} {126}},\ \bibinfo {pages} {021103} (\bibinfo {year} {2021}{\natexlab{a}})},\ \Eprint {http://arxiv.org/abs/2007.03583} {arXiv:2007.03583 [astro-ph.HE]} \BibitemShut {NoStop}%
\bibitem [{\citenamefont {Phukon}\ \emph {et~al.}(2021)\citenamefont {Phukon}, \citenamefont {Baltus}, \citenamefont {Caudill}, \citenamefont {Clesse}, \citenamefont {Depasse}, \citenamefont {Fays}, \citenamefont {Fong}, \citenamefont {Kapadia}, \citenamefont {Magee},\ and\ \citenamefont {Tanasijczuk}}]{Phukon:2021cus}%
  \BibitemOpen
  \bibfield  {author} {\bibinfo {author} {\bibfnamefont {K.~S.}\ \bibnamefont {Phukon}}, \bibinfo {author} {\bibfnamefont {G.}~\bibnamefont {Baltus}}, \bibinfo {author} {\bibfnamefont {S.}~\bibnamefont {Caudill}}, \bibinfo {author} {\bibfnamefont {S.}~\bibnamefont {Clesse}}, \bibinfo {author} {\bibfnamefont {A.}~\bibnamefont {Depasse}}, \bibinfo {author} {\bibfnamefont {M.}~\bibnamefont {Fays}}, \bibinfo {author} {\bibfnamefont {H.}~\bibnamefont {Fong}}, \bibinfo {author} {\bibfnamefont {S.~J.}\ \bibnamefont {Kapadia}}, \bibinfo {author} {\bibfnamefont {R.}~\bibnamefont {Magee}}, \ and\ \bibinfo {author} {\bibfnamefont {A.~J.}\ \bibnamefont {Tanasijczuk}},\ }\href@noop {} {\  (\bibinfo {year} {2021})},\ \Eprint {http://arxiv.org/abs/2105.11449} {arXiv:2105.11449 [astro-ph.CO]} \BibitemShut {NoStop}%
\bibitem [{\citenamefont {Nitz}\ and\ \citenamefont {Wang}(2021{\natexlab{b}})}]{Nitz:2021mzz}%
  \BibitemOpen
  \bibfield  {author} {\bibinfo {author} {\bibfnamefont {A.~H.}\ \bibnamefont {Nitz}}\ and\ \bibinfo {author} {\bibfnamefont {Y.-F.}\ \bibnamefont {Wang}},\ }\href {\doibase 10.3847/1538-4357/ac01d9} {\  (\bibinfo {year} {2021}{\natexlab{b}}),\ 10.3847/1538-4357/ac01d9},\ \Eprint {http://arxiv.org/abs/2102.00868} {arXiv:2102.00868 [astro-ph.HE]} \BibitemShut {NoStop}%
\bibitem [{\citenamefont {Nitz}\ and\ \citenamefont {Wang}(2021{\natexlab{c}})}]{Nitz:2021vqh}%
  \BibitemOpen
  \bibfield  {author} {\bibinfo {author} {\bibfnamefont {A.~H.}\ \bibnamefont {Nitz}}\ and\ \bibinfo {author} {\bibfnamefont {Y.-F.}\ \bibnamefont {Wang}},\ }\href {\doibase 10.1103/PhysRevLett.127.151101} {\bibfield  {journal} {\bibinfo  {journal} {Phys. Rev. Lett.}\ }\textbf {\bibinfo {volume} {127}},\ \bibinfo {pages} {151101} (\bibinfo {year} {2021}{\natexlab{c}})},\ \Eprint {http://arxiv.org/abs/2106.08979} {arXiv:2106.08979 [astro-ph.HE]} \BibitemShut {NoStop}%
\bibitem [{\citenamefont {Nitz}\ and\ \citenamefont {Wang}(2022)}]{Nitz:2022ltl}%
  \BibitemOpen
  \bibfield  {author} {\bibinfo {author} {\bibfnamefont {A.~H.}\ \bibnamefont {Nitz}}\ and\ \bibinfo {author} {\bibfnamefont {Y.-F.}\ \bibnamefont {Wang}},\ }\href {\doibase 10.1103/PhysRevD.106.023024} {\bibfield  {journal} {\bibinfo  {journal} {Phys. Rev. D}\ }\textbf {\bibinfo {volume} {106}},\ \bibinfo {pages} {023024} (\bibinfo {year} {2022})},\ \Eprint {http://arxiv.org/abs/2202.11024} {arXiv:2202.11024 [astro-ph.HE]} \BibitemShut {NoStop}%
\bibitem [{\citenamefont {Miller}\ \emph {et~al.}(2021)\citenamefont {Miller}, \citenamefont {Clesse}, \citenamefont {De~Lillo}, \citenamefont {Bruno}, \citenamefont {Depasse},\ and\ \citenamefont {Tanasijczuk}}]{Miller:2020kmv}%
  \BibitemOpen
  \bibfield  {author} {\bibinfo {author} {\bibfnamefont {A.~L.}\ \bibnamefont {Miller}}, \bibinfo {author} {\bibfnamefont {S.}~\bibnamefont {Clesse}}, \bibinfo {author} {\bibfnamefont {F.}~\bibnamefont {De~Lillo}}, \bibinfo {author} {\bibfnamefont {G.}~\bibnamefont {Bruno}}, \bibinfo {author} {\bibfnamefont {A.}~\bibnamefont {Depasse}}, \ and\ \bibinfo {author} {\bibfnamefont {A.}~\bibnamefont {Tanasijczuk}},\ }\href {\doibase 10.1016/j.dark.2021.100836} {\bibfield  {journal} {\bibinfo  {journal} {Phys. Dark Univ.}\ }\textbf {\bibinfo {volume} {32}},\ \bibinfo {pages} {100836} (\bibinfo {year} {2021})},\ \Eprint {http://arxiv.org/abs/2012.12983} {arXiv:2012.12983 [astro-ph.HE]} \BibitemShut {NoStop}%
\bibitem [{\citenamefont {Miller}\ \emph {et~al.}(2022)\citenamefont {Miller}, \citenamefont {Aggarwal}, \citenamefont {Clesse},\ and\ \citenamefont {De~Lillo}}]{Miller:2021knj}%
  \BibitemOpen
  \bibfield  {author} {\bibinfo {author} {\bibfnamefont {A.~L.}\ \bibnamefont {Miller}}, \bibinfo {author} {\bibfnamefont {N.}~\bibnamefont {Aggarwal}}, \bibinfo {author} {\bibfnamefont {S.}~\bibnamefont {Clesse}}, \ and\ \bibinfo {author} {\bibfnamefont {F.}~\bibnamefont {De~Lillo}},\ }\href {\doibase 10.1103/PhysRevD.105.062008} {\bibfield  {journal} {\bibinfo  {journal} {Phys. Rev. D}\ }\textbf {\bibinfo {volume} {105}},\ \bibinfo {pages} {062008} (\bibinfo {year} {2022})},\ \Eprint {http://arxiv.org/abs/2110.06188} {arXiv:2110.06188 [gr-qc]} \BibitemShut {NoStop}%
\bibitem [{\citenamefont {Morras}\ \emph {et~al.}(2023)\citenamefont {Morras} \emph {et~al.}}]{Morras:2023jvb}%
  \BibitemOpen
  \bibfield  {author} {\bibinfo {author} {\bibfnamefont {G.}~\bibnamefont {Morras}} \emph {et~al.},\ }\href {\doibase 10.1016/j.dark.2023.101285} {\bibfield  {journal} {\bibinfo  {journal} {Phys. Dark Univ.}\ }\textbf {\bibinfo {volume} {42}},\ \bibinfo {pages} {101285} (\bibinfo {year} {2023})},\ \Eprint {http://arxiv.org/abs/2301.11619} {arXiv:2301.11619 [gr-qc]} \BibitemShut {NoStop}%
\bibitem [{\citenamefont {Mukherjee}\ and\ \citenamefont {Silk}(2021)}]{Mukherjee:2021ags}%
  \BibitemOpen
  \bibfield  {author} {\bibinfo {author} {\bibfnamefont {S.}~\bibnamefont {Mukherjee}}\ and\ \bibinfo {author} {\bibfnamefont {J.}~\bibnamefont {Silk}},\ }\href {\doibase 10.1093/mnras/stab1932} {\bibfield  {journal} {\bibinfo  {journal} {Mon. Not. Roy. Astron. Soc.}\ }\textbf {\bibinfo {volume} {506}},\ \bibinfo {pages} {3977} (\bibinfo {year} {2021})},\ \Eprint {http://arxiv.org/abs/2105.11139} {arXiv:2105.11139 [gr-qc]} \BibitemShut {NoStop}%
\bibitem [{\citenamefont {Mukherjee}\ \emph {et~al.}(2022)\citenamefont {Mukherjee}, \citenamefont {Meinema},\ and\ \citenamefont {Silk}}]{Mukherjee:2021itf}%
  \BibitemOpen
  \bibfield  {author} {\bibinfo {author} {\bibfnamefont {S.}~\bibnamefont {Mukherjee}}, \bibinfo {author} {\bibfnamefont {M.~S.~P.}\ \bibnamefont {Meinema}}, \ and\ \bibinfo {author} {\bibfnamefont {J.}~\bibnamefont {Silk}},\ }\href {\doibase 10.1093/mnras/stab3756} {\bibfield  {journal} {\bibinfo  {journal} {Mon. Not. Roy. Astron. Soc.}\ }\textbf {\bibinfo {volume} {510}},\ \bibinfo {pages} {6218} (\bibinfo {year} {2022})},\ \Eprint {http://arxiv.org/abs/2107.02181} {arXiv:2107.02181 [astro-ph.CO]} \BibitemShut {NoStop}%
\bibitem [{\citenamefont {Andres-Carcasona}\ \emph {et~al.}(2023)\citenamefont {Andres-Carcasona}, \citenamefont {Menendez-Vazquez}, \citenamefont {Martinez},\ and\ \citenamefont {Mir}}]{Andres-Carcasona:2022prl}%
  \BibitemOpen
  \bibfield  {author} {\bibinfo {author} {\bibfnamefont {M.}~\bibnamefont {Andres-Carcasona}}, \bibinfo {author} {\bibfnamefont {A.}~\bibnamefont {Menendez-Vazquez}}, \bibinfo {author} {\bibfnamefont {M.}~\bibnamefont {Martinez}}, \ and\ \bibinfo {author} {\bibfnamefont {L.~M.}\ \bibnamefont {Mir}},\ }\href {\doibase 10.1103/PhysRevD.107.082003} {\bibfield  {journal} {\bibinfo  {journal} {Phys. Rev. D}\ }\textbf {\bibinfo {volume} {107}},\ \bibinfo {pages} {082003} (\bibinfo {year} {2023})},\ \Eprint {http://arxiv.org/abs/2212.02829} {arXiv:2212.02829 [gr-qc]} \BibitemShut {NoStop}%
\bibitem [{\citenamefont {Miller}\ \emph {et~al.}(2024)\citenamefont {Miller}, \citenamefont {Aggarwal}, \citenamefont {Clesse}, \citenamefont {De~Lillo}, \citenamefont {Sachdev}, \citenamefont {Astone}, \citenamefont {Palomba}, \citenamefont {Piccinni},\ and\ \citenamefont {Pierini}}]{Miller:2024fpo}%
  \BibitemOpen
  \bibfield  {author} {\bibinfo {author} {\bibfnamefont {A.~L.}\ \bibnamefont {Miller}}, \bibinfo {author} {\bibfnamefont {N.}~\bibnamefont {Aggarwal}}, \bibinfo {author} {\bibfnamefont {S.}~\bibnamefont {Clesse}}, \bibinfo {author} {\bibfnamefont {F.}~\bibnamefont {De~Lillo}}, \bibinfo {author} {\bibfnamefont {S.}~\bibnamefont {Sachdev}}, \bibinfo {author} {\bibfnamefont {P.}~\bibnamefont {Astone}}, \bibinfo {author} {\bibfnamefont {C.}~\bibnamefont {Palomba}}, \bibinfo {author} {\bibfnamefont {O.~J.}\ \bibnamefont {Piccinni}}, \ and\ \bibinfo {author} {\bibfnamefont {L.}~\bibnamefont {Pierini}},\ }\href@noop {} {\  (\bibinfo {year} {2024})},\ \Eprint {http://arxiv.org/abs/2402.19468} {arXiv:2402.19468 [gr-qc]} \BibitemShut {NoStop}%
\bibitem [{\citenamefont {Prunier}\ \emph {et~al.}(2023)\citenamefont {Prunier}, \citenamefont {Morr\'as}, \citenamefont {Siles}, \citenamefont {Clesse}, \citenamefont {Garc\'\i{}a-Bellido},\ and\ \citenamefont {Ruiz~Morales}}]{Prunier:2023cyv}%
  \BibitemOpen
  \bibfield  {author} {\bibinfo {author} {\bibfnamefont {M.}~\bibnamefont {Prunier}}, \bibinfo {author} {\bibfnamefont {G.}~\bibnamefont {Morr\'as}}, \bibinfo {author} {\bibfnamefont {J.~F. N.~n.}\ \bibnamefont {Siles}}, \bibinfo {author} {\bibfnamefont {S.}~\bibnamefont {Clesse}}, \bibinfo {author} {\bibfnamefont {J.}~\bibnamefont {Garc\'\i{}a-Bellido}}, \ and\ \bibinfo {author} {\bibfnamefont {E.}~\bibnamefont {Ruiz~Morales}},\ }\href@noop {} {\  (\bibinfo {year} {2023})},\ \Eprint {http://arxiv.org/abs/2311.16085} {arXiv:2311.16085 [gr-qc]} \BibitemShut {NoStop}%
\bibitem [{\citenamefont {H\"utsi}\ \emph {et~al.}(2021)\citenamefont {H\"utsi}, \citenamefont {Raidal}, \citenamefont {Vaskonen},\ and\ \citenamefont {Veerm\"ae}}]{Hutsi:2020sol}%
  \BibitemOpen
  \bibfield  {author} {\bibinfo {author} {\bibfnamefont {G.}~\bibnamefont {H\"utsi}}, \bibinfo {author} {\bibfnamefont {M.}~\bibnamefont {Raidal}}, \bibinfo {author} {\bibfnamefont {V.}~\bibnamefont {Vaskonen}}, \ and\ \bibinfo {author} {\bibfnamefont {H.}~\bibnamefont {Veerm\"ae}},\ }\href {\doibase 10.1088/1475-7516/2021/03/068} {\bibfield  {journal} {\bibinfo  {journal} {JCAP}\ }\textbf {\bibinfo {volume} {03}},\ \bibinfo {pages} {068} (\bibinfo {year} {2021})},\ \Eprint {http://arxiv.org/abs/2012.02786} {arXiv:2012.02786 [astro-ph.CO]} \BibitemShut {NoStop}%
\bibitem [{\citenamefont {Hall}\ \emph {et~al.}(2020)\citenamefont {Hall}, \citenamefont {Gow},\ and\ \citenamefont {Byrnes}}]{Hall:2020daa}%
  \BibitemOpen
  \bibfield  {author} {\bibinfo {author} {\bibfnamefont {A.}~\bibnamefont {Hall}}, \bibinfo {author} {\bibfnamefont {A.~D.}\ \bibnamefont {Gow}}, \ and\ \bibinfo {author} {\bibfnamefont {C.~T.}\ \bibnamefont {Byrnes}},\ }\href {\doibase 10.1103/PhysRevD.102.123524} {\bibfield  {journal} {\bibinfo  {journal} {Phys. Rev. D}\ }\textbf {\bibinfo {volume} {102}},\ \bibinfo {pages} {123524} (\bibinfo {year} {2020})},\ \Eprint {http://arxiv.org/abs/2008.13704} {arXiv:2008.13704 [astro-ph.CO]} \BibitemShut {NoStop}%
\bibitem [{\citenamefont {Wong}\ \emph {et~al.}(2021)\citenamefont {Wong}, \citenamefont {Franciolini}, \citenamefont {De~Luca}, \citenamefont {Baibhav}, \citenamefont {Berti}, \citenamefont {Pani},\ and\ \citenamefont {Riotto}}]{Wong:2020yig}%
  \BibitemOpen
  \bibfield  {author} {\bibinfo {author} {\bibfnamefont {K.~W.~K.}\ \bibnamefont {Wong}}, \bibinfo {author} {\bibfnamefont {G.}~\bibnamefont {Franciolini}}, \bibinfo {author} {\bibfnamefont {V.}~\bibnamefont {De~Luca}}, \bibinfo {author} {\bibfnamefont {V.}~\bibnamefont {Baibhav}}, \bibinfo {author} {\bibfnamefont {E.}~\bibnamefont {Berti}}, \bibinfo {author} {\bibfnamefont {P.}~\bibnamefont {Pani}}, \ and\ \bibinfo {author} {\bibfnamefont {A.}~\bibnamefont {Riotto}},\ }\href {\doibase 10.1103/PhysRevD.103.023026} {\bibfield  {journal} {\bibinfo  {journal} {Phys. Rev. D}\ }\textbf {\bibinfo {volume} {103}},\ \bibinfo {pages} {023026} (\bibinfo {year} {2021})},\ \Eprint {http://arxiv.org/abs/2011.01865} {arXiv:2011.01865 [gr-qc]} \BibitemShut {NoStop}%
\bibitem [{\citenamefont {Franciolini}\ \emph {et~al.}(2022{\natexlab{a}})\citenamefont {Franciolini}, \citenamefont {Baibhav}, \citenamefont {De~Luca}, \citenamefont {Ng}, \citenamefont {Wong}, \citenamefont {Berti}, \citenamefont {Pani}, \citenamefont {Riotto},\ and\ \citenamefont {Vitale}}]{Franciolini:2021tla}%
  \BibitemOpen
  \bibfield  {author} {\bibinfo {author} {\bibfnamefont {G.}~\bibnamefont {Franciolini}}, \bibinfo {author} {\bibfnamefont {V.}~\bibnamefont {Baibhav}}, \bibinfo {author} {\bibfnamefont {V.}~\bibnamefont {De~Luca}}, \bibinfo {author} {\bibfnamefont {K.~K.~Y.}\ \bibnamefont {Ng}}, \bibinfo {author} {\bibfnamefont {K.~W.~K.}\ \bibnamefont {Wong}}, \bibinfo {author} {\bibfnamefont {E.}~\bibnamefont {Berti}}, \bibinfo {author} {\bibfnamefont {P.}~\bibnamefont {Pani}}, \bibinfo {author} {\bibfnamefont {A.}~\bibnamefont {Riotto}}, \ and\ \bibinfo {author} {\bibfnamefont {S.}~\bibnamefont {Vitale}},\ }\href {\doibase 10.1103/PhysRevD.105.083526} {\bibfield  {journal} {\bibinfo  {journal} {Phys. Rev. D}\ }\textbf {\bibinfo {volume} {105}},\ \bibinfo {pages} {083526} (\bibinfo {year} {2022}{\natexlab{a}})},\ \Eprint {http://arxiv.org/abs/2105.03349} {arXiv:2105.03349 [gr-qc]} \BibitemShut {NoStop}%
\bibitem [{\citenamefont {De~Luca}\ \emph {et~al.}(2021)\citenamefont {De~Luca}, \citenamefont {Franciolini}, \citenamefont {Pani},\ and\ \citenamefont {Riotto}}]{DeLuca:2021wjr}%
  \BibitemOpen
  \bibfield  {author} {\bibinfo {author} {\bibfnamefont {V.}~\bibnamefont {De~Luca}}, \bibinfo {author} {\bibfnamefont {G.}~\bibnamefont {Franciolini}}, \bibinfo {author} {\bibfnamefont {P.}~\bibnamefont {Pani}}, \ and\ \bibinfo {author} {\bibfnamefont {A.}~\bibnamefont {Riotto}},\ }\href {\doibase 10.1088/1475-7516/2021/05/003} {\bibfield  {journal} {\bibinfo  {journal} {JCAP}\ }\textbf {\bibinfo {volume} {05}},\ \bibinfo {pages} {003} (\bibinfo {year} {2021})},\ \Eprint {http://arxiv.org/abs/2102.03809} {arXiv:2102.03809 [astro-ph.CO]} \BibitemShut {NoStop}%
\bibitem [{\citenamefont {Abbott}\ \emph {et~al.}(2020)\citenamefont {Abbott} \emph {et~al.}}]{LIGOScientific:2020iuh}%
  \BibitemOpen
  \bibfield  {author} {\bibinfo {author} {\bibfnamefont {R.}~\bibnamefont {Abbott}} \emph {et~al.} (\bibinfo {collaboration} {LIGO Scientific, Virgo}),\ }\href {\doibase 10.1103/PhysRevLett.125.101102} {\bibfield  {journal} {\bibinfo  {journal} {Phys. Rev. Lett.}\ }\textbf {\bibinfo {volume} {125}},\ \bibinfo {pages} {101102} (\bibinfo {year} {2020})},\ \Eprint {http://arxiv.org/abs/2009.01075} {arXiv:2009.01075 [gr-qc]} \BibitemShut {NoStop}%
\bibitem [{LIG(2024)}]{LIGOScientific:2024elc}%
  \BibitemOpen
  \href@noop {} {\  (\bibinfo {year} {2024})},\ \Eprint {http://arxiv.org/abs/2404.04248} {arXiv:2404.04248 [astro-ph.HE]} \BibitemShut {NoStop}%
\bibitem [{\citenamefont {Raidal}\ \emph {et~al.}(2017)\citenamefont {Raidal}, \citenamefont {Vaskonen},\ and\ \citenamefont {Veerm\"ae}}]{Raidal:2017mfl}%
  \BibitemOpen
  \bibfield  {author} {\bibinfo {author} {\bibfnamefont {M.}~\bibnamefont {Raidal}}, \bibinfo {author} {\bibfnamefont {V.}~\bibnamefont {Vaskonen}}, \ and\ \bibinfo {author} {\bibfnamefont {H.}~\bibnamefont {Veerm\"ae}},\ }\href {\doibase 10.1088/1475-7516/2017/09/037} {\bibfield  {journal} {\bibinfo  {journal} {JCAP}\ }\textbf {\bibinfo {volume} {09}},\ \bibinfo {pages} {037} (\bibinfo {year} {2017})},\ \Eprint {http://arxiv.org/abs/1707.01480} {arXiv:1707.01480 [astro-ph.CO]} \BibitemShut {NoStop}%
\bibitem [{\citenamefont {Raidal}\ \emph {et~al.}(2019)\citenamefont {Raidal}, \citenamefont {Spethmann}, \citenamefont {Vaskonen},\ and\ \citenamefont {Veerm\"ae}}]{Raidal:2018bbj}%
  \BibitemOpen
  \bibfield  {author} {\bibinfo {author} {\bibfnamefont {M.}~\bibnamefont {Raidal}}, \bibinfo {author} {\bibfnamefont {C.}~\bibnamefont {Spethmann}}, \bibinfo {author} {\bibfnamefont {V.}~\bibnamefont {Vaskonen}}, \ and\ \bibinfo {author} {\bibfnamefont {H.}~\bibnamefont {Veerm\"ae}},\ }\href {\doibase 10.1088/1475-7516/2019/02/018} {\bibfield  {journal} {\bibinfo  {journal} {JCAP}\ }\textbf {\bibinfo {volume} {02}},\ \bibinfo {pages} {018} (\bibinfo {year} {2019})},\ \Eprint {http://arxiv.org/abs/1812.01930} {arXiv:1812.01930 [astro-ph.CO]} \BibitemShut {NoStop}%
\bibitem [{\citenamefont {Vaskonen}\ and\ \citenamefont {Veerm\"ae}(2020)}]{Vaskonen:2019jpv}%
  \BibitemOpen
  \bibfield  {author} {\bibinfo {author} {\bibfnamefont {V.}~\bibnamefont {Vaskonen}}\ and\ \bibinfo {author} {\bibfnamefont {H.}~\bibnamefont {Veerm\"ae}},\ }\href {\doibase 10.1103/PhysRevD.101.043015} {\bibfield  {journal} {\bibinfo  {journal} {Phys. Rev. D}\ }\textbf {\bibinfo {volume} {101}},\ \bibinfo {pages} {043015} (\bibinfo {year} {2020})},\ \Eprint {http://arxiv.org/abs/1908.09752} {arXiv:1908.09752 [astro-ph.CO]} \BibitemShut {NoStop}%
\bibitem [{\citenamefont {Romero-Rodriguez}\ \emph {et~al.}(2022)\citenamefont {Romero-Rodriguez}, \citenamefont {Martinez}, \citenamefont {Pujol\`as}, \citenamefont {Sakellariadou},\ and\ \citenamefont {Vaskonen}}]{Romero-Rodriguez:2021aws}%
  \BibitemOpen
  \bibfield  {author} {\bibinfo {author} {\bibfnamefont {A.}~\bibnamefont {Romero-Rodriguez}}, \bibinfo {author} {\bibfnamefont {M.}~\bibnamefont {Martinez}}, \bibinfo {author} {\bibfnamefont {O.}~\bibnamefont {Pujol\`as}}, \bibinfo {author} {\bibfnamefont {M.}~\bibnamefont {Sakellariadou}}, \ and\ \bibinfo {author} {\bibfnamefont {V.}~\bibnamefont {Vaskonen}},\ }\href {\doibase 10.1103/PhysRevLett.128.051301} {\bibfield  {journal} {\bibinfo  {journal} {Phys. Rev. Lett.}\ }\textbf {\bibinfo {volume} {128}},\ \bibinfo {pages} {051301} (\bibinfo {year} {2022})},\ \Eprint {http://arxiv.org/abs/2107.11660} {arXiv:2107.11660 [gr-qc]} \BibitemShut {NoStop}%
\bibitem [{\citenamefont {Franciolini}\ \emph {et~al.}(2022{\natexlab{b}})\citenamefont {Franciolini}, \citenamefont {Musco}, \citenamefont {Pani},\ and\ \citenamefont {Urbano}}]{Franciolini:2022tfm}%
  \BibitemOpen
  \bibfield  {author} {\bibinfo {author} {\bibfnamefont {G.}~\bibnamefont {Franciolini}}, \bibinfo {author} {\bibfnamefont {I.}~\bibnamefont {Musco}}, \bibinfo {author} {\bibfnamefont {P.}~\bibnamefont {Pani}}, \ and\ \bibinfo {author} {\bibfnamefont {A.}~\bibnamefont {Urbano}},\ }\href {\doibase 10.1103/PhysRevD.106.123526} {\bibfield  {journal} {\bibinfo  {journal} {Phys. Rev. D}\ }\textbf {\bibinfo {volume} {106}},\ \bibinfo {pages} {123526} (\bibinfo {year} {2022}{\natexlab{b}})},\ \Eprint {http://arxiv.org/abs/2209.05959} {arXiv:2209.05959 [astro-ph.CO]} \BibitemShut {NoStop}%
\bibitem [{\citenamefont {Musco}\ \emph {et~al.}(2009)\citenamefont {Musco}, \citenamefont {Miller},\ and\ \citenamefont {Polnarev}}]{Musco:2008hv}%
  \BibitemOpen
  \bibfield  {author} {\bibinfo {author} {\bibfnamefont {I.}~\bibnamefont {Musco}}, \bibinfo {author} {\bibfnamefont {J.~C.}\ \bibnamefont {Miller}}, \ and\ \bibinfo {author} {\bibfnamefont {A.~G.}\ \bibnamefont {Polnarev}},\ }\href {\doibase 10.1088/0264-9381/26/23/235001} {\bibfield  {journal} {\bibinfo  {journal} {Class. Quant. Grav.}\ }\textbf {\bibinfo {volume} {26}},\ \bibinfo {pages} {235001} (\bibinfo {year} {2009})},\ \Eprint {http://arxiv.org/abs/0811.1452} {arXiv:0811.1452 [gr-qc]} \BibitemShut {NoStop}%
\bibitem [{\citenamefont {Niemeyer}\ and\ \citenamefont {Jedamzik}(1999)}]{Niemeyer:1999ak}%
  \BibitemOpen
  \bibfield  {author} {\bibinfo {author} {\bibfnamefont {J.~C.}\ \bibnamefont {Niemeyer}}\ and\ \bibinfo {author} {\bibfnamefont {K.}~\bibnamefont {Jedamzik}},\ }\href {\doibase 10.1103/PhysRevD.59.124013} {\bibfield  {journal} {\bibinfo  {journal} {Phys. Rev. D}\ }\textbf {\bibinfo {volume} {59}},\ \bibinfo {pages} {124013} (\bibinfo {year} {1999})},\ \Eprint {http://arxiv.org/abs/astro-ph/9901292} {arXiv:astro-ph/9901292} \BibitemShut {NoStop}%
\bibitem [{\citenamefont {Yokoyama}(1998)}]{Yokoyama:1998xd}%
  \BibitemOpen
  \bibfield  {author} {\bibinfo {author} {\bibfnamefont {J.}~\bibnamefont {Yokoyama}},\ }\href {\doibase 10.1103/PhysRevD.58.107502} {\bibfield  {journal} {\bibinfo  {journal} {Phys. Rev. D}\ }\textbf {\bibinfo {volume} {58}},\ \bibinfo {pages} {107502} (\bibinfo {year} {1998})},\ \Eprint {http://arxiv.org/abs/gr-qc/9804041} {arXiv:gr-qc/9804041} \BibitemShut {NoStop}%
\bibitem [{\citenamefont {Jedamzik}(1997)}]{Jedamzik:1996mr}%
  \BibitemOpen
  \bibfield  {author} {\bibinfo {author} {\bibfnamefont {K.}~\bibnamefont {Jedamzik}},\ }\href {\doibase 10.1103/PhysRevD.55.R5871} {\bibfield  {journal} {\bibinfo  {journal} {Phys. Rev. D}\ }\textbf {\bibinfo {volume} {55}},\ \bibinfo {pages} {5871} (\bibinfo {year} {1997})},\ \Eprint {http://arxiv.org/abs/astro-ph/9605152} {arXiv:astro-ph/9605152} \BibitemShut {NoStop}%
\bibitem [{\citenamefont {Byrnes}\ \emph {et~al.}(2018)\citenamefont {Byrnes}, \citenamefont {Hindmarsh}, \citenamefont {Young},\ and\ \citenamefont {Hawkins}}]{Byrnes:2018clq}%
  \BibitemOpen
  \bibfield  {author} {\bibinfo {author} {\bibfnamefont {C.~T.}\ \bibnamefont {Byrnes}}, \bibinfo {author} {\bibfnamefont {M.}~\bibnamefont {Hindmarsh}}, \bibinfo {author} {\bibfnamefont {S.}~\bibnamefont {Young}}, \ and\ \bibinfo {author} {\bibfnamefont {M.~R.~S.}\ \bibnamefont {Hawkins}},\ }\href {\doibase 10.1088/1475-7516/2018/08/041} {\bibfield  {journal} {\bibinfo  {journal} {JCAP}\ }\textbf {\bibinfo {volume} {08}},\ \bibinfo {pages} {041} (\bibinfo {year} {2018})},\ \Eprint {http://arxiv.org/abs/1801.06138} {arXiv:1801.06138 [astro-ph.CO]} \BibitemShut {NoStop}%
\bibitem [{\citenamefont {Musco}\ \emph {et~al.}(2024)\citenamefont {Musco}, \citenamefont {Jedamzik},\ and\ \citenamefont {Young}}]{Musco:2023dak}%
  \BibitemOpen
  \bibfield  {author} {\bibinfo {author} {\bibfnamefont {I.}~\bibnamefont {Musco}}, \bibinfo {author} {\bibfnamefont {K.}~\bibnamefont {Jedamzik}}, \ and\ \bibinfo {author} {\bibfnamefont {S.}~\bibnamefont {Young}},\ }\href {\doibase 10.1103/PhysRevD.109.083506} {\bibfield  {journal} {\bibinfo  {journal} {Phys. Rev. D}\ }\textbf {\bibinfo {volume} {109}},\ \bibinfo {pages} {083506} (\bibinfo {year} {2024})},\ \Eprint {http://arxiv.org/abs/2303.07980} {arXiv:2303.07980 [astro-ph.CO]} \BibitemShut {NoStop}%
\bibitem [{\citenamefont {Carr}\ \emph {et~al.}(2021{\natexlab{b}})\citenamefont {Carr}, \citenamefont {Clesse}, \citenamefont {Garc\'\i{}a-Bellido},\ and\ \citenamefont {K\"uhnel}}]{Carr:2019kxo}%
  \BibitemOpen
  \bibfield  {author} {\bibinfo {author} {\bibfnamefont {B.}~\bibnamefont {Carr}}, \bibinfo {author} {\bibfnamefont {S.}~\bibnamefont {Clesse}}, \bibinfo {author} {\bibfnamefont {J.}~\bibnamefont {Garc\'\i{}a-Bellido}}, \ and\ \bibinfo {author} {\bibfnamefont {F.}~\bibnamefont {K\"uhnel}},\ }\href {\doibase 10.1016/j.dark.2020.100755} {\bibfield  {journal} {\bibinfo  {journal} {Phys. Dark Univ.}\ }\textbf {\bibinfo {volume} {31}},\ \bibinfo {pages} {100755} (\bibinfo {year} {2021}{\natexlab{b}})},\ \Eprint {http://arxiv.org/abs/1906.08217} {arXiv:1906.08217 [astro-ph.CO]} \BibitemShut {NoStop}%
\bibitem [{\citenamefont {Clesse}\ and\ \citenamefont {Garcia-Bellido}(2022)}]{Clesse:2020ghq}%
  \BibitemOpen
  \bibfield  {author} {\bibinfo {author} {\bibfnamefont {S.}~\bibnamefont {Clesse}}\ and\ \bibinfo {author} {\bibfnamefont {J.}~\bibnamefont {Garcia-Bellido}},\ }\href {\doibase 10.1016/j.dark.2022.101111} {\bibfield  {journal} {\bibinfo  {journal} {Phys. Dark Univ.}\ }\textbf {\bibinfo {volume} {38}},\ \bibinfo {pages} {101111} (\bibinfo {year} {2022})},\ \Eprint {http://arxiv.org/abs/2007.06481} {arXiv:2007.06481 [astro-ph.CO]} \BibitemShut {NoStop}%
\bibitem [{\citenamefont {Bagui}\ and\ \citenamefont {Clesse}(2022)}]{Bagui:2021dqi}%
  \BibitemOpen
  \bibfield  {author} {\bibinfo {author} {\bibfnamefont {E.}~\bibnamefont {Bagui}}\ and\ \bibinfo {author} {\bibfnamefont {S.}~\bibnamefont {Clesse}},\ }\href {\doibase 10.1016/j.dark.2022.101115} {\bibfield  {journal} {\bibinfo  {journal} {Phys. Dark Univ.}\ }\textbf {\bibinfo {volume} {38}},\ \bibinfo {pages} {101115} (\bibinfo {year} {2022})},\ \Eprint {http://arxiv.org/abs/2110.07487} {arXiv:2110.07487 [astro-ph.CO]} \BibitemShut {NoStop}%
\bibitem [{\citenamefont {Escriv\`a}\ \emph {et~al.}(2023)\citenamefont {Escriv\`a}, \citenamefont {Bagui},\ and\ \citenamefont {Clesse}}]{Escriva:2022bwe}%
  \BibitemOpen
  \bibfield  {author} {\bibinfo {author} {\bibfnamefont {A.}~\bibnamefont {Escriv\`a}}, \bibinfo {author} {\bibfnamefont {E.}~\bibnamefont {Bagui}}, \ and\ \bibinfo {author} {\bibfnamefont {S.}~\bibnamefont {Clesse}},\ }\href {\doibase 10.1088/1475-7516/2023/05/004} {\bibfield  {journal} {\bibinfo  {journal} {JCAP}\ }\textbf {\bibinfo {volume} {05}},\ \bibinfo {pages} {004} (\bibinfo {year} {2023})},\ \Eprint {http://arxiv.org/abs/2209.06196} {arXiv:2209.06196 [astro-ph.CO]} \BibitemShut {NoStop}%
\bibitem [{\citenamefont {Carr}\ \emph {et~al.}(2024)\citenamefont {Carr}, \citenamefont {Clesse}, \citenamefont {Garcia-Bellido}, \citenamefont {Hawkins},\ and\ \citenamefont {Kuhnel}}]{Carr:2023tpt}%
  \BibitemOpen
  \bibfield  {author} {\bibinfo {author} {\bibfnamefont {B.}~\bibnamefont {Carr}}, \bibinfo {author} {\bibfnamefont {S.}~\bibnamefont {Clesse}}, \bibinfo {author} {\bibfnamefont {J.}~\bibnamefont {Garcia-Bellido}}, \bibinfo {author} {\bibfnamefont {M.}~\bibnamefont {Hawkins}}, \ and\ \bibinfo {author} {\bibfnamefont {F.}~\bibnamefont {Kuhnel}},\ }\href {\doibase 10.1016/j.physrep.2023.11.005} {\bibfield  {journal} {\bibinfo  {journal} {Phys. Rept.}\ }\textbf {\bibinfo {volume} {1054}},\ \bibinfo {pages} {1} (\bibinfo {year} {2024})},\ \Eprint {http://arxiv.org/abs/2306.03903} {arXiv:2306.03903 [astro-ph.CO]} \BibitemShut {NoStop}%
\bibitem [{\citenamefont {Abbott}\ \emph {et~al.}(2021{\natexlab{b}})\citenamefont {Abbott} \emph {et~al.}}]{LIGOScientific:2020kqk}%
  \BibitemOpen
  \bibfield  {author} {\bibinfo {author} {\bibfnamefont {R.}~\bibnamefont {Abbott}} \emph {et~al.} (\bibinfo {collaboration} {LIGO Scientific, Virgo}),\ }\href {\doibase 10.3847/2041-8213/abe949} {\bibfield  {journal} {\bibinfo  {journal} {Astrophys. J. Lett.}\ }\textbf {\bibinfo {volume} {913}},\ \bibinfo {pages} {L7} (\bibinfo {year} {2021}{\natexlab{b}})},\ \Eprint {http://arxiv.org/abs/2010.14533} {arXiv:2010.14533 [astro-ph.HE]} \BibitemShut {NoStop}%
\bibitem [{\citenamefont {Thrane}\ and\ \citenamefont {Talbot}(2019)}]{Thrane:2018qnx}%
  \BibitemOpen
  \bibfield  {author} {\bibinfo {author} {\bibfnamefont {E.}~\bibnamefont {Thrane}}\ and\ \bibinfo {author} {\bibfnamefont {C.}~\bibnamefont {Talbot}},\ }\href {\doibase 10.1017/pasa.2019.2} {\bibfield  {journal} {\bibinfo  {journal} {Publ. Astron. Soc. Austral.}\ }\textbf {\bibinfo {volume} {36}},\ \bibinfo {pages} {e010} (\bibinfo {year} {2019})},\ \bibinfo {note} {[Erratum: Publ.Astron.Soc.Austral. 37, e036 (2020)]},\ \Eprint {http://arxiv.org/abs/1809.02293} {arXiv:1809.02293 [astro-ph.IM]} \BibitemShut {NoStop}%
\bibitem [{\citenamefont {Mandel}\ \emph {et~al.}(2019)\citenamefont {Mandel}, \citenamefont {Farr},\ and\ \citenamefont {Gair}}]{Mandel:2018mve}%
  \BibitemOpen
  \bibfield  {author} {\bibinfo {author} {\bibfnamefont {I.}~\bibnamefont {Mandel}}, \bibinfo {author} {\bibfnamefont {W.~M.}\ \bibnamefont {Farr}}, \ and\ \bibinfo {author} {\bibfnamefont {J.~R.}\ \bibnamefont {Gair}},\ }\href {\doibase 10.1093/mnras/stz896} {\bibfield  {journal} {\bibinfo  {journal} {Mon. Not. Roy. Astron. Soc.}\ }\textbf {\bibinfo {volume} {486}},\ \bibinfo {pages} {1086} (\bibinfo {year} {2019})},\ \Eprint {http://arxiv.org/abs/1809.02063} {arXiv:1809.02063 [physics.data-an]} \BibitemShut {NoStop}%
\bibitem [{\citenamefont {Mastrogiovanni}\ \emph {et~al.}(2024)\citenamefont {Mastrogiovanni}, \citenamefont {Pierra}, \citenamefont {Perri\`es}, \citenamefont {Laghi}, \citenamefont {Caneva~Santoro}, \citenamefont {Ghosh}, \citenamefont {Gray}, \citenamefont {Karathanasis},\ and\ \citenamefont {Leyde}}]{Mastrogiovanni:2023zbw}%
  \BibitemOpen
  \bibfield  {author} {\bibinfo {author} {\bibfnamefont {S.}~\bibnamefont {Mastrogiovanni}}, \bibinfo {author} {\bibfnamefont {G.}~\bibnamefont {Pierra}}, \bibinfo {author} {\bibfnamefont {S.}~\bibnamefont {Perri\`es}}, \bibinfo {author} {\bibfnamefont {D.}~\bibnamefont {Laghi}}, \bibinfo {author} {\bibfnamefont {G.}~\bibnamefont {Caneva~Santoro}}, \bibinfo {author} {\bibfnamefont {A.}~\bibnamefont {Ghosh}}, \bibinfo {author} {\bibfnamefont {R.}~\bibnamefont {Gray}}, \bibinfo {author} {\bibfnamefont {C.}~\bibnamefont {Karathanasis}}, \ and\ \bibinfo {author} {\bibfnamefont {K.}~\bibnamefont {Leyde}},\ }\href {\doibase 10.1051/0004-6361/202347007} {\bibfield  {journal} {\bibinfo  {journal} {Astron. Astrophys.}\ }\textbf {\bibinfo {volume} {682}},\ \bibinfo {pages} {A167} (\bibinfo {year} {2024})},\ \Eprint {http://arxiv.org/abs/2305.17973} {arXiv:2305.17973 [astro-ph.CO]} \BibitemShut {NoStop}%
\bibitem [{\citenamefont {Talbot}\ and\ \citenamefont {Golomb}(2023)}]{Talbot:2023pex}%
  \BibitemOpen
  \bibfield  {author} {\bibinfo {author} {\bibfnamefont {C.}~\bibnamefont {Talbot}}\ and\ \bibinfo {author} {\bibfnamefont {J.}~\bibnamefont {Golomb}},\ }\href {\doibase 10.1093/mnras/stad2968} {\bibfield  {journal} {\bibinfo  {journal} {Mon. Not. Roy. Astron. Soc.}\ }\textbf {\bibinfo {volume} {526}},\ \bibinfo {pages} {3495} (\bibinfo {year} {2023})},\ \Eprint {http://arxiv.org/abs/2304.06138} {arXiv:2304.06138 [astro-ph.IM]} \BibitemShut {NoStop}%
\bibitem [{\citenamefont {Abbott}\ \emph {et~al.}(2023{\natexlab{c}})\citenamefont {Abbott} \emph {et~al.}}]{LIGOScientific:2021aug}%
  \BibitemOpen
  \bibfield  {author} {\bibinfo {author} {\bibfnamefont {R.}~\bibnamefont {Abbott}} \emph {et~al.} (\bibinfo {collaboration} {LIGO Scientific, Virgo, KAGRA}),\ }\href {\doibase 10.3847/1538-4357/ac74bb} {\bibfield  {journal} {\bibinfo  {journal} {Astrophys. J.}\ }\textbf {\bibinfo {volume} {949}},\ \bibinfo {pages} {76} (\bibinfo {year} {2023}{\natexlab{c}})},\ \Eprint {http://arxiv.org/abs/2111.03604} {arXiv:2111.03604 [astro-ph.CO]} \BibitemShut {NoStop}%
\bibitem [{\citenamefont {Ajith}\ \emph {et~al.}(2008)\citenamefont {Ajith} \emph {et~al.}}]{Ajith:2007kx}%
  \BibitemOpen
  \bibfield  {author} {\bibinfo {author} {\bibfnamefont {P.}~\bibnamefont {Ajith}} \emph {et~al.},\ }\href {\doibase 10.1103/PhysRevD.77.104017} {\bibfield  {journal} {\bibinfo  {journal} {Phys. Rev. D}\ }\textbf {\bibinfo {volume} {77}},\ \bibinfo {pages} {104017} (\bibinfo {year} {2008})},\ \bibinfo {note} {[Erratum: Phys.Rev.D 79, 129901 (2009)]},\ \Eprint {http://arxiv.org/abs/0710.2335} {arXiv:0710.2335 [gr-qc]} \BibitemShut {NoStop}%
\bibitem [{\citenamefont {Abbott}\ \emph {et~al.}(2016{\natexlab{b}})\citenamefont {Abbott} \emph {et~al.}}]{LIGOScientific:2016kwr}%
  \BibitemOpen
  \bibfield  {author} {\bibinfo {author} {\bibfnamefont {B.~P.}\ \bibnamefont {Abbott}} \emph {et~al.} (\bibinfo {collaboration} {LIGO Scientific, Virgo}),\ }\href {\doibase 10.3847/2041-8205/833/1/L1} {\bibfield  {journal} {\bibinfo  {journal} {Astrophys. J. Lett.}\ }\textbf {\bibinfo {volume} {833}},\ \bibinfo {pages} {L1} (\bibinfo {year} {2016}{\natexlab{b}})},\ \Eprint {http://arxiv.org/abs/1602.03842} {arXiv:1602.03842 [astro-ph.HE]} \BibitemShut {NoStop}%
\bibitem [{\citenamefont {Abbott}\ \emph {et~al.}(2019{\natexlab{c}})\citenamefont {Abbott} \emph {et~al.}}]{LIGOScientific:2018jsj}%
  \BibitemOpen
  \bibfield  {author} {\bibinfo {author} {\bibfnamefont {B.~P.}\ \bibnamefont {Abbott}} \emph {et~al.} (\bibinfo {collaboration} {LIGO Scientific, Virgo}),\ }\href {\doibase 10.3847/2041-8213/ab3800} {\bibfield  {journal} {\bibinfo  {journal} {Astrophys. J. Lett.}\ }\textbf {\bibinfo {volume} {882}},\ \bibinfo {pages} {L24} (\bibinfo {year} {2019}{\natexlab{c}})},\ \Eprint {http://arxiv.org/abs/1811.12940} {arXiv:1811.12940 [astro-ph.HE]} \BibitemShut {NoStop}%
\bibitem [{\citenamefont {Farr}(2019)}]{Farr:2019rap}%
  \BibitemOpen
  \bibfield  {author} {\bibinfo {author} {\bibfnamefont {W.~M.}\ \bibnamefont {Farr}},\ }\href {\doibase 10.3847/2515-5172/ab1d5f} {\bibfield  {journal} {\bibinfo  {journal} {Research Notes of the AAS}\ }\textbf {\bibinfo {volume} {3}},\ \bibinfo {pages} {66} (\bibinfo {year} {2019})},\ \Eprint {http://arxiv.org/abs/1904.10879} {arXiv:1904.10879 [astro-ph.IM]} \BibitemShut {NoStop}%
\bibitem [{\citenamefont {Vaskonen}\ and\ \citenamefont {Veerm\"ae}(2021)}]{Vaskonen:2020lbd}%
  \BibitemOpen
  \bibfield  {author} {\bibinfo {author} {\bibfnamefont {V.}~\bibnamefont {Vaskonen}}\ and\ \bibinfo {author} {\bibfnamefont {H.}~\bibnamefont {Veerm\"ae}},\ }\href {\doibase 10.1103/PhysRevLett.126.051303} {\bibfield  {journal} {\bibinfo  {journal} {Phys. Rev. Lett.}\ }\textbf {\bibinfo {volume} {126}},\ \bibinfo {pages} {051303} (\bibinfo {year} {2021})},\ \Eprint {http://arxiv.org/abs/2009.07832} {arXiv:2009.07832 [astro-ph.CO]} \BibitemShut {NoStop}%
\bibitem [{\citenamefont {Raidal}\ \emph {et~al.}(2024)\citenamefont {Raidal}, \citenamefont {Vaskonen},\ and\ \citenamefont {Veerm\"ae}}]{Raidal:2024bmm}%
  \BibitemOpen
  \bibfield  {author} {\bibinfo {author} {\bibfnamefont {M.}~\bibnamefont {Raidal}}, \bibinfo {author} {\bibfnamefont {V.}~\bibnamefont {Vaskonen}}, \ and\ \bibinfo {author} {\bibfnamefont {H.}~\bibnamefont {Veerm\"ae}},\ }\href@noop {} {\  (\bibinfo {year} {2024})},\ \Eprint {http://arxiv.org/abs/2404.08416} {arXiv:2404.08416 [astro-ph.CO]} \BibitemShut {NoStop}%
\bibitem [{\citenamefont {Nakamura}\ \emph {et~al.}(1997)\citenamefont {Nakamura}, \citenamefont {Sasaki}, \citenamefont {Tanaka},\ and\ \citenamefont {Thorne}}]{Nakamura:1997sm}%
  \BibitemOpen
  \bibfield  {author} {\bibinfo {author} {\bibfnamefont {T.}~\bibnamefont {Nakamura}}, \bibinfo {author} {\bibfnamefont {M.}~\bibnamefont {Sasaki}}, \bibinfo {author} {\bibfnamefont {T.}~\bibnamefont {Tanaka}}, \ and\ \bibinfo {author} {\bibfnamefont {K.~S.}\ \bibnamefont {Thorne}},\ }\href {\doibase 10.1086/310886} {\bibfield  {journal} {\bibinfo  {journal} {Astrophys. J. Lett.}\ }\textbf {\bibinfo {volume} {487}},\ \bibinfo {pages} {L139} (\bibinfo {year} {1997})},\ \Eprint {http://arxiv.org/abs/astro-ph/9708060} {arXiv:astro-ph/9708060} \BibitemShut {NoStop}%
\bibitem [{\citenamefont {Ioka}\ \emph {et~al.}(1998)\citenamefont {Ioka}, \citenamefont {Chiba}, \citenamefont {Tanaka},\ and\ \citenamefont {Nakamura}}]{Ioka:1998nz}%
  \BibitemOpen
  \bibfield  {author} {\bibinfo {author} {\bibfnamefont {K.}~\bibnamefont {Ioka}}, \bibinfo {author} {\bibfnamefont {T.}~\bibnamefont {Chiba}}, \bibinfo {author} {\bibfnamefont {T.}~\bibnamefont {Tanaka}}, \ and\ \bibinfo {author} {\bibfnamefont {T.}~\bibnamefont {Nakamura}},\ }\href {\doibase 10.1103/PhysRevD.58.063003} {\bibfield  {journal} {\bibinfo  {journal} {Phys. Rev. D}\ }\textbf {\bibinfo {volume} {58}},\ \bibinfo {pages} {063003} (\bibinfo {year} {1998})},\ \Eprint {http://arxiv.org/abs/astro-ph/9807018} {arXiv:astro-ph/9807018} \BibitemShut {NoStop}%
\bibitem [{\citenamefont {Ali-Ha\"\i{}moud}\ \emph {et~al.}(2017)\citenamefont {Ali-Ha\"\i{}moud}, \citenamefont {Kovetz},\ and\ \citenamefont {Kamionkowski}}]{Ali-Haimoud:2017rtz}%
  \BibitemOpen
  \bibfield  {author} {\bibinfo {author} {\bibfnamefont {Y.}~\bibnamefont {Ali-Ha\"\i{}moud}}, \bibinfo {author} {\bibfnamefont {E.~D.}\ \bibnamefont {Kovetz}}, \ and\ \bibinfo {author} {\bibfnamefont {M.}~\bibnamefont {Kamionkowski}},\ }\href {\doibase 10.1103/PhysRevD.96.123523} {\bibfield  {journal} {\bibinfo  {journal} {Phys. Rev. D}\ }\textbf {\bibinfo {volume} {96}},\ \bibinfo {pages} {123523} (\bibinfo {year} {2017})},\ \Eprint {http://arxiv.org/abs/1709.06576} {arXiv:1709.06576 [astro-ph.CO]} \BibitemShut {NoStop}%
\bibitem [{\citenamefont {Mouri}\ and\ \citenamefont {Taniguchi}(2002)}]{Mouri:2002mc}%
  \BibitemOpen
  \bibfield  {author} {\bibinfo {author} {\bibfnamefont {H.}~\bibnamefont {Mouri}}\ and\ \bibinfo {author} {\bibfnamefont {Y.}~\bibnamefont {Taniguchi}},\ }\href {\doibase 10.1086/339472} {\bibfield  {journal} {\bibinfo  {journal} {Astrophys. J. Lett.}\ }\textbf {\bibinfo {volume} {566}},\ \bibinfo {pages} {L17} (\bibinfo {year} {2002})},\ \Eprint {http://arxiv.org/abs/astro-ph/0201102} {arXiv:astro-ph/0201102} \BibitemShut {NoStop}%
\bibitem [{\citenamefont {Bird}\ \emph {et~al.}(2016)\citenamefont {Bird}, \citenamefont {Cholis}, \citenamefont {Mu\~noz}, \citenamefont {Ali-Ha\"\i{}moud}, \citenamefont {Kamionkowski}, \citenamefont {Kovetz}, \citenamefont {Raccanelli},\ and\ \citenamefont {Riess}}]{Bird:2016dcv}%
  \BibitemOpen
  \bibfield  {author} {\bibinfo {author} {\bibfnamefont {S.}~\bibnamefont {Bird}}, \bibinfo {author} {\bibfnamefont {I.}~\bibnamefont {Cholis}}, \bibinfo {author} {\bibfnamefont {J.~B.}\ \bibnamefont {Mu\~noz}}, \bibinfo {author} {\bibfnamefont {Y.}~\bibnamefont {Ali-Ha\"\i{}moud}}, \bibinfo {author} {\bibfnamefont {M.}~\bibnamefont {Kamionkowski}}, \bibinfo {author} {\bibfnamefont {E.~D.}\ \bibnamefont {Kovetz}}, \bibinfo {author} {\bibfnamefont {A.}~\bibnamefont {Raccanelli}}, \ and\ \bibinfo {author} {\bibfnamefont {A.~G.}\ \bibnamefont {Riess}},\ }\href {\doibase 10.1103/PhysRevLett.116.201301} {\bibfield  {journal} {\bibinfo  {journal} {Phys. Rev. Lett.}\ }\textbf {\bibinfo {volume} {116}},\ \bibinfo {pages} {201301} (\bibinfo {year} {2016})},\ \Eprint {http://arxiv.org/abs/1603.00464} {arXiv:1603.00464 [astro-ph.CO]} \BibitemShut {NoStop}%
\bibitem [{\citenamefont {Franciolini}\ \emph {et~al.}(2022{\natexlab{c}})\citenamefont {Franciolini}, \citenamefont {Kritos}, \citenamefont {Berti},\ and\ \citenamefont {Silk}}]{Franciolini:2022ewd}%
  \BibitemOpen
  \bibfield  {author} {\bibinfo {author} {\bibfnamefont {G.}~\bibnamefont {Franciolini}}, \bibinfo {author} {\bibfnamefont {K.}~\bibnamefont {Kritos}}, \bibinfo {author} {\bibfnamefont {E.}~\bibnamefont {Berti}}, \ and\ \bibinfo {author} {\bibfnamefont {J.}~\bibnamefont {Silk}},\ }\href {\doibase 10.1103/PhysRevD.106.083529} {\bibfield  {journal} {\bibinfo  {journal} {Phys. Rev. D}\ }\textbf {\bibinfo {volume} {106}},\ \bibinfo {pages} {083529} (\bibinfo {year} {2022}{\natexlab{c}})},\ \Eprint {http://arxiv.org/abs/2205.15340} {arXiv:2205.15340 [astro-ph.CO]} \BibitemShut {NoStop}%
\bibitem [{\citenamefont {De~Luca}\ \emph {et~al.}(2020)\citenamefont {De~Luca}, \citenamefont {Desjacques}, \citenamefont {Franciolini},\ and\ \citenamefont {Riotto}}]{DeLuca:2020jug}%
  \BibitemOpen
  \bibfield  {author} {\bibinfo {author} {\bibfnamefont {V.}~\bibnamefont {De~Luca}}, \bibinfo {author} {\bibfnamefont {V.}~\bibnamefont {Desjacques}}, \bibinfo {author} {\bibfnamefont {G.}~\bibnamefont {Franciolini}}, \ and\ \bibinfo {author} {\bibfnamefont {A.}~\bibnamefont {Riotto}},\ }\href {\doibase 10.1088/1475-7516/2020/11/028} {\bibfield  {journal} {\bibinfo  {journal} {JCAP}\ }\textbf {\bibinfo {volume} {11}},\ \bibinfo {pages} {028} (\bibinfo {year} {2020})},\ \Eprint {http://arxiv.org/abs/2009.04731} {arXiv:2009.04731 [astro-ph.CO]} \BibitemShut {NoStop}%
\bibitem [{\citenamefont {{Stone}}\ and\ \citenamefont {{Leigh}}(2019)}]{2019Natur.576..406S}%
  \BibitemOpen
  \bibfield  {author} {\bibinfo {author} {\bibfnamefont {N.~C.}\ \bibnamefont {{Stone}}}\ and\ \bibinfo {author} {\bibfnamefont {N.~W.~C.}\ \bibnamefont {{Leigh}}},\ }\href {\doibase 10.1038/s41586-019-1833-8} {\bibfield  {journal} {\bibinfo  {journal} {\nat}\ }\textbf {\bibinfo {volume} {576}},\ \bibinfo {pages} {406} (\bibinfo {year} {2019})},\ \Eprint {http://arxiv.org/abs/1909.05272} {arXiv:1909.05272 [astro-ph.GA]} \BibitemShut {NoStop}%
\bibitem [{\citenamefont {De~Luca}\ \emph {et~al.}(2022)\citenamefont {De~Luca}, \citenamefont {Franciolini}, \citenamefont {Riotto},\ and\ \citenamefont {Veerm\"ae}}]{DeLuca:2022uvz}%
  \BibitemOpen
  \bibfield  {author} {\bibinfo {author} {\bibfnamefont {V.}~\bibnamefont {De~Luca}}, \bibinfo {author} {\bibfnamefont {G.}~\bibnamefont {Franciolini}}, \bibinfo {author} {\bibfnamefont {A.}~\bibnamefont {Riotto}}, \ and\ \bibinfo {author} {\bibfnamefont {H.}~\bibnamefont {Veerm\"ae}},\ }\href {\doibase 10.1103/PhysRevLett.129.191302} {\bibfield  {journal} {\bibinfo  {journal} {Phys. Rev. Lett.}\ }\textbf {\bibinfo {volume} {129}},\ \bibinfo {pages} {191302} (\bibinfo {year} {2022})},\ \Eprint {http://arxiv.org/abs/2208.01683} {arXiv:2208.01683 [astro-ph.CO]} \BibitemShut {NoStop}%
\bibitem [{\citenamefont {Carr}\ \emph {et~al.}(2017{\natexlab{a}})\citenamefont {Carr}, \citenamefont {Raidal}, \citenamefont {Tenkanen}, \citenamefont {Vaskonen},\ and\ \citenamefont {Veerm\"ae}}]{Carr:2017jsz}%
  \BibitemOpen
  \bibfield  {author} {\bibinfo {author} {\bibfnamefont {B.}~\bibnamefont {Carr}}, \bibinfo {author} {\bibfnamefont {M.}~\bibnamefont {Raidal}}, \bibinfo {author} {\bibfnamefont {T.}~\bibnamefont {Tenkanen}}, \bibinfo {author} {\bibfnamefont {V.}~\bibnamefont {Vaskonen}}, \ and\ \bibinfo {author} {\bibfnamefont {H.}~\bibnamefont {Veerm\"ae}},\ }\href {\doibase 10.1103/PhysRevD.96.023514} {\bibfield  {journal} {\bibinfo  {journal} {Phys. Rev. D}\ }\textbf {\bibinfo {volume} {96}},\ \bibinfo {pages} {023514} (\bibinfo {year} {2017}{\natexlab{a}})},\ \Eprint {http://arxiv.org/abs/1705.05567} {arXiv:1705.05567 [astro-ph.CO]} \BibitemShut {NoStop}%
\bibitem [{\citenamefont {Choptuik}(1993)}]{Choptuik:1992jv}%
  \BibitemOpen
  \bibfield  {author} {\bibinfo {author} {\bibfnamefont {M.~W.}\ \bibnamefont {Choptuik}},\ }\href {\doibase 10.1103/PhysRevLett.70.9} {\bibfield  {journal} {\bibinfo  {journal} {Phys. Rev. Lett.}\ }\textbf {\bibinfo {volume} {70}},\ \bibinfo {pages} {9} (\bibinfo {year} {1993})}\BibitemShut {NoStop}%
\bibitem [{\citenamefont {Niemeyer}\ and\ \citenamefont {Jedamzik}(1998)}]{Niemeyer:1997mt}%
  \BibitemOpen
  \bibfield  {author} {\bibinfo {author} {\bibfnamefont {J.~C.}\ \bibnamefont {Niemeyer}}\ and\ \bibinfo {author} {\bibfnamefont {K.}~\bibnamefont {Jedamzik}},\ }\href {\doibase 10.1103/PhysRevLett.80.5481} {\bibfield  {journal} {\bibinfo  {journal} {Phys. Rev. Lett.}\ }\textbf {\bibinfo {volume} {80}},\ \bibinfo {pages} {5481} (\bibinfo {year} {1998})},\ \Eprint {http://arxiv.org/abs/astro-ph/9709072} {arXiv:astro-ph/9709072} \BibitemShut {NoStop}%
\bibitem [{\citenamefont {Bugaev}\ and\ \citenamefont {Klimai}(2013)}]{Bugaev:2013vba}%
  \BibitemOpen
  \bibfield  {author} {\bibinfo {author} {\bibfnamefont {E.~V.}\ \bibnamefont {Bugaev}}\ and\ \bibinfo {author} {\bibfnamefont {P.~A.}\ \bibnamefont {Klimai}},\ }\href {\doibase 10.1142/S021827181350034X} {\bibfield  {journal} {\bibinfo  {journal} {Int. J. Mod. Phys. D}\ }\textbf {\bibinfo {volume} {22}},\ \bibinfo {pages} {1350034} (\bibinfo {year} {2013})},\ \Eprint {http://arxiv.org/abs/1303.3146} {arXiv:1303.3146 [astro-ph.CO]} \BibitemShut {NoStop}%
\bibitem [{\citenamefont {Nakama}\ \emph {et~al.}(2017)\citenamefont {Nakama}, \citenamefont {Silk},\ and\ \citenamefont {Kamionkowski}}]{Nakama:2016gzw}%
  \BibitemOpen
  \bibfield  {author} {\bibinfo {author} {\bibfnamefont {T.}~\bibnamefont {Nakama}}, \bibinfo {author} {\bibfnamefont {J.}~\bibnamefont {Silk}}, \ and\ \bibinfo {author} {\bibfnamefont {M.}~\bibnamefont {Kamionkowski}},\ }\href {\doibase 10.1103/PhysRevD.95.043511} {\bibfield  {journal} {\bibinfo  {journal} {Phys. Rev. D}\ }\textbf {\bibinfo {volume} {95}},\ \bibinfo {pages} {043511} (\bibinfo {year} {2017})},\ \Eprint {http://arxiv.org/abs/1612.06264} {arXiv:1612.06264 [astro-ph.CO]} \BibitemShut {NoStop}%
\bibitem [{\citenamefont {Byrnes}\ \emph {et~al.}(2012)\citenamefont {Byrnes}, \citenamefont {Copeland},\ and\ \citenamefont {Green}}]{Byrnes:2012yx}%
  \BibitemOpen
  \bibfield  {author} {\bibinfo {author} {\bibfnamefont {C.~T.}\ \bibnamefont {Byrnes}}, \bibinfo {author} {\bibfnamefont {E.~J.}\ \bibnamefont {Copeland}}, \ and\ \bibinfo {author} {\bibfnamefont {A.~M.}\ \bibnamefont {Green}},\ }\href {\doibase 10.1103/PhysRevD.86.043512} {\bibfield  {journal} {\bibinfo  {journal} {Phys. Rev. D}\ }\textbf {\bibinfo {volume} {86}},\ \bibinfo {pages} {043512} (\bibinfo {year} {2012})},\ \Eprint {http://arxiv.org/abs/1206.4188} {arXiv:1206.4188 [astro-ph.CO]} \BibitemShut {NoStop}%
\bibitem [{\citenamefont {Young}\ and\ \citenamefont {Byrnes}(2013)}]{Young:2013oia}%
  \BibitemOpen
  \bibfield  {author} {\bibinfo {author} {\bibfnamefont {S.}~\bibnamefont {Young}}\ and\ \bibinfo {author} {\bibfnamefont {C.~T.}\ \bibnamefont {Byrnes}},\ }\href {\doibase 10.1088/1475-7516/2013/08/052} {\bibfield  {journal} {\bibinfo  {journal} {JCAP}\ }\textbf {\bibinfo {volume} {08}},\ \bibinfo {pages} {052} (\bibinfo {year} {2013})},\ \Eprint {http://arxiv.org/abs/1307.4995} {arXiv:1307.4995 [astro-ph.CO]} \BibitemShut {NoStop}%
\bibitem [{\citenamefont {Yoo}\ \emph {et~al.}(2018)\citenamefont {Yoo}, \citenamefont {Harada}, \citenamefont {Garriga},\ and\ \citenamefont {Kohri}}]{Yoo:2018kvb}%
  \BibitemOpen
  \bibfield  {author} {\bibinfo {author} {\bibfnamefont {C.-M.}\ \bibnamefont {Yoo}}, \bibinfo {author} {\bibfnamefont {T.}~\bibnamefont {Harada}}, \bibinfo {author} {\bibfnamefont {J.}~\bibnamefont {Garriga}}, \ and\ \bibinfo {author} {\bibfnamefont {K.}~\bibnamefont {Kohri}},\ }\href {\doibase 10.1093/ptep/pty120} {\bibfield  {journal} {\bibinfo  {journal} {PTEP}\ }\textbf {\bibinfo {volume} {2018}},\ \bibinfo {pages} {123E01} (\bibinfo {year} {2018})},\ \Eprint {http://arxiv.org/abs/1805.03946} {arXiv:1805.03946 [astro-ph.CO]} \BibitemShut {NoStop}%
\bibitem [{\citenamefont {Kawasaki}\ and\ \citenamefont {Nakatsuka}(2019)}]{Kawasaki:2019mbl}%
  \BibitemOpen
  \bibfield  {author} {\bibinfo {author} {\bibfnamefont {M.}~\bibnamefont {Kawasaki}}\ and\ \bibinfo {author} {\bibfnamefont {H.}~\bibnamefont {Nakatsuka}},\ }\href {\doibase 10.1103/PhysRevD.99.123501} {\bibfield  {journal} {\bibinfo  {journal} {Phys. Rev. D}\ }\textbf {\bibinfo {volume} {99}},\ \bibinfo {pages} {123501} (\bibinfo {year} {2019})},\ \Eprint {http://arxiv.org/abs/1903.02994} {arXiv:1903.02994 [astro-ph.CO]} \BibitemShut {NoStop}%
\bibitem [{\citenamefont {Yoo}\ \emph {et~al.}(2019)\citenamefont {Yoo}, \citenamefont {Gong},\ and\ \citenamefont {Yokoyama}}]{Yoo:2019pma}%
  \BibitemOpen
  \bibfield  {author} {\bibinfo {author} {\bibfnamefont {C.-M.}\ \bibnamefont {Yoo}}, \bibinfo {author} {\bibfnamefont {J.-O.}\ \bibnamefont {Gong}}, \ and\ \bibinfo {author} {\bibfnamefont {S.}~\bibnamefont {Yokoyama}},\ }\href {\doibase 10.1088/1475-7516/2019/09/033} {\bibfield  {journal} {\bibinfo  {journal} {JCAP}\ }\textbf {\bibinfo {volume} {09}},\ \bibinfo {pages} {033} (\bibinfo {year} {2019})},\ \Eprint {http://arxiv.org/abs/1906.06790} {arXiv:1906.06790 [astro-ph.CO]} \BibitemShut {NoStop}%
\bibitem [{\citenamefont {Riccardi}\ \emph {et~al.}(2021)\citenamefont {Riccardi}, \citenamefont {Taoso},\ and\ \citenamefont {Urbano}}]{Riccardi:2021rlf}%
  \BibitemOpen
  \bibfield  {author} {\bibinfo {author} {\bibfnamefont {F.}~\bibnamefont {Riccardi}}, \bibinfo {author} {\bibfnamefont {M.}~\bibnamefont {Taoso}}, \ and\ \bibinfo {author} {\bibfnamefont {A.}~\bibnamefont {Urbano}},\ }\href {\doibase 10.1088/1475-7516/2021/08/060} {\bibfield  {journal} {\bibinfo  {journal} {JCAP}\ }\textbf {\bibinfo {volume} {08}},\ \bibinfo {pages} {060} (\bibinfo {year} {2021})},\ \Eprint {http://arxiv.org/abs/2102.04084} {arXiv:2102.04084 [astro-ph.CO]} \BibitemShut {NoStop}%
\bibitem [{\citenamefont {Taoso}\ and\ \citenamefont {Urbano}(2021)}]{Taoso:2021uvl}%
  \BibitemOpen
  \bibfield  {author} {\bibinfo {author} {\bibfnamefont {M.}~\bibnamefont {Taoso}}\ and\ \bibinfo {author} {\bibfnamefont {A.}~\bibnamefont {Urbano}},\ }\href {\doibase 10.1088/1475-7516/2021/08/016} {\bibfield  {journal} {\bibinfo  {journal} {JCAP}\ }\textbf {\bibinfo {volume} {08}},\ \bibinfo {pages} {016} (\bibinfo {year} {2021})},\ \Eprint {http://arxiv.org/abs/2102.03610} {arXiv:2102.03610 [astro-ph.CO]} \BibitemShut {NoStop}%
\bibitem [{\citenamefont {Meng}\ \emph {et~al.}(2022)\citenamefont {Meng}, \citenamefont {Yuan},\ and\ \citenamefont {Huang}}]{Meng:2022ixx}%
  \BibitemOpen
  \bibfield  {author} {\bibinfo {author} {\bibfnamefont {D.-S.}\ \bibnamefont {Meng}}, \bibinfo {author} {\bibfnamefont {C.}~\bibnamefont {Yuan}}, \ and\ \bibinfo {author} {\bibfnamefont {Q.-g.}\ \bibnamefont {Huang}},\ }\href {\doibase 10.1103/PhysRevD.106.063508} {\bibfield  {journal} {\bibinfo  {journal} {Phys. Rev. D}\ }\textbf {\bibinfo {volume} {106}},\ \bibinfo {pages} {063508} (\bibinfo {year} {2022})},\ \Eprint {http://arxiv.org/abs/2207.07668} {arXiv:2207.07668 [astro-ph.CO]} \BibitemShut {NoStop}%
\bibitem [{\citenamefont {Escriv\`a}\ \emph {et~al.}(2022)\citenamefont {Escriv\`a}, \citenamefont {Tada}, \citenamefont {Yokoyama},\ and\ \citenamefont {Yoo}}]{Escriva:2022pnz}%
  \BibitemOpen
  \bibfield  {author} {\bibinfo {author} {\bibfnamefont {A.}~\bibnamefont {Escriv\`a}}, \bibinfo {author} {\bibfnamefont {Y.}~\bibnamefont {Tada}}, \bibinfo {author} {\bibfnamefont {S.}~\bibnamefont {Yokoyama}}, \ and\ \bibinfo {author} {\bibfnamefont {C.-M.}\ \bibnamefont {Yoo}},\ }\href {\doibase 10.1088/1475-7516/2022/05/012} {\bibfield  {journal} {\bibinfo  {journal} {JCAP}\ }\textbf {\bibinfo {volume} {05}},\ \bibinfo {pages} {012} (\bibinfo {year} {2022})},\ \Eprint {http://arxiv.org/abs/2202.01028} {arXiv:2202.01028 [astro-ph.CO]} \BibitemShut {NoStop}%
\bibitem [{\citenamefont {Young}\ \emph {et~al.}(2014)\citenamefont {Young}, \citenamefont {Byrnes},\ and\ \citenamefont {Sasaki}}]{Young:2014ana}%
  \BibitemOpen
  \bibfield  {author} {\bibinfo {author} {\bibfnamefont {S.}~\bibnamefont {Young}}, \bibinfo {author} {\bibfnamefont {C.~T.}\ \bibnamefont {Byrnes}}, \ and\ \bibinfo {author} {\bibfnamefont {M.}~\bibnamefont {Sasaki}},\ }\href {\doibase 10.1088/1475-7516/2014/07/045} {\bibfield  {journal} {\bibinfo  {journal} {JCAP}\ }\textbf {\bibinfo {volume} {07}},\ \bibinfo {pages} {045} (\bibinfo {year} {2014})},\ \Eprint {http://arxiv.org/abs/1405.7023} {arXiv:1405.7023 [gr-qc]} \BibitemShut {NoStop}%
\bibitem [{\citenamefont {Shibata}\ and\ \citenamefont {Sasaki}(1999)}]{Shibata:1999zs}%
  \BibitemOpen
  \bibfield  {author} {\bibinfo {author} {\bibfnamefont {M.}~\bibnamefont {Shibata}}\ and\ \bibinfo {author} {\bibfnamefont {M.}~\bibnamefont {Sasaki}},\ }\href {\doibase 10.1103/PhysRevD.60.084002} {\bibfield  {journal} {\bibinfo  {journal} {Phys. Rev. D}\ }\textbf {\bibinfo {volume} {60}},\ \bibinfo {pages} {084002} (\bibinfo {year} {1999})},\ \Eprint {http://arxiv.org/abs/gr-qc/9905064} {arXiv:gr-qc/9905064} \BibitemShut {NoStop}%
\bibitem [{\citenamefont {Musco}(2019)}]{Musco:2018rwt}%
  \BibitemOpen
  \bibfield  {author} {\bibinfo {author} {\bibfnamefont {I.}~\bibnamefont {Musco}},\ }\href {\doibase 10.1103/PhysRevD.100.123524} {\bibfield  {journal} {\bibinfo  {journal} {Phys. Rev. D}\ }\textbf {\bibinfo {volume} {100}},\ \bibinfo {pages} {123524} (\bibinfo {year} {2019})},\ \Eprint {http://arxiv.org/abs/1809.02127} {arXiv:1809.02127 [gr-qc]} \BibitemShut {NoStop}%
\bibitem [{\citenamefont {Young}(2022)}]{Young:2022phe}%
  \BibitemOpen
  \bibfield  {author} {\bibinfo {author} {\bibfnamefont {S.}~\bibnamefont {Young}},\ }\href {\doibase 10.1088/1475-7516/2022/05/037} {\bibfield  {journal} {\bibinfo  {journal} {JCAP}\ }\textbf {\bibinfo {volume} {05}},\ \bibinfo {pages} {037} (\bibinfo {year} {2022})},\ \Eprint {http://arxiv.org/abs/2201.13345} {arXiv:2201.13345 [astro-ph.CO]} \BibitemShut {NoStop}%
\bibitem [{\citenamefont {Byrnes}\ \emph {et~al.}(2019)\citenamefont {Byrnes}, \citenamefont {Cole},\ and\ \citenamefont {Patil}}]{Byrnes:2018txb}%
  \BibitemOpen
  \bibfield  {author} {\bibinfo {author} {\bibfnamefont {C.~T.}\ \bibnamefont {Byrnes}}, \bibinfo {author} {\bibfnamefont {P.~S.}\ \bibnamefont {Cole}}, \ and\ \bibinfo {author} {\bibfnamefont {S.~P.}\ \bibnamefont {Patil}},\ }\href {\doibase 10.1088/1475-7516/2019/06/028} {\bibfield  {journal} {\bibinfo  {journal} {JCAP}\ }\textbf {\bibinfo {volume} {06}},\ \bibinfo {pages} {028} (\bibinfo {year} {2019})},\ \Eprint {http://arxiv.org/abs/1811.11158} {arXiv:1811.11158 [astro-ph.CO]} \BibitemShut {NoStop}%
\bibitem [{\citenamefont {Karam}\ \emph {et~al.}(2023{\natexlab{a}})\citenamefont {Karam}, \citenamefont {Koivunen}, \citenamefont {Tomberg}, \citenamefont {Vaskonen},\ and\ \citenamefont {Veerm\"ae}}]{Karam:2022nym}%
  \BibitemOpen
  \bibfield  {author} {\bibinfo {author} {\bibfnamefont {A.}~\bibnamefont {Karam}}, \bibinfo {author} {\bibfnamefont {N.}~\bibnamefont {Koivunen}}, \bibinfo {author} {\bibfnamefont {E.}~\bibnamefont {Tomberg}}, \bibinfo {author} {\bibfnamefont {V.}~\bibnamefont {Vaskonen}}, \ and\ \bibinfo {author} {\bibfnamefont {H.}~\bibnamefont {Veerm\"ae}},\ }\href {\doibase 10.1088/1475-7516/2023/03/013} {\bibfield  {journal} {\bibinfo  {journal} {JCAP}\ }\textbf {\bibinfo {volume} {03}},\ \bibinfo {pages} {013} (\bibinfo {year} {2023}{\natexlab{a}})},\ \Eprint {http://arxiv.org/abs/2205.13540} {arXiv:2205.13540 [astro-ph.CO]} \BibitemShut {NoStop}%
\bibitem [{\citenamefont {Harada}\ \emph {et~al.}(2015)\citenamefont {Harada}, \citenamefont {Yoo}, \citenamefont {Nakama},\ and\ \citenamefont {Koga}}]{Harada:2015yda}%
  \BibitemOpen
  \bibfield  {author} {\bibinfo {author} {\bibfnamefont {T.}~\bibnamefont {Harada}}, \bibinfo {author} {\bibfnamefont {C.-M.}\ \bibnamefont {Yoo}}, \bibinfo {author} {\bibfnamefont {T.}~\bibnamefont {Nakama}}, \ and\ \bibinfo {author} {\bibfnamefont {Y.}~\bibnamefont {Koga}},\ }\href {\doibase 10.1103/PhysRevD.91.084057} {\bibfield  {journal} {\bibinfo  {journal} {Phys. Rev. D}\ }\textbf {\bibinfo {volume} {91}},\ \bibinfo {pages} {084057} (\bibinfo {year} {2015})},\ \Eprint {http://arxiv.org/abs/1503.03934} {arXiv:1503.03934 [gr-qc]} \BibitemShut {NoStop}%
\bibitem [{\citenamefont {De~Luca}\ \emph {et~al.}(2019)\citenamefont {De~Luca}, \citenamefont {Franciolini}, \citenamefont {Kehagias}, \citenamefont {Peloso}, \citenamefont {Riotto},\ and\ \citenamefont {\"Unal}}]{DeLuca:2019qsy}%
  \BibitemOpen
  \bibfield  {author} {\bibinfo {author} {\bibfnamefont {V.}~\bibnamefont {De~Luca}}, \bibinfo {author} {\bibfnamefont {G.}~\bibnamefont {Franciolini}}, \bibinfo {author} {\bibfnamefont {A.}~\bibnamefont {Kehagias}}, \bibinfo {author} {\bibfnamefont {M.}~\bibnamefont {Peloso}}, \bibinfo {author} {\bibfnamefont {A.}~\bibnamefont {Riotto}}, \ and\ \bibinfo {author} {\bibfnamefont {C.}~\bibnamefont {\"Unal}},\ }\href {\doibase 10.1088/1475-7516/2019/07/048} {\bibfield  {journal} {\bibinfo  {journal} {JCAP}\ }\textbf {\bibinfo {volume} {07}},\ \bibinfo {pages} {048} (\bibinfo {year} {2019})},\ \Eprint {http://arxiv.org/abs/1904.00970} {arXiv:1904.00970 [astro-ph.CO]} \BibitemShut {NoStop}%
\bibitem [{\citenamefont {Young}\ \emph {et~al.}(2019)\citenamefont {Young}, \citenamefont {Musco},\ and\ \citenamefont {Byrnes}}]{Young:2019yug}%
  \BibitemOpen
  \bibfield  {author} {\bibinfo {author} {\bibfnamefont {S.}~\bibnamefont {Young}}, \bibinfo {author} {\bibfnamefont {I.}~\bibnamefont {Musco}}, \ and\ \bibinfo {author} {\bibfnamefont {C.~T.}\ \bibnamefont {Byrnes}},\ }\href {\doibase 10.1088/1475-7516/2019/11/012} {\bibfield  {journal} {\bibinfo  {journal} {JCAP}\ }\textbf {\bibinfo {volume} {11}},\ \bibinfo {pages} {012} (\bibinfo {year} {2019})},\ \Eprint {http://arxiv.org/abs/1904.00984} {arXiv:1904.00984 [astro-ph.CO]} \BibitemShut {NoStop}%
\bibitem [{\citenamefont {Germani}\ and\ \citenamefont {Sheth}(2020)}]{Germani:2019zez}%
  \BibitemOpen
  \bibfield  {author} {\bibinfo {author} {\bibfnamefont {C.}~\bibnamefont {Germani}}\ and\ \bibinfo {author} {\bibfnamefont {R.~K.}\ \bibnamefont {Sheth}},\ }\href {\doibase 10.1103/PhysRevD.101.063520} {\bibfield  {journal} {\bibinfo  {journal} {Phys. Rev. D}\ }\textbf {\bibinfo {volume} {101}},\ \bibinfo {pages} {063520} (\bibinfo {year} {2020})},\ \Eprint {http://arxiv.org/abs/1912.07072} {arXiv:1912.07072 [astro-ph.CO]} \BibitemShut {NoStop}%
\bibitem [{\citenamefont {Polnarev}\ and\ \citenamefont {Musco}(2007)}]{Polnarev:2006aa}%
  \BibitemOpen
  \bibfield  {author} {\bibinfo {author} {\bibfnamefont {A.~G.}\ \bibnamefont {Polnarev}}\ and\ \bibinfo {author} {\bibfnamefont {I.}~\bibnamefont {Musco}},\ }\href {\doibase 10.1088/0264-9381/24/6/003} {\bibfield  {journal} {\bibinfo  {journal} {Class. Quant. Grav.}\ }\textbf {\bibinfo {volume} {24}},\ \bibinfo {pages} {1405} (\bibinfo {year} {2007})},\ \Eprint {http://arxiv.org/abs/gr-qc/0605122} {arXiv:gr-qc/0605122} \BibitemShut {NoStop}%
\bibitem [{\citenamefont {Garcia-Bellido}\ and\ \citenamefont {Ruiz~Morales}(2017)}]{Garcia-Bellido:2017mdw}%
  \BibitemOpen
  \bibfield  {author} {\bibinfo {author} {\bibfnamefont {J.}~\bibnamefont {Garcia-Bellido}}\ and\ \bibinfo {author} {\bibfnamefont {E.}~\bibnamefont {Ruiz~Morales}},\ }\href {\doibase 10.1016/j.dark.2017.09.007} {\bibfield  {journal} {\bibinfo  {journal} {Phys. Dark Univ.}\ }\textbf {\bibinfo {volume} {18}},\ \bibinfo {pages} {47} (\bibinfo {year} {2017})},\ \Eprint {http://arxiv.org/abs/1702.03901} {arXiv:1702.03901 [astro-ph.CO]} \BibitemShut {NoStop}%
\bibitem [{\citenamefont {Pi}\ \emph {et~al.}(2018)\citenamefont {Pi}, \citenamefont {Zhang}, \citenamefont {Huang},\ and\ \citenamefont {Sasaki}}]{Pi:2017gih}%
  \BibitemOpen
  \bibfield  {author} {\bibinfo {author} {\bibfnamefont {S.}~\bibnamefont {Pi}}, \bibinfo {author} {\bibfnamefont {Y.-l.}\ \bibnamefont {Zhang}}, \bibinfo {author} {\bibfnamefont {Q.-G.}\ \bibnamefont {Huang}}, \ and\ \bibinfo {author} {\bibfnamefont {M.}~\bibnamefont {Sasaki}},\ }\href {\doibase 10.1088/1475-7516/2018/05/042} {\bibfield  {journal} {\bibinfo  {journal} {JCAP}\ }\textbf {\bibinfo {volume} {05}},\ \bibinfo {pages} {042} (\bibinfo {year} {2018})},\ \Eprint {http://arxiv.org/abs/1712.09896} {arXiv:1712.09896 [astro-ph.CO]} \BibitemShut {NoStop}%
\bibitem [{\citenamefont {Kannike}\ \emph {et~al.}(2017)\citenamefont {Kannike}, \citenamefont {Marzola}, \citenamefont {Raidal},\ and\ \citenamefont {Veerm\"ae}}]{Kannike:2017bxn}%
  \BibitemOpen
  \bibfield  {author} {\bibinfo {author} {\bibfnamefont {K.}~\bibnamefont {Kannike}}, \bibinfo {author} {\bibfnamefont {L.}~\bibnamefont {Marzola}}, \bibinfo {author} {\bibfnamefont {M.}~\bibnamefont {Raidal}}, \ and\ \bibinfo {author} {\bibfnamefont {H.}~\bibnamefont {Veerm\"ae}},\ }\href {\doibase 10.1088/1475-7516/2017/09/020} {\bibfield  {journal} {\bibinfo  {journal} {JCAP}\ }\textbf {\bibinfo {volume} {09}},\ \bibinfo {pages} {020} (\bibinfo {year} {2017})},\ \Eprint {http://arxiv.org/abs/1705.06225} {arXiv:1705.06225 [astro-ph.CO]} \BibitemShut {NoStop}%
\bibitem [{\citenamefont {Ballesteros}\ \emph {et~al.}(2020)\citenamefont {Ballesteros}, \citenamefont {Rey}, \citenamefont {Taoso},\ and\ \citenamefont {Urbano}}]{Ballesteros:2020qam}%
  \BibitemOpen
  \bibfield  {author} {\bibinfo {author} {\bibfnamefont {G.}~\bibnamefont {Ballesteros}}, \bibinfo {author} {\bibfnamefont {J.}~\bibnamefont {Rey}}, \bibinfo {author} {\bibfnamefont {M.}~\bibnamefont {Taoso}}, \ and\ \bibinfo {author} {\bibfnamefont {A.}~\bibnamefont {Urbano}},\ }\href {\doibase 10.1088/1475-7516/2020/07/025} {\bibfield  {journal} {\bibinfo  {journal} {JCAP}\ }\textbf {\bibinfo {volume} {07}},\ \bibinfo {pages} {025} (\bibinfo {year} {2020})},\ \Eprint {http://arxiv.org/abs/2001.08220} {arXiv:2001.08220 [astro-ph.CO]} \BibitemShut {NoStop}%
\bibitem [{\citenamefont {Inomata}\ \emph {et~al.}(2017)\citenamefont {Inomata}, \citenamefont {Kawasaki}, \citenamefont {Mukaida}, \citenamefont {Tada},\ and\ \citenamefont {Yanagida}}]{Inomata:2016rbd}%
  \BibitemOpen
  \bibfield  {author} {\bibinfo {author} {\bibfnamefont {K.}~\bibnamefont {Inomata}}, \bibinfo {author} {\bibfnamefont {M.}~\bibnamefont {Kawasaki}}, \bibinfo {author} {\bibfnamefont {K.}~\bibnamefont {Mukaida}}, \bibinfo {author} {\bibfnamefont {Y.}~\bibnamefont {Tada}}, \ and\ \bibinfo {author} {\bibfnamefont {T.~T.}\ \bibnamefont {Yanagida}},\ }\href {\doibase 10.1103/PhysRevD.95.123510} {\bibfield  {journal} {\bibinfo  {journal} {Phys. Rev. D}\ }\textbf {\bibinfo {volume} {95}},\ \bibinfo {pages} {123510} (\bibinfo {year} {2017})},\ \Eprint {http://arxiv.org/abs/1611.06130} {arXiv:1611.06130 [astro-ph.CO]} \BibitemShut {NoStop}%
\bibitem [{\citenamefont {Iacconi}\ \emph {et~al.}(2022)\citenamefont {Iacconi}, \citenamefont {Assadullahi}, \citenamefont {Fasiello},\ and\ \citenamefont {Wands}}]{Iacconi:2021ltm}%
  \BibitemOpen
  \bibfield  {author} {\bibinfo {author} {\bibfnamefont {L.}~\bibnamefont {Iacconi}}, \bibinfo {author} {\bibfnamefont {H.}~\bibnamefont {Assadullahi}}, \bibinfo {author} {\bibfnamefont {M.}~\bibnamefont {Fasiello}}, \ and\ \bibinfo {author} {\bibfnamefont {D.}~\bibnamefont {Wands}},\ }\href {\doibase 10.1088/1475-7516/2022/06/007} {\bibfield  {journal} {\bibinfo  {journal} {JCAP}\ }\textbf {\bibinfo {volume} {06}},\ \bibinfo {pages} {007} (\bibinfo {year} {2022})},\ \Eprint {http://arxiv.org/abs/2112.05092} {arXiv:2112.05092 [astro-ph.CO]} \BibitemShut {NoStop}%
\bibitem [{\citenamefont {Kawai}\ and\ \citenamefont {Kim}(2021)}]{Kawai:2021edk}%
  \BibitemOpen
  \bibfield  {author} {\bibinfo {author} {\bibfnamefont {S.}~\bibnamefont {Kawai}}\ and\ \bibinfo {author} {\bibfnamefont {J.}~\bibnamefont {Kim}},\ }\href {\doibase 10.1103/PhysRevD.104.083545} {\bibfield  {journal} {\bibinfo  {journal} {Phys. Rev. D}\ }\textbf {\bibinfo {volume} {104}},\ \bibinfo {pages} {083545} (\bibinfo {year} {2021})},\ \Eprint {http://arxiv.org/abs/2108.01340} {arXiv:2108.01340 [astro-ph.CO]} \BibitemShut {NoStop}%
\bibitem [{\citenamefont {Bhaumik}\ and\ \citenamefont {Jain}(2020)}]{Bhaumik:2019tvl}%
  \BibitemOpen
  \bibfield  {author} {\bibinfo {author} {\bibfnamefont {N.}~\bibnamefont {Bhaumik}}\ and\ \bibinfo {author} {\bibfnamefont {R.~K.}\ \bibnamefont {Jain}},\ }\href {\doibase 10.1088/1475-7516/2020/01/037} {\bibfield  {journal} {\bibinfo  {journal} {JCAP}\ }\textbf {\bibinfo {volume} {01}},\ \bibinfo {pages} {037} (\bibinfo {year} {2020})},\ \Eprint {http://arxiv.org/abs/1907.04125} {arXiv:1907.04125 [astro-ph.CO]} \BibitemShut {NoStop}%
\bibitem [{\citenamefont {Cheong}\ \emph {et~al.}(2021)\citenamefont {Cheong}, \citenamefont {Lee},\ and\ \citenamefont {Park}}]{Cheong:2019vzl}%
  \BibitemOpen
  \bibfield  {author} {\bibinfo {author} {\bibfnamefont {D.~Y.}\ \bibnamefont {Cheong}}, \bibinfo {author} {\bibfnamefont {S.~M.}\ \bibnamefont {Lee}}, \ and\ \bibinfo {author} {\bibfnamefont {S.~C.}\ \bibnamefont {Park}},\ }\href {\doibase 10.1088/1475-7516/2021/01/032} {\bibfield  {journal} {\bibinfo  {journal} {JCAP}\ }\textbf {\bibinfo {volume} {01}},\ \bibinfo {pages} {032} (\bibinfo {year} {2021})},\ \Eprint {http://arxiv.org/abs/1912.12032} {arXiv:1912.12032 [hep-ph]} \BibitemShut {NoStop}%
\bibitem [{\citenamefont {Inomata}\ \emph {et~al.}(2018)\citenamefont {Inomata}, \citenamefont {Kawasaki}, \citenamefont {Mukaida},\ and\ \citenamefont {Yanagida}}]{Inomata:2018cht}%
  \BibitemOpen
  \bibfield  {author} {\bibinfo {author} {\bibfnamefont {K.}~\bibnamefont {Inomata}}, \bibinfo {author} {\bibfnamefont {M.}~\bibnamefont {Kawasaki}}, \bibinfo {author} {\bibfnamefont {K.}~\bibnamefont {Mukaida}}, \ and\ \bibinfo {author} {\bibfnamefont {T.~T.}\ \bibnamefont {Yanagida}},\ }\href {\doibase 10.1103/PhysRevD.97.043514} {\bibfield  {journal} {\bibinfo  {journal} {Phys. Rev. D}\ }\textbf {\bibinfo {volume} {97}},\ \bibinfo {pages} {043514} (\bibinfo {year} {2018})},\ \Eprint {http://arxiv.org/abs/1711.06129} {arXiv:1711.06129 [astro-ph.CO]} \BibitemShut {NoStop}%
\bibitem [{\citenamefont {Dalianis}\ \emph {et~al.}(2019)\citenamefont {Dalianis}, \citenamefont {Kehagias},\ and\ \citenamefont {Tringas}}]{Dalianis:2018frf}%
  \BibitemOpen
  \bibfield  {author} {\bibinfo {author} {\bibfnamefont {I.}~\bibnamefont {Dalianis}}, \bibinfo {author} {\bibfnamefont {A.}~\bibnamefont {Kehagias}}, \ and\ \bibinfo {author} {\bibfnamefont {G.}~\bibnamefont {Tringas}},\ }\href {\doibase 10.1088/1475-7516/2019/01/037} {\bibfield  {journal} {\bibinfo  {journal} {JCAP}\ }\textbf {\bibinfo {volume} {01}},\ \bibinfo {pages} {037} (\bibinfo {year} {2019})},\ \Eprint {http://arxiv.org/abs/1805.09483} {arXiv:1805.09483 [astro-ph.CO]} \BibitemShut {NoStop}%
\bibitem [{\citenamefont {Motohashi}\ \emph {et~al.}(2020)\citenamefont {Motohashi}, \citenamefont {Mukohyama},\ and\ \citenamefont {Oliosi}}]{Motohashi:2019rhu}%
  \BibitemOpen
  \bibfield  {author} {\bibinfo {author} {\bibfnamefont {H.}~\bibnamefont {Motohashi}}, \bibinfo {author} {\bibfnamefont {S.}~\bibnamefont {Mukohyama}}, \ and\ \bibinfo {author} {\bibfnamefont {M.}~\bibnamefont {Oliosi}},\ }\href {\doibase 10.1088/1475-7516/2020/03/002} {\bibfield  {journal} {\bibinfo  {journal} {JCAP}\ }\textbf {\bibinfo {volume} {03}},\ \bibinfo {pages} {002} (\bibinfo {year} {2020})},\ \Eprint {http://arxiv.org/abs/1910.13235} {arXiv:1910.13235 [gr-qc]} \BibitemShut {NoStop}%
\bibitem [{\citenamefont {Hertzberg}\ and\ \citenamefont {Yamada}(2018)}]{Hertzberg:2017dkh}%
  \BibitemOpen
  \bibfield  {author} {\bibinfo {author} {\bibfnamefont {M.~P.}\ \bibnamefont {Hertzberg}}\ and\ \bibinfo {author} {\bibfnamefont {M.}~\bibnamefont {Yamada}},\ }\href {\doibase 10.1103/PhysRevD.97.083509} {\bibfield  {journal} {\bibinfo  {journal} {Phys. Rev. D}\ }\textbf {\bibinfo {volume} {97}},\ \bibinfo {pages} {083509} (\bibinfo {year} {2018})},\ \Eprint {http://arxiv.org/abs/1712.09750} {arXiv:1712.09750 [astro-ph.CO]} \BibitemShut {NoStop}%
\bibitem [{\citenamefont {Ballesteros}\ and\ \citenamefont {Taoso}(2018)}]{Ballesteros:2017fsr}%
  \BibitemOpen
  \bibfield  {author} {\bibinfo {author} {\bibfnamefont {G.}~\bibnamefont {Ballesteros}}\ and\ \bibinfo {author} {\bibfnamefont {M.}~\bibnamefont {Taoso}},\ }\href {\doibase 10.1103/PhysRevD.97.023501} {\bibfield  {journal} {\bibinfo  {journal} {Phys. Rev. D}\ }\textbf {\bibinfo {volume} {97}},\ \bibinfo {pages} {023501} (\bibinfo {year} {2018})},\ \Eprint {http://arxiv.org/abs/1709.05565} {arXiv:1709.05565 [hep-ph]} \BibitemShut {NoStop}%
\bibitem [{\citenamefont {Rasanen}\ and\ \citenamefont {Tomberg}(2019)}]{Rasanen:2018fom}%
  \BibitemOpen
  \bibfield  {author} {\bibinfo {author} {\bibfnamefont {S.}~\bibnamefont {Rasanen}}\ and\ \bibinfo {author} {\bibfnamefont {E.}~\bibnamefont {Tomberg}},\ }\href {\doibase 10.1088/1475-7516/2019/01/038} {\bibfield  {journal} {\bibinfo  {journal} {JCAP}\ }\textbf {\bibinfo {volume} {01}},\ \bibinfo {pages} {038} (\bibinfo {year} {2019})},\ \Eprint {http://arxiv.org/abs/1810.12608} {arXiv:1810.12608 [astro-ph.CO]} \BibitemShut {NoStop}%
\bibitem [{\citenamefont {Balaji}\ \emph {et~al.}(2022)\citenamefont {Balaji}, \citenamefont {Silk},\ and\ \citenamefont {Wu}}]{Balaji:2022rsy}%
  \BibitemOpen
  \bibfield  {author} {\bibinfo {author} {\bibfnamefont {S.}~\bibnamefont {Balaji}}, \bibinfo {author} {\bibfnamefont {J.}~\bibnamefont {Silk}}, \ and\ \bibinfo {author} {\bibfnamefont {Y.-P.}\ \bibnamefont {Wu}},\ }\href {\doibase 10.1088/1475-7516/2022/06/008} {\bibfield  {journal} {\bibinfo  {journal} {JCAP}\ }\textbf {\bibinfo {volume} {06}},\ \bibinfo {pages} {008} (\bibinfo {year} {2022})},\ \Eprint {http://arxiv.org/abs/2202.00700} {arXiv:2202.00700 [astro-ph.CO]} \BibitemShut {NoStop}%
\bibitem [{\citenamefont {Frolovsky}\ and\ \citenamefont {Ketov}(2023)}]{Frolovsky:2023hqd}%
  \BibitemOpen
  \bibfield  {author} {\bibinfo {author} {\bibfnamefont {D.}~\bibnamefont {Frolovsky}}\ and\ \bibinfo {author} {\bibfnamefont {S.~V.}\ \bibnamefont {Ketov}},\ }\href {\doibase 10.3390/universe9060294} {\bibfield  {journal} {\bibinfo  {journal} {Universe}\ }\textbf {\bibinfo {volume} {9}},\ \bibinfo {pages} {294} (\bibinfo {year} {2023})},\ \Eprint {http://arxiv.org/abs/2304.12558} {arXiv:2304.12558 [astro-ph.CO]} \BibitemShut {NoStop}%
\bibitem [{\citenamefont {Dimopoulos}(2017)}]{Dimopoulos:2017ged}%
  \BibitemOpen
  \bibfield  {author} {\bibinfo {author} {\bibfnamefont {K.}~\bibnamefont {Dimopoulos}},\ }\href {\doibase 10.1016/j.physletb.2017.10.066} {\bibfield  {journal} {\bibinfo  {journal} {Phys. Lett. B}\ }\textbf {\bibinfo {volume} {775}},\ \bibinfo {pages} {262} (\bibinfo {year} {2017})},\ \Eprint {http://arxiv.org/abs/1707.05644} {arXiv:1707.05644 [hep-ph]} \BibitemShut {NoStop}%
\bibitem [{\citenamefont {Germani}\ and\ \citenamefont {Prokopec}(2017)}]{Germani:2017bcs}%
  \BibitemOpen
  \bibfield  {author} {\bibinfo {author} {\bibfnamefont {C.}~\bibnamefont {Germani}}\ and\ \bibinfo {author} {\bibfnamefont {T.}~\bibnamefont {Prokopec}},\ }\href {\doibase 10.1016/j.dark.2017.09.001} {\bibfield  {journal} {\bibinfo  {journal} {Phys. Dark Univ.}\ }\textbf {\bibinfo {volume} {18}},\ \bibinfo {pages} {6} (\bibinfo {year} {2017})},\ \Eprint {http://arxiv.org/abs/1706.04226} {arXiv:1706.04226 [astro-ph.CO]} \BibitemShut {NoStop}%
\bibitem [{\citenamefont {Choudhury}\ and\ \citenamefont {Mazumdar}(2014)}]{Choudhury:2013woa}%
  \BibitemOpen
  \bibfield  {author} {\bibinfo {author} {\bibfnamefont {S.}~\bibnamefont {Choudhury}}\ and\ \bibinfo {author} {\bibfnamefont {A.}~\bibnamefont {Mazumdar}},\ }\href {\doibase 10.1016/j.physletb.2014.04.050} {\bibfield  {journal} {\bibinfo  {journal} {Phys. Lett. B}\ }\textbf {\bibinfo {volume} {733}},\ \bibinfo {pages} {270} (\bibinfo {year} {2014})},\ \Eprint {http://arxiv.org/abs/1307.5119} {arXiv:1307.5119 [astro-ph.CO]} \BibitemShut {NoStop}%
\bibitem [{\citenamefont {Ragavendra}\ and\ \citenamefont {Sriramkumar}(2023)}]{Ragavendra:2023ret}%
  \BibitemOpen
  \bibfield  {author} {\bibinfo {author} {\bibfnamefont {H.~V.}\ \bibnamefont {Ragavendra}}\ and\ \bibinfo {author} {\bibfnamefont {L.}~\bibnamefont {Sriramkumar}},\ }\href {\doibase 10.3390/galaxies11010034} {\bibfield  {journal} {\bibinfo  {journal} {Galaxies}\ }\textbf {\bibinfo {volume} {11}},\ \bibinfo {pages} {34} (\bibinfo {year} {2023})},\ \Eprint {http://arxiv.org/abs/2301.08887} {arXiv:2301.08887 [astro-ph.CO]} \BibitemShut {NoStop}%
\bibitem [{\citenamefont {Cheng}\ \emph {et~al.}(2022)\citenamefont {Cheng}, \citenamefont {Lee},\ and\ \citenamefont {Ng}}]{Cheng:2021lif}%
  \BibitemOpen
  \bibfield  {author} {\bibinfo {author} {\bibfnamefont {S.-L.}\ \bibnamefont {Cheng}}, \bibinfo {author} {\bibfnamefont {D.-S.}\ \bibnamefont {Lee}}, \ and\ \bibinfo {author} {\bibfnamefont {K.-W.}\ \bibnamefont {Ng}},\ }\href {\doibase 10.1016/j.physletb.2022.136956} {\bibfield  {journal} {\bibinfo  {journal} {Phys. Lett. B}\ }\textbf {\bibinfo {volume} {827}},\ \bibinfo {pages} {136956} (\bibinfo {year} {2022})},\ \Eprint {http://arxiv.org/abs/2106.09275} {arXiv:2106.09275 [astro-ph.CO]} \BibitemShut {NoStop}%
\bibitem [{\citenamefont {Franciolini}\ \emph {et~al.}(2023{\natexlab{a}})\citenamefont {Franciolini}, \citenamefont {Iovino}, \citenamefont {Taoso},\ and\ \citenamefont {Urbano}}]{Franciolini:2023lgy}%
  \BibitemOpen
  \bibfield  {author} {\bibinfo {author} {\bibfnamefont {G.}~\bibnamefont {Franciolini}}, \bibinfo {author} {\bibfnamefont {A.}~\bibnamefont {Iovino}, \bibfnamefont {Junior.}}, \bibinfo {author} {\bibfnamefont {M.}~\bibnamefont {Taoso}}, \ and\ \bibinfo {author} {\bibfnamefont {A.}~\bibnamefont {Urbano}},\ }\href@noop {} {\  (\bibinfo {year} {2023}{\natexlab{a}})},\ \Eprint {http://arxiv.org/abs/2305.03491} {arXiv:2305.03491 [astro-ph.CO]} \BibitemShut {NoStop}%
\bibitem [{\citenamefont {Karam}\ \emph {et~al.}(2023{\natexlab{b}})\citenamefont {Karam}, \citenamefont {Koivunen}, \citenamefont {Tomberg}, \citenamefont {Racioppi},\ and\ \citenamefont {Veerm\"ae}}]{Karam:2023haj}%
  \BibitemOpen
  \bibfield  {author} {\bibinfo {author} {\bibfnamefont {A.}~\bibnamefont {Karam}}, \bibinfo {author} {\bibfnamefont {N.}~\bibnamefont {Koivunen}}, \bibinfo {author} {\bibfnamefont {E.}~\bibnamefont {Tomberg}}, \bibinfo {author} {\bibfnamefont {A.}~\bibnamefont {Racioppi}}, \ and\ \bibinfo {author} {\bibfnamefont {H.}~\bibnamefont {Veerm\"ae}},\ }\href {\doibase 10.1088/1475-7516/2023/09/002} {\bibfield  {journal} {\bibinfo  {journal} {JCAP}\ }\textbf {\bibinfo {volume} {09}},\ \bibinfo {pages} {002} (\bibinfo {year} {2023}{\natexlab{b}})},\ \Eprint {http://arxiv.org/abs/2305.09630} {arXiv:2305.09630 [astro-ph.CO]} \BibitemShut {NoStop}%
\bibitem [{\citenamefont {Mishra}\ \emph {et~al.}(2023)\citenamefont {Mishra}, \citenamefont {Copeland},\ and\ \citenamefont {Green}}]{Mishra:2023lhe}%
  \BibitemOpen
  \bibfield  {author} {\bibinfo {author} {\bibfnamefont {S.~S.}\ \bibnamefont {Mishra}}, \bibinfo {author} {\bibfnamefont {E.~J.}\ \bibnamefont {Copeland}}, \ and\ \bibinfo {author} {\bibfnamefont {A.~M.}\ \bibnamefont {Green}},\ }\href {\doibase 10.1088/1475-7516/2023/09/005} {\bibfield  {journal} {\bibinfo  {journal} {JCAP}\ }\textbf {\bibinfo {volume} {09}},\ \bibinfo {pages} {005} (\bibinfo {year} {2023})},\ \Eprint {http://arxiv.org/abs/2303.17375} {arXiv:2303.17375 [astro-ph.CO]} \BibitemShut {NoStop}%
\bibitem [{\citenamefont {Cole}\ \emph {et~al.}(2023)\citenamefont {Cole}, \citenamefont {Gow}, \citenamefont {Byrnes},\ and\ \citenamefont {Patil}}]{Cole:2023wyx}%
  \BibitemOpen
  \bibfield  {author} {\bibinfo {author} {\bibfnamefont {P.~S.}\ \bibnamefont {Cole}}, \bibinfo {author} {\bibfnamefont {A.~D.}\ \bibnamefont {Gow}}, \bibinfo {author} {\bibfnamefont {C.~T.}\ \bibnamefont {Byrnes}}, \ and\ \bibinfo {author} {\bibfnamefont {S.~P.}\ \bibnamefont {Patil}},\ }\href {\doibase 10.1088/1475-7516/2023/08/031} {\bibfield  {journal} {\bibinfo  {journal} {JCAP}\ }\textbf {\bibinfo {volume} {08}},\ \bibinfo {pages} {031} (\bibinfo {year} {2023})},\ \Eprint {http://arxiv.org/abs/2304.01997} {arXiv:2304.01997 [astro-ph.CO]} \BibitemShut {NoStop}%
\bibitem [{\citenamefont {Frosina}\ and\ \citenamefont {Urbano}(2023)}]{Frosina:2023nxu}%
  \BibitemOpen
  \bibfield  {author} {\bibinfo {author} {\bibfnamefont {L.}~\bibnamefont {Frosina}}\ and\ \bibinfo {author} {\bibfnamefont {A.}~\bibnamefont {Urbano}},\ }\href {\doibase 10.1103/PhysRevD.108.103544} {\bibfield  {journal} {\bibinfo  {journal} {Phys. Rev. D}\ }\textbf {\bibinfo {volume} {108}},\ \bibinfo {pages} {103544} (\bibinfo {year} {2023})},\ \Eprint {http://arxiv.org/abs/2308.06915} {arXiv:2308.06915 [astro-ph.CO]} \BibitemShut {NoStop}%
\bibitem [{\citenamefont {Franciolini}\ and\ \citenamefont {Urbano}(2022)}]{Franciolini:2022pav}%
  \BibitemOpen
  \bibfield  {author} {\bibinfo {author} {\bibfnamefont {G.}~\bibnamefont {Franciolini}}\ and\ \bibinfo {author} {\bibfnamefont {A.}~\bibnamefont {Urbano}},\ }\href {\doibase 10.1103/PhysRevD.106.123519} {\bibfield  {journal} {\bibinfo  {journal} {Phys. Rev. D}\ }\textbf {\bibinfo {volume} {106}},\ \bibinfo {pages} {123519} (\bibinfo {year} {2022})},\ \Eprint {http://arxiv.org/abs/2207.10056} {arXiv:2207.10056 [astro-ph.CO]} \BibitemShut {NoStop}%
\bibitem [{\citenamefont {Choudhury}\ \emph {et~al.}(2024)\citenamefont {Choudhury}, \citenamefont {Karde}, \citenamefont {Panda},\ and\ \citenamefont {Sami}}]{Choudhury:2024one}%
  \BibitemOpen
  \bibfield  {author} {\bibinfo {author} {\bibfnamefont {S.}~\bibnamefont {Choudhury}}, \bibinfo {author} {\bibfnamefont {A.}~\bibnamefont {Karde}}, \bibinfo {author} {\bibfnamefont {S.}~\bibnamefont {Panda}}, \ and\ \bibinfo {author} {\bibfnamefont {M.}~\bibnamefont {Sami}},\ }\href@noop {} {\  (\bibinfo {year} {2024})},\ \Eprint {http://arxiv.org/abs/2401.10925} {arXiv:2401.10925 [astro-ph.CO]} \BibitemShut {NoStop}%
\bibitem [{\citenamefont {Wang}\ \emph {et~al.}(2024)\citenamefont {Wang}, \citenamefont {Zhang},\ and\ \citenamefont {Sasaki}}]{Wang:2024vfv}%
  \BibitemOpen
  \bibfield  {author} {\bibinfo {author} {\bibfnamefont {X.}~\bibnamefont {Wang}}, \bibinfo {author} {\bibfnamefont {Y.-l.}\ \bibnamefont {Zhang}}, \ and\ \bibinfo {author} {\bibfnamefont {M.}~\bibnamefont {Sasaki}},\ }\href@noop {} {\  (\bibinfo {year} {2024})},\ \Eprint {http://arxiv.org/abs/2404.02492} {arXiv:2404.02492 [astro-ph.CO]} \BibitemShut {NoStop}%
\bibitem [{\citenamefont {Stamou}(2021)}]{Stamou:2021qdk}%
  \BibitemOpen
  \bibfield  {author} {\bibinfo {author} {\bibfnamefont {I.~D.}\ \bibnamefont {Stamou}},\ }\href {\doibase 10.1103/PhysRevD.103.083512} {\bibfield  {journal} {\bibinfo  {journal} {Phys. Rev. D}\ }\textbf {\bibinfo {volume} {103}},\ \bibinfo {pages} {083512} (\bibinfo {year} {2021})},\ \Eprint {http://arxiv.org/abs/2104.08654} {arXiv:2104.08654 [hep-ph]} \BibitemShut {NoStop}%
\bibitem [{\citenamefont {Stamou}(2024)}]{Stamou:2024lqf}%
  \BibitemOpen
  \bibfield  {author} {\bibinfo {author} {\bibfnamefont {I.}~\bibnamefont {Stamou}},\ }\href@noop {} {\  (\bibinfo {year} {2024})},\ \Eprint {http://arxiv.org/abs/2404.14321} {arXiv:2404.14321 [astro-ph.CO]} \BibitemShut {NoStop}%
\bibitem [{\citenamefont {Heydari}\ and\ \citenamefont {Karami}(2022{\natexlab{a}})}]{Heydari:2021gea}%
  \BibitemOpen
  \bibfield  {author} {\bibinfo {author} {\bibfnamefont {S.}~\bibnamefont {Heydari}}\ and\ \bibinfo {author} {\bibfnamefont {K.}~\bibnamefont {Karami}},\ }\href {\doibase 10.1140/epjc/s10052-022-10036-2} {\bibfield  {journal} {\bibinfo  {journal} {Eur. Phys. J. C}\ }\textbf {\bibinfo {volume} {82}},\ \bibinfo {pages} {83} (\bibinfo {year} {2022}{\natexlab{a}})},\ \Eprint {http://arxiv.org/abs/2107.10550} {arXiv:2107.10550 [gr-qc]} \BibitemShut {NoStop}%
\bibitem [{\citenamefont {Heydari}\ and\ \citenamefont {Karami}(2022{\natexlab{b}})}]{Heydari:2021qsr}%
  \BibitemOpen
  \bibfield  {author} {\bibinfo {author} {\bibfnamefont {S.}~\bibnamefont {Heydari}}\ and\ \bibinfo {author} {\bibfnamefont {K.}~\bibnamefont {Karami}},\ }\href {\doibase 10.1088/1475-7516/2022/03/033} {\bibfield  {journal} {\bibinfo  {journal} {JCAP}\ }\textbf {\bibinfo {volume} {03}},\ \bibinfo {pages} {033} (\bibinfo {year} {2022}{\natexlab{b}})},\ \Eprint {http://arxiv.org/abs/2111.00494} {arXiv:2111.00494 [gr-qc]} \BibitemShut {NoStop}%
\bibitem [{\citenamefont {Heydari}\ and\ \citenamefont {Karami}(2024)}]{Heydari:2023xts}%
  \BibitemOpen
  \bibfield  {author} {\bibinfo {author} {\bibfnamefont {S.}~\bibnamefont {Heydari}}\ and\ \bibinfo {author} {\bibfnamefont {K.}~\bibnamefont {Karami}},\ }\href {\doibase 10.1088/1475-7516/2024/02/047} {\bibfield  {journal} {\bibinfo  {journal} {JCAP}\ }\textbf {\bibinfo {volume} {02}},\ \bibinfo {pages} {047} (\bibinfo {year} {2024})},\ \Eprint {http://arxiv.org/abs/2309.01239} {arXiv:2309.01239 [astro-ph.CO]} \BibitemShut {NoStop}%
\bibitem [{\citenamefont {Atal}\ \emph {et~al.}(2019)\citenamefont {Atal}, \citenamefont {Garriga},\ and\ \citenamefont {Marcos-Caballero}}]{Atal:2019cdz}%
  \BibitemOpen
  \bibfield  {author} {\bibinfo {author} {\bibfnamefont {V.}~\bibnamefont {Atal}}, \bibinfo {author} {\bibfnamefont {J.}~\bibnamefont {Garriga}}, \ and\ \bibinfo {author} {\bibfnamefont {A.}~\bibnamefont {Marcos-Caballero}},\ }\href {\doibase 10.1088/1475-7516/2019/09/073} {\bibfield  {journal} {\bibinfo  {journal} {JCAP}\ }\textbf {\bibinfo {volume} {09}},\ \bibinfo {pages} {073} (\bibinfo {year} {2019})},\ \Eprint {http://arxiv.org/abs/1905.13202} {arXiv:1905.13202 [astro-ph.CO]} \BibitemShut {NoStop}%
\bibitem [{\citenamefont {Tomberg}(2023)}]{Tomberg:2023kli}%
  \BibitemOpen
  \bibfield  {author} {\bibinfo {author} {\bibfnamefont {E.}~\bibnamefont {Tomberg}},\ }\href {\doibase 10.1103/PhysRevD.108.043502} {\bibfield  {journal} {\bibinfo  {journal} {Phys. Rev. D}\ }\textbf {\bibinfo {volume} {108}},\ \bibinfo {pages} {043502} (\bibinfo {year} {2023})},\ \Eprint {http://arxiv.org/abs/2304.10903} {arXiv:2304.10903 [astro-ph.CO]} \BibitemShut {NoStop}%
\bibitem [{\citenamefont {Biagetti}\ \emph {et~al.}(2018)\citenamefont {Biagetti}, \citenamefont {Franciolini}, \citenamefont {Kehagias},\ and\ \citenamefont {Riotto}}]{Biagetti:2018pjj}%
  \BibitemOpen
  \bibfield  {author} {\bibinfo {author} {\bibfnamefont {M.}~\bibnamefont {Biagetti}}, \bibinfo {author} {\bibfnamefont {G.}~\bibnamefont {Franciolini}}, \bibinfo {author} {\bibfnamefont {A.}~\bibnamefont {Kehagias}}, \ and\ \bibinfo {author} {\bibfnamefont {A.}~\bibnamefont {Riotto}},\ }\href {\doibase 10.1088/1475-7516/2018/07/032} {\bibfield  {journal} {\bibinfo  {journal} {JCAP}\ }\textbf {\bibinfo {volume} {07}},\ \bibinfo {pages} {032} (\bibinfo {year} {2018})},\ \Eprint {http://arxiv.org/abs/1804.07124} {arXiv:1804.07124 [astro-ph.CO]} \BibitemShut {NoStop}%
\bibitem [{\citenamefont {Ezquiaga}\ and\ \citenamefont {Garc\'\i{}a-Bellido}(2018)}]{Ezquiaga:2018gbw}%
  \BibitemOpen
  \bibfield  {author} {\bibinfo {author} {\bibfnamefont {J.~M.}\ \bibnamefont {Ezquiaga}}\ and\ \bibinfo {author} {\bibfnamefont {J.}~\bibnamefont {Garc\'\i{}a-Bellido}},\ }\href {\doibase 10.1088/1475-7516/2018/08/018} {\bibfield  {journal} {\bibinfo  {journal} {JCAP}\ }\textbf {\bibinfo {volume} {08}},\ \bibinfo {pages} {018} (\bibinfo {year} {2018})},\ \Eprint {http://arxiv.org/abs/1805.06731} {arXiv:1805.06731 [astro-ph.CO]} \BibitemShut {NoStop}%
\bibitem [{\citenamefont {Ezquiaga}\ \emph {et~al.}(2020)\citenamefont {Ezquiaga}, \citenamefont {Garc\'\i{}a-Bellido},\ and\ \citenamefont {Vennin}}]{Ezquiaga:2019ftu}%
  \BibitemOpen
  \bibfield  {author} {\bibinfo {author} {\bibfnamefont {J.~M.}\ \bibnamefont {Ezquiaga}}, \bibinfo {author} {\bibfnamefont {J.}~\bibnamefont {Garc\'\i{}a-Bellido}}, \ and\ \bibinfo {author} {\bibfnamefont {V.}~\bibnamefont {Vennin}},\ }\href {\doibase 10.1088/1475-7516/2020/03/029} {\bibfield  {journal} {\bibinfo  {journal} {JCAP}\ }\textbf {\bibinfo {volume} {03}},\ \bibinfo {pages} {029} (\bibinfo {year} {2020})},\ \Eprint {http://arxiv.org/abs/1912.05399} {arXiv:1912.05399 [astro-ph.CO]} \BibitemShut {NoStop}%
\bibitem [{\citenamefont {Enqvist}\ and\ \citenamefont {Sloth}(2002)}]{Enqvist:2001zp}%
  \BibitemOpen
  \bibfield  {author} {\bibinfo {author} {\bibfnamefont {K.}~\bibnamefont {Enqvist}}\ and\ \bibinfo {author} {\bibfnamefont {M.~S.}\ \bibnamefont {Sloth}},\ }\href {\doibase 10.1016/S0550-3213(02)00043-3} {\bibfield  {journal} {\bibinfo  {journal} {Nucl. Phys. B}\ }\textbf {\bibinfo {volume} {626}},\ \bibinfo {pages} {395} (\bibinfo {year} {2002})},\ \Eprint {http://arxiv.org/abs/hep-ph/0109214} {arXiv:hep-ph/0109214} \BibitemShut {NoStop}%
\bibitem [{\citenamefont {Lyth}\ and\ \citenamefont {Wands}(2002)}]{Lyth:2001nq}%
  \BibitemOpen
  \bibfield  {author} {\bibinfo {author} {\bibfnamefont {D.~H.}\ \bibnamefont {Lyth}}\ and\ \bibinfo {author} {\bibfnamefont {D.}~\bibnamefont {Wands}},\ }\href {\doibase 10.1016/S0370-2693(01)01366-1} {\bibfield  {journal} {\bibinfo  {journal} {Phys. Lett. B}\ }\textbf {\bibinfo {volume} {524}},\ \bibinfo {pages} {5} (\bibinfo {year} {2002})},\ \Eprint {http://arxiv.org/abs/hep-ph/0110002} {arXiv:hep-ph/0110002} \BibitemShut {NoStop}%
\bibitem [{\citenamefont {Sloth}(2003)}]{Sloth:2002xn}%
  \BibitemOpen
  \bibfield  {author} {\bibinfo {author} {\bibfnamefont {M.~S.}\ \bibnamefont {Sloth}},\ }\href {\doibase 10.1016/S0550-3213(03)00114-7} {\bibfield  {journal} {\bibinfo  {journal} {Nucl. Phys. B}\ }\textbf {\bibinfo {volume} {656}},\ \bibinfo {pages} {239} (\bibinfo {year} {2003})},\ \Eprint {http://arxiv.org/abs/hep-ph/0208241} {arXiv:hep-ph/0208241} \BibitemShut {NoStop}%
\bibitem [{\citenamefont {Lyth}\ \emph {et~al.}(2003)\citenamefont {Lyth}, \citenamefont {Ungarelli},\ and\ \citenamefont {Wands}}]{Lyth:2002my}%
  \BibitemOpen
  \bibfield  {author} {\bibinfo {author} {\bibfnamefont {D.~H.}\ \bibnamefont {Lyth}}, \bibinfo {author} {\bibfnamefont {C.}~\bibnamefont {Ungarelli}}, \ and\ \bibinfo {author} {\bibfnamefont {D.}~\bibnamefont {Wands}},\ }\href {\doibase 10.1103/PhysRevD.67.023503} {\bibfield  {journal} {\bibinfo  {journal} {Phys. Rev. D}\ }\textbf {\bibinfo {volume} {67}},\ \bibinfo {pages} {023503} (\bibinfo {year} {2003})},\ \Eprint {http://arxiv.org/abs/astro-ph/0208055} {arXiv:astro-ph/0208055} \BibitemShut {NoStop}%
\bibitem [{\citenamefont {Dimopoulos}\ \emph {et~al.}(2003)\citenamefont {Dimopoulos}, \citenamefont {Lazarides}, \citenamefont {Lyth},\ and\ \citenamefont {Ruiz~de Austri}}]{Dimopoulos:2003ii}%
  \BibitemOpen
  \bibfield  {author} {\bibinfo {author} {\bibfnamefont {K.}~\bibnamefont {Dimopoulos}}, \bibinfo {author} {\bibfnamefont {G.}~\bibnamefont {Lazarides}}, \bibinfo {author} {\bibfnamefont {D.}~\bibnamefont {Lyth}}, \ and\ \bibinfo {author} {\bibfnamefont {R.}~\bibnamefont {Ruiz~de Austri}},\ }\href {\doibase 10.1088/1126-6708/2003/05/057} {\bibfield  {journal} {\bibinfo  {journal} {JHEP}\ }\textbf {\bibinfo {volume} {05}},\ \bibinfo {pages} {057} (\bibinfo {year} {2003})},\ \Eprint {http://arxiv.org/abs/hep-ph/0303154} {arXiv:hep-ph/0303154} \BibitemShut {NoStop}%
\bibitem [{\citenamefont {Kohri}\ \emph {et~al.}(2013)\citenamefont {Kohri}, \citenamefont {Lin},\ and\ \citenamefont {Matsuda}}]{Kohri:2012yw}%
  \BibitemOpen
  \bibfield  {author} {\bibinfo {author} {\bibfnamefont {K.}~\bibnamefont {Kohri}}, \bibinfo {author} {\bibfnamefont {C.-M.}\ \bibnamefont {Lin}}, \ and\ \bibinfo {author} {\bibfnamefont {T.}~\bibnamefont {Matsuda}},\ }\href {\doibase 10.1103/PhysRevD.87.103527} {\bibfield  {journal} {\bibinfo  {journal} {Phys. Rev. D}\ }\textbf {\bibinfo {volume} {87}},\ \bibinfo {pages} {103527} (\bibinfo {year} {2013})},\ \Eprint {http://arxiv.org/abs/1211.2371} {arXiv:1211.2371 [hep-ph]} \BibitemShut {NoStop}%
\bibitem [{\citenamefont {Kawasaki}\ \emph {et~al.}(2013{\natexlab{a}})\citenamefont {Kawasaki}, \citenamefont {Kitajima},\ and\ \citenamefont {Yanagida}}]{Kawasaki:2012wr}%
  \BibitemOpen
  \bibfield  {author} {\bibinfo {author} {\bibfnamefont {M.}~\bibnamefont {Kawasaki}}, \bibinfo {author} {\bibfnamefont {N.}~\bibnamefont {Kitajima}}, \ and\ \bibinfo {author} {\bibfnamefont {T.~T.}\ \bibnamefont {Yanagida}},\ }\href {\doibase 10.1103/PhysRevD.87.063519} {\bibfield  {journal} {\bibinfo  {journal} {Phys. Rev. D}\ }\textbf {\bibinfo {volume} {87}},\ \bibinfo {pages} {063519} (\bibinfo {year} {2013}{\natexlab{a}})},\ \Eprint {http://arxiv.org/abs/1207.2550} {arXiv:1207.2550 [hep-ph]} \BibitemShut {NoStop}%
\bibitem [{\citenamefont {Kawasaki}\ \emph {et~al.}(2013{\natexlab{b}})\citenamefont {Kawasaki}, \citenamefont {Kitajima},\ and\ \citenamefont {Yokoyama}}]{Kawasaki:2013xsa}%
  \BibitemOpen
  \bibfield  {author} {\bibinfo {author} {\bibfnamefont {M.}~\bibnamefont {Kawasaki}}, \bibinfo {author} {\bibfnamefont {N.}~\bibnamefont {Kitajima}}, \ and\ \bibinfo {author} {\bibfnamefont {S.}~\bibnamefont {Yokoyama}},\ }\href {\doibase 10.1088/1475-7516/2013/08/042} {\bibfield  {journal} {\bibinfo  {journal} {JCAP}\ }\textbf {\bibinfo {volume} {08}},\ \bibinfo {pages} {042} (\bibinfo {year} {2013}{\natexlab{b}})},\ \Eprint {http://arxiv.org/abs/1305.4464} {arXiv:1305.4464 [astro-ph.CO]} \BibitemShut {NoStop}%
\bibitem [{\citenamefont {Carr}\ \emph {et~al.}(2017{\natexlab{b}})\citenamefont {Carr}, \citenamefont {Tenkanen},\ and\ \citenamefont {Vaskonen}}]{Carr:2017edp}%
  \BibitemOpen
  \bibfield  {author} {\bibinfo {author} {\bibfnamefont {B.}~\bibnamefont {Carr}}, \bibinfo {author} {\bibfnamefont {T.}~\bibnamefont {Tenkanen}}, \ and\ \bibinfo {author} {\bibfnamefont {V.}~\bibnamefont {Vaskonen}},\ }\href {\doibase 10.1103/PhysRevD.96.063507} {\bibfield  {journal} {\bibinfo  {journal} {Phys. Rev. D}\ }\textbf {\bibinfo {volume} {96}},\ \bibinfo {pages} {063507} (\bibinfo {year} {2017}{\natexlab{b}})},\ \Eprint {http://arxiv.org/abs/1706.03746} {arXiv:1706.03746 [astro-ph.CO]} \BibitemShut {NoStop}%
\bibitem [{\citenamefont {Ando}\ \emph {et~al.}(2018{\natexlab{a}})\citenamefont {Ando}, \citenamefont {Inomata}, \citenamefont {Kawasaki}, \citenamefont {Mukaida},\ and\ \citenamefont {Yanagida}}]{Ando:2017veq}%
  \BibitemOpen
  \bibfield  {author} {\bibinfo {author} {\bibfnamefont {K.}~\bibnamefont {Ando}}, \bibinfo {author} {\bibfnamefont {K.}~\bibnamefont {Inomata}}, \bibinfo {author} {\bibfnamefont {M.}~\bibnamefont {Kawasaki}}, \bibinfo {author} {\bibfnamefont {K.}~\bibnamefont {Mukaida}}, \ and\ \bibinfo {author} {\bibfnamefont {T.~T.}\ \bibnamefont {Yanagida}},\ }\href {\doibase 10.1103/PhysRevD.97.123512} {\bibfield  {journal} {\bibinfo  {journal} {Phys. Rev. D}\ }\textbf {\bibinfo {volume} {97}},\ \bibinfo {pages} {123512} (\bibinfo {year} {2018}{\natexlab{a}})},\ \Eprint {http://arxiv.org/abs/1711.08956} {arXiv:1711.08956 [astro-ph.CO]} \BibitemShut {NoStop}%
\bibitem [{\citenamefont {Ando}\ \emph {et~al.}(2018{\natexlab{b}})\citenamefont {Ando}, \citenamefont {Kawasaki},\ and\ \citenamefont {Nakatsuka}}]{Ando:2018nge}%
  \BibitemOpen
  \bibfield  {author} {\bibinfo {author} {\bibfnamefont {K.}~\bibnamefont {Ando}}, \bibinfo {author} {\bibfnamefont {M.}~\bibnamefont {Kawasaki}}, \ and\ \bibinfo {author} {\bibfnamefont {H.}~\bibnamefont {Nakatsuka}},\ }\href {\doibase 10.1103/PhysRevD.98.083508} {\bibfield  {journal} {\bibinfo  {journal} {Phys. Rev. D}\ }\textbf {\bibinfo {volume} {98}},\ \bibinfo {pages} {083508} (\bibinfo {year} {2018}{\natexlab{b}})},\ \Eprint {http://arxiv.org/abs/1805.07757} {arXiv:1805.07757 [astro-ph.CO]} \BibitemShut {NoStop}%
\bibitem [{\citenamefont {Chen}\ and\ \citenamefont {Cai}(2019)}]{Chen:2019zza}%
  \BibitemOpen
  \bibfield  {author} {\bibinfo {author} {\bibfnamefont {C.}~\bibnamefont {Chen}}\ and\ \bibinfo {author} {\bibfnamefont {Y.-F.}\ \bibnamefont {Cai}},\ }\href {\doibase 10.1088/1475-7516/2019/10/068} {\bibfield  {journal} {\bibinfo  {journal} {JCAP}\ }\textbf {\bibinfo {volume} {10}},\ \bibinfo {pages} {068} (\bibinfo {year} {2019})},\ \Eprint {http://arxiv.org/abs/1908.03942} {arXiv:1908.03942 [astro-ph.CO]} \BibitemShut {NoStop}%
\bibitem [{\citenamefont {Liu}\ and\ \citenamefont {Prokopec}(2021)}]{Liu:2020zzv}%
  \BibitemOpen
  \bibfield  {author} {\bibinfo {author} {\bibfnamefont {L.-H.}\ \bibnamefont {Liu}}\ and\ \bibinfo {author} {\bibfnamefont {T.}~\bibnamefont {Prokopec}},\ }\href {\doibase 10.1088/1475-7516/2021/06/033} {\bibfield  {journal} {\bibinfo  {journal} {JCAP}\ }\textbf {\bibinfo {volume} {06}},\ \bibinfo {pages} {033} (\bibinfo {year} {2021})},\ \Eprint {http://arxiv.org/abs/2005.11069} {arXiv:2005.11069 [astro-ph.CO]} \BibitemShut {NoStop}%
\bibitem [{\citenamefont {Pi}\ and\ \citenamefont {Sasaki}(2023{\natexlab{a}})}]{Pi:2021dft}%
  \BibitemOpen
  \bibfield  {author} {\bibinfo {author} {\bibfnamefont {S.}~\bibnamefont {Pi}}\ and\ \bibinfo {author} {\bibfnamefont {M.}~\bibnamefont {Sasaki}},\ }\href {\doibase 10.1103/PhysRevD.108.L101301} {\bibfield  {journal} {\bibinfo  {journal} {Phys. Rev. D}\ }\textbf {\bibinfo {volume} {108}},\ \bibinfo {pages} {L101301} (\bibinfo {year} {2023}{\natexlab{a}})},\ \Eprint {http://arxiv.org/abs/2112.12680} {arXiv:2112.12680 [astro-ph.CO]} \BibitemShut {NoStop}%
\bibitem [{\citenamefont {Cai}\ \emph {et~al.}(2021)\citenamefont {Cai}, \citenamefont {Chen},\ and\ \citenamefont {Fu}}]{Cai:2021wzd}%
  \BibitemOpen
  \bibfield  {author} {\bibinfo {author} {\bibfnamefont {R.-G.}\ \bibnamefont {Cai}}, \bibinfo {author} {\bibfnamefont {C.}~\bibnamefont {Chen}}, \ and\ \bibinfo {author} {\bibfnamefont {C.}~\bibnamefont {Fu}},\ }\href {\doibase 10.1103/PhysRevD.104.083537} {\bibfield  {journal} {\bibinfo  {journal} {Phys. Rev. D}\ }\textbf {\bibinfo {volume} {104}},\ \bibinfo {pages} {083537} (\bibinfo {year} {2021})},\ \Eprint {http://arxiv.org/abs/2108.03422} {arXiv:2108.03422 [astro-ph.CO]} \BibitemShut {NoStop}%
\bibitem [{\citenamefont {Liu}(2023)}]{Liu:2021rgq}%
  \BibitemOpen
  \bibfield  {author} {\bibinfo {author} {\bibfnamefont {L.-H.}\ \bibnamefont {Liu}},\ }\href {\doibase 10.1088/1674-1137/ac9d28} {\bibfield  {journal} {\bibinfo  {journal} {Chin. Phys. C}\ }\textbf {\bibinfo {volume} {47}},\ \bibinfo {pages} {015105} (\bibinfo {year} {2023})},\ \Eprint {http://arxiv.org/abs/2107.07310} {arXiv:2107.07310 [astro-ph.CO]} \BibitemShut {NoStop}%
\bibitem [{\citenamefont {Chen}\ \emph {et~al.}(2023)\citenamefont {Chen}, \citenamefont {Ghoshal}, \citenamefont {Lalak}, \citenamefont {Luo},\ and\ \citenamefont {Naskar}}]{Chen:2023lou}%
  \BibitemOpen
  \bibfield  {author} {\bibinfo {author} {\bibfnamefont {C.}~\bibnamefont {Chen}}, \bibinfo {author} {\bibfnamefont {A.}~\bibnamefont {Ghoshal}}, \bibinfo {author} {\bibfnamefont {Z.}~\bibnamefont {Lalak}}, \bibinfo {author} {\bibfnamefont {Y.}~\bibnamefont {Luo}}, \ and\ \bibinfo {author} {\bibfnamefont {A.}~\bibnamefont {Naskar}},\ }\href {\doibase 10.1088/1475-7516/2023/08/041} {\bibfield  {journal} {\bibinfo  {journal} {JCAP}\ }\textbf {\bibinfo {volume} {08}},\ \bibinfo {pages} {041} (\bibinfo {year} {2023})},\ \Eprint {http://arxiv.org/abs/2305.12325} {arXiv:2305.12325 [astro-ph.CO]} \BibitemShut {NoStop}%
\bibitem [{\citenamefont {Torrado}\ \emph {et~al.}(2018)\citenamefont {Torrado}, \citenamefont {Byrnes}, \citenamefont {Hardwick}, \citenamefont {Vennin},\ and\ \citenamefont {Wands}}]{Torrado:2017qtr}%
  \BibitemOpen
  \bibfield  {author} {\bibinfo {author} {\bibfnamefont {J.}~\bibnamefont {Torrado}}, \bibinfo {author} {\bibfnamefont {C.~T.}\ \bibnamefont {Byrnes}}, \bibinfo {author} {\bibfnamefont {R.~J.}\ \bibnamefont {Hardwick}}, \bibinfo {author} {\bibfnamefont {V.}~\bibnamefont {Vennin}}, \ and\ \bibinfo {author} {\bibfnamefont {D.}~\bibnamefont {Wands}},\ }\href {\doibase 10.1103/PhysRevD.98.063525} {\bibfield  {journal} {\bibinfo  {journal} {Phys. Rev. D}\ }\textbf {\bibinfo {volume} {98}},\ \bibinfo {pages} {063525} (\bibinfo {year} {2018})},\ \Eprint {http://arxiv.org/abs/1712.05364} {arXiv:1712.05364 [astro-ph.CO]} \BibitemShut {NoStop}%
\bibitem [{\citenamefont {Wilkins}\ and\ \citenamefont {Cable}(2024)}]{Wilkins:2023asp}%
  \BibitemOpen
  \bibfield  {author} {\bibinfo {author} {\bibfnamefont {A.}~\bibnamefont {Wilkins}}\ and\ \bibinfo {author} {\bibfnamefont {A.}~\bibnamefont {Cable}},\ }\href {\doibase 10.1088/1475-7516/2024/02/026} {\bibfield  {journal} {\bibinfo  {journal} {JCAP}\ }\textbf {\bibinfo {volume} {02}},\ \bibinfo {pages} {026} (\bibinfo {year} {2024})},\ \Eprint {http://arxiv.org/abs/2306.09232} {arXiv:2306.09232 [astro-ph.CO]} \BibitemShut {NoStop}%
\bibitem [{\citenamefont {Ferrante}\ \emph {et~al.}(2023{\natexlab{a}})\citenamefont {Ferrante}, \citenamefont {Franciolini}, \citenamefont {Iovino},\ and\ \citenamefont {Urbano}}]{Ferrante:2023bgz}%
  \BibitemOpen
  \bibfield  {author} {\bibinfo {author} {\bibfnamefont {G.}~\bibnamefont {Ferrante}}, \bibinfo {author} {\bibfnamefont {G.}~\bibnamefont {Franciolini}}, \bibinfo {author} {\bibfnamefont {A.}~\bibnamefont {Iovino}, \bibfnamefont {Junior.}}, \ and\ \bibinfo {author} {\bibfnamefont {A.}~\bibnamefont {Urbano}},\ }\href {\doibase 10.1088/1475-7516/2023/06/057} {\bibfield  {journal} {\bibinfo  {journal} {JCAP}\ }\textbf {\bibinfo {volume} {06}},\ \bibinfo {pages} {057} (\bibinfo {year} {2023}{\natexlab{a}})},\ \Eprint {http://arxiv.org/abs/2305.13382} {arXiv:2305.13382 [astro-ph.CO]} \BibitemShut {NoStop}%
\bibitem [{\citenamefont {Inomata}\ \emph {et~al.}(2024)\citenamefont {Inomata}, \citenamefont {Kawasaki}, \citenamefont {Mukaida},\ and\ \citenamefont {Yanagida}}]{Inomata:2023drn}%
  \BibitemOpen
  \bibfield  {author} {\bibinfo {author} {\bibfnamefont {K.}~\bibnamefont {Inomata}}, \bibinfo {author} {\bibfnamefont {M.}~\bibnamefont {Kawasaki}}, \bibinfo {author} {\bibfnamefont {K.}~\bibnamefont {Mukaida}}, \ and\ \bibinfo {author} {\bibfnamefont {T.~T.}\ \bibnamefont {Yanagida}},\ }\href {\doibase 10.1103/PhysRevD.109.043508} {\bibfield  {journal} {\bibinfo  {journal} {Phys. Rev. D}\ }\textbf {\bibinfo {volume} {109}},\ \bibinfo {pages} {043508} (\bibinfo {year} {2024})},\ \Eprint {http://arxiv.org/abs/2309.11398} {arXiv:2309.11398 [astro-ph.CO]} \BibitemShut {NoStop}%
\bibitem [{\citenamefont {Sasaki}\ \emph {et~al.}(2006)\citenamefont {Sasaki}, \citenamefont {Valiviita},\ and\ \citenamefont {Wands}}]{Sasaki:2006kq}%
  \BibitemOpen
  \bibfield  {author} {\bibinfo {author} {\bibfnamefont {M.}~\bibnamefont {Sasaki}}, \bibinfo {author} {\bibfnamefont {J.}~\bibnamefont {Valiviita}}, \ and\ \bibinfo {author} {\bibfnamefont {D.}~\bibnamefont {Wands}},\ }\href {\doibase 10.1103/PhysRevD.74.103003} {\bibfield  {journal} {\bibinfo  {journal} {Phys. Rev. D}\ }\textbf {\bibinfo {volume} {74}},\ \bibinfo {pages} {103003} (\bibinfo {year} {2006})},\ \Eprint {http://arxiv.org/abs/astro-ph/0607627} {arXiv:astro-ph/0607627} \BibitemShut {NoStop}%
\bibitem [{\citenamefont {Pi}\ and\ \citenamefont {Sasaki}(2023{\natexlab{b}})}]{Pi:2022ysn}%
  \BibitemOpen
  \bibfield  {author} {\bibinfo {author} {\bibfnamefont {S.}~\bibnamefont {Pi}}\ and\ \bibinfo {author} {\bibfnamefont {M.}~\bibnamefont {Sasaki}},\ }\href {\doibase 10.1103/PhysRevLett.131.011002} {\bibfield  {journal} {\bibinfo  {journal} {Phys. Rev. Lett.}\ }\textbf {\bibinfo {volume} {131}},\ \bibinfo {pages} {011002} (\bibinfo {year} {2023}{\natexlab{b}})},\ \Eprint {http://arxiv.org/abs/2211.13932} {arXiv:2211.13932 [astro-ph.CO]} \BibitemShut {NoStop}%
\bibitem [{\citenamefont {Enqvist}\ and\ \citenamefont {Takahashi}(2008)}]{Enqvist:2008gk}%
  \BibitemOpen
  \bibfield  {author} {\bibinfo {author} {\bibfnamefont {K.}~\bibnamefont {Enqvist}}\ and\ \bibinfo {author} {\bibfnamefont {T.}~\bibnamefont {Takahashi}},\ }\href {\doibase 10.1088/1475-7516/2008/09/012} {\bibfield  {journal} {\bibinfo  {journal} {JCAP}\ }\textbf {\bibinfo {volume} {09}},\ \bibinfo {pages} {012} (\bibinfo {year} {2008})},\ \Eprint {http://arxiv.org/abs/0807.3069} {arXiv:0807.3069 [astro-ph]} \BibitemShut {NoStop}%
\bibitem [{\citenamefont {Enqvist}\ \emph {et~al.}(2009)\citenamefont {Enqvist}, \citenamefont {Nurmi}, \citenamefont {Rigopoulos}, \citenamefont {Taanila},\ and\ \citenamefont {Takahashi}}]{Enqvist:2009zf}%
  \BibitemOpen
  \bibfield  {author} {\bibinfo {author} {\bibfnamefont {K.}~\bibnamefont {Enqvist}}, \bibinfo {author} {\bibfnamefont {S.}~\bibnamefont {Nurmi}}, \bibinfo {author} {\bibfnamefont {G.}~\bibnamefont {Rigopoulos}}, \bibinfo {author} {\bibfnamefont {O.}~\bibnamefont {Taanila}}, \ and\ \bibinfo {author} {\bibfnamefont {T.}~\bibnamefont {Takahashi}},\ }\href {\doibase 10.1088/1475-7516/2009/11/003} {\bibfield  {journal} {\bibinfo  {journal} {JCAP}\ }\textbf {\bibinfo {volume} {11}},\ \bibinfo {pages} {003} (\bibinfo {year} {2009})},\ \Eprint {http://arxiv.org/abs/0906.3126} {arXiv:0906.3126 [astro-ph.CO]} \BibitemShut {NoStop}%
\bibitem [{\citenamefont {Huang}(2008)}]{Huang:2008zj}%
  \BibitemOpen
  \bibfield  {author} {\bibinfo {author} {\bibfnamefont {Q.-G.}\ \bibnamefont {Huang}},\ }\href {\doibase 10.1088/1475-7516/2008/11/005} {\bibfield  {journal} {\bibinfo  {journal} {JCAP}\ }\textbf {\bibinfo {volume} {11}},\ \bibinfo {pages} {005} (\bibinfo {year} {2008})},\ \Eprint {http://arxiv.org/abs/0808.1793} {arXiv:0808.1793 [hep-th]} \BibitemShut {NoStop}%
\bibitem [{\citenamefont {Fonseca}\ and\ \citenamefont {Wands}(2011)}]{Fonseca:2011aa}%
  \BibitemOpen
  \bibfield  {author} {\bibinfo {author} {\bibfnamefont {J.}~\bibnamefont {Fonseca}}\ and\ \bibinfo {author} {\bibfnamefont {D.}~\bibnamefont {Wands}},\ }\href {\doibase 10.1103/PhysRevD.83.064025} {\bibfield  {journal} {\bibinfo  {journal} {Phys. Rev. D}\ }\textbf {\bibinfo {volume} {83}},\ \bibinfo {pages} {064025} (\bibinfo {year} {2011})},\ \Eprint {http://arxiv.org/abs/1101.1254} {arXiv:1101.1254 [astro-ph.CO]} \BibitemShut {NoStop}%
\bibitem [{\citenamefont {Byrnes}\ \emph {et~al.}(2011)\citenamefont {Byrnes}, \citenamefont {Enqvist}, \citenamefont {Nurmi},\ and\ \citenamefont {Takahashi}}]{Byrnes:2011gh}%
  \BibitemOpen
  \bibfield  {author} {\bibinfo {author} {\bibfnamefont {C.~T.}\ \bibnamefont {Byrnes}}, \bibinfo {author} {\bibfnamefont {K.}~\bibnamefont {Enqvist}}, \bibinfo {author} {\bibfnamefont {S.}~\bibnamefont {Nurmi}}, \ and\ \bibinfo {author} {\bibfnamefont {T.}~\bibnamefont {Takahashi}},\ }\href {\doibase 10.1088/1475-7516/2011/11/011} {\bibfield  {journal} {\bibinfo  {journal} {JCAP}\ }\textbf {\bibinfo {volume} {11}},\ \bibinfo {pages} {011} (\bibinfo {year} {2011})},\ \Eprint {http://arxiv.org/abs/1108.2708} {arXiv:1108.2708 [astro-ph.CO]} \BibitemShut {NoStop}%
\bibitem [{\citenamefont {Kobayashi}\ and\ \citenamefont {Takahashi}(2012)}]{Kobayashi:2012ba}%
  \BibitemOpen
  \bibfield  {author} {\bibinfo {author} {\bibfnamefont {T.}~\bibnamefont {Kobayashi}}\ and\ \bibinfo {author} {\bibfnamefont {T.}~\bibnamefont {Takahashi}},\ }\href {\doibase 10.1088/1475-7516/2012/06/004} {\bibfield  {journal} {\bibinfo  {journal} {JCAP}\ }\textbf {\bibinfo {volume} {06}},\ \bibinfo {pages} {004} (\bibinfo {year} {2012})},\ \Eprint {http://arxiv.org/abs/1203.3011} {arXiv:1203.3011 [astro-ph.CO]} \BibitemShut {NoStop}%
\bibitem [{\citenamefont {Liu}\ \emph {et~al.}(2021)\citenamefont {Liu}, \citenamefont {Liang}, \citenamefont {Zhou}, \citenamefont {Liu}, \citenamefont {Xu},\ and\ \citenamefont {Li}}]{Liu:2020zlr}%
  \BibitemOpen
  \bibfield  {author} {\bibinfo {author} {\bibfnamefont {L.-H.}\ \bibnamefont {Liu}}, \bibinfo {author} {\bibfnamefont {B.}~\bibnamefont {Liang}}, \bibinfo {author} {\bibfnamefont {Y.-C.}\ \bibnamefont {Zhou}}, \bibinfo {author} {\bibfnamefont {X.-D.}\ \bibnamefont {Liu}}, \bibinfo {author} {\bibfnamefont {W.-L.}\ \bibnamefont {Xu}}, \ and\ \bibinfo {author} {\bibfnamefont {A.-C.}\ \bibnamefont {Li}},\ }\href {\doibase 10.1103/PhysRevD.103.063515} {\bibfield  {journal} {\bibinfo  {journal} {Phys. Rev. D}\ }\textbf {\bibinfo {volume} {103}},\ \bibinfo {pages} {063515} (\bibinfo {year} {2021})},\ \Eprint {http://arxiv.org/abs/2007.08278} {arXiv:2007.08278 [astro-ph.CO]} \BibitemShut {NoStop}%
\bibitem [{\citenamefont {Hooper}\ \emph {et~al.}(2024)\citenamefont {Hooper}, \citenamefont {Ireland}, \citenamefont {Krnjaic},\ and\ \citenamefont {Stebbins}}]{Hooper:2023nnl}%
  \BibitemOpen
  \bibfield  {author} {\bibinfo {author} {\bibfnamefont {D.}~\bibnamefont {Hooper}}, \bibinfo {author} {\bibfnamefont {A.}~\bibnamefont {Ireland}}, \bibinfo {author} {\bibfnamefont {G.}~\bibnamefont {Krnjaic}}, \ and\ \bibinfo {author} {\bibfnamefont {A.}~\bibnamefont {Stebbins}},\ }\href {\doibase 10.1088/1475-7516/2024/04/021} {\bibfield  {journal} {\bibinfo  {journal} {JCAP}\ }\textbf {\bibinfo {volume} {04}},\ \bibinfo {pages} {021} (\bibinfo {year} {2024})},\ \Eprint {http://arxiv.org/abs/2308.00756} {arXiv:2308.00756 [astro-ph.CO]} \BibitemShut {NoStop}%
\bibitem [{\citenamefont {Ferrante}\ \emph {et~al.}(2023{\natexlab{b}})\citenamefont {Ferrante}, \citenamefont {Franciolini}, \citenamefont {Iovino},\ and\ \citenamefont {Urbano}}]{Ferrante:2022mui}%
  \BibitemOpen
  \bibfield  {author} {\bibinfo {author} {\bibfnamefont {G.}~\bibnamefont {Ferrante}}, \bibinfo {author} {\bibfnamefont {G.}~\bibnamefont {Franciolini}}, \bibinfo {author} {\bibfnamefont {A.}~\bibnamefont {Iovino}, \bibfnamefont {Junior.}}, \ and\ \bibinfo {author} {\bibfnamefont {A.}~\bibnamefont {Urbano}},\ }\href {\doibase 10.1103/PhysRevD.107.043520} {\bibfield  {journal} {\bibinfo  {journal} {Phys. Rev. D}\ }\textbf {\bibinfo {volume} {107}},\ \bibinfo {pages} {043520} (\bibinfo {year} {2023}{\natexlab{b}})},\ \Eprint {http://arxiv.org/abs/2211.01728} {arXiv:2211.01728 [astro-ph.CO]} \BibitemShut {NoStop}%
\bibitem [{\citenamefont {Gow}\ \emph {et~al.}(2023)\citenamefont {Gow}, \citenamefont {Assadullahi}, \citenamefont {Jackson}, \citenamefont {Koyama}, \citenamefont {Vennin},\ and\ \citenamefont {Wands}}]{Gow:2022jfb}%
  \BibitemOpen
  \bibfield  {author} {\bibinfo {author} {\bibfnamefont {A.~D.}\ \bibnamefont {Gow}}, \bibinfo {author} {\bibfnamefont {H.}~\bibnamefont {Assadullahi}}, \bibinfo {author} {\bibfnamefont {J.~H.~P.}\ \bibnamefont {Jackson}}, \bibinfo {author} {\bibfnamefont {K.}~\bibnamefont {Koyama}}, \bibinfo {author} {\bibfnamefont {V.}~\bibnamefont {Vennin}}, \ and\ \bibinfo {author} {\bibfnamefont {D.}~\bibnamefont {Wands}},\ }\href {\doibase 10.1209/0295-5075/acd417} {\bibfield  {journal} {\bibinfo  {journal} {EPL}\ }\textbf {\bibinfo {volume} {142}},\ \bibinfo {pages} {49001} (\bibinfo {year} {2023})},\ \Eprint {http://arxiv.org/abs/2211.08348} {arXiv:2211.08348 [astro-ph.CO]} \BibitemShut {NoStop}%
\bibitem [{\citenamefont {Musco}\ \emph {et~al.}(2021)\citenamefont {Musco}, \citenamefont {De~Luca}, \citenamefont {Franciolini},\ and\ \citenamefont {Riotto}}]{Musco:2020jjb}%
  \BibitemOpen
  \bibfield  {author} {\bibinfo {author} {\bibfnamefont {I.}~\bibnamefont {Musco}}, \bibinfo {author} {\bibfnamefont {V.}~\bibnamefont {De~Luca}}, \bibinfo {author} {\bibfnamefont {G.}~\bibnamefont {Franciolini}}, \ and\ \bibinfo {author} {\bibfnamefont {A.}~\bibnamefont {Riotto}},\ }\href {\doibase 10.1103/PhysRevD.103.063538} {\bibfield  {journal} {\bibinfo  {journal} {Phys. Rev. D}\ }\textbf {\bibinfo {volume} {103}},\ \bibinfo {pages} {063538} (\bibinfo {year} {2021})},\ \Eprint {http://arxiv.org/abs/2011.03014} {arXiv:2011.03014 [astro-ph.CO]} \BibitemShut {NoStop}%
\bibitem [{\citenamefont {Ianniccari}\ \emph {et~al.}(2024{\natexlab{a}})\citenamefont {Ianniccari}, \citenamefont {Iovino}, \citenamefont {Kehagias}, \citenamefont {Perrone},\ and\ \citenamefont {Riotto}}]{Ianniccari:2024ltb}%
  \BibitemOpen
  \bibfield  {author} {\bibinfo {author} {\bibfnamefont {A.}~\bibnamefont {Ianniccari}}, \bibinfo {author} {\bibfnamefont {A.~J.}\ \bibnamefont {Iovino}}, \bibinfo {author} {\bibfnamefont {A.}~\bibnamefont {Kehagias}}, \bibinfo {author} {\bibfnamefont {D.}~\bibnamefont {Perrone}}, \ and\ \bibinfo {author} {\bibfnamefont {A.}~\bibnamefont {Riotto}},\ }\href@noop {} {\  (\bibinfo {year} {2024}{\natexlab{a}})},\ \Eprint {http://arxiv.org/abs/2404.02801} {arXiv:2404.02801 [astro-ph.CO]} \BibitemShut {NoStop}%
\bibitem [{\citenamefont {Ianniccari}\ \emph {et~al.}(2024{\natexlab{b}})\citenamefont {Ianniccari}, \citenamefont {Iovino}, \citenamefont {Kehagias}, \citenamefont {Perrone},\ and\ \citenamefont {Riotto}}]{Ianniccari:2024bkh}%
  \BibitemOpen
  \bibfield  {author} {\bibinfo {author} {\bibfnamefont {A.}~\bibnamefont {Ianniccari}}, \bibinfo {author} {\bibfnamefont {A.~J.}\ \bibnamefont {Iovino}}, \bibinfo {author} {\bibfnamefont {A.}~\bibnamefont {Kehagias}}, \bibinfo {author} {\bibfnamefont {D.}~\bibnamefont {Perrone}}, \ and\ \bibinfo {author} {\bibfnamefont {A.}~\bibnamefont {Riotto}},\ }\href@noop {} {\  (\bibinfo {year} {2024}{\natexlab{b}})},\ \Eprint {http://arxiv.org/abs/2402.11033} {arXiv:2402.11033 [astro-ph.CO]} \BibitemShut {NoStop}%
\bibitem [{\citenamefont {Mroz}\ \emph {et~al.}(2024{\natexlab{a}})\citenamefont {Mroz} \emph {et~al.}}]{Mroz:2024mse}%
  \BibitemOpen
  \bibfield  {author} {\bibinfo {author} {\bibfnamefont {P.}~\bibnamefont {Mroz}} \emph {et~al.},\ }\href@noop {} {\  (\bibinfo {year} {2024}{\natexlab{a}})},\ \Eprint {http://arxiv.org/abs/2403.02386} {arXiv:2403.02386 [astro-ph.GA]} \BibitemShut {NoStop}%
\bibitem [{\citenamefont {Mroz}\ \emph {et~al.}(2024{\natexlab{b}})\citenamefont {Mroz} \emph {et~al.}}]{Mroz:2024wag}%
  \BibitemOpen
  \bibfield  {author} {\bibinfo {author} {\bibfnamefont {P.}~\bibnamefont {Mroz}} \emph {et~al.},\ }\href@noop {} {\  (\bibinfo {year} {2024}{\natexlab{b}})},\ \Eprint {http://arxiv.org/abs/2403.02398} {arXiv:2403.02398 [astro-ph.GA]} \BibitemShut {NoStop}%
\bibitem [{\citenamefont {Koushiappas}\ and\ \citenamefont {Loeb}(2017)}]{Koushiappas:2017chw}%
  \BibitemOpen
  \bibfield  {author} {\bibinfo {author} {\bibfnamefont {S.~M.}\ \bibnamefont {Koushiappas}}\ and\ \bibinfo {author} {\bibfnamefont {A.}~\bibnamefont {Loeb}},\ }\href {\doibase 10.1103/PhysRevLett.119.041102} {\bibfield  {journal} {\bibinfo  {journal} {Phys. Rev. Lett.}\ }\textbf {\bibinfo {volume} {119}},\ \bibinfo {pages} {041102} (\bibinfo {year} {2017})},\ \Eprint {http://arxiv.org/abs/1704.01668} {arXiv:1704.01668 [astro-ph.GA]} \BibitemShut {NoStop}%
\bibitem [{\citenamefont {Agius}\ \emph {et~al.}(2024)\citenamefont {Agius}, \citenamefont {Essig}, \citenamefont {Gaggero}, \citenamefont {Scarcella}, \citenamefont {Suczewski},\ and\ \citenamefont {Valli}}]{Agius:2024ecw}%
  \BibitemOpen
  \bibfield  {author} {\bibinfo {author} {\bibfnamefont {D.}~\bibnamefont {Agius}}, \bibinfo {author} {\bibfnamefont {R.}~\bibnamefont {Essig}}, \bibinfo {author} {\bibfnamefont {D.}~\bibnamefont {Gaggero}}, \bibinfo {author} {\bibfnamefont {F.}~\bibnamefont {Scarcella}}, \bibinfo {author} {\bibfnamefont {G.}~\bibnamefont {Suczewski}}, \ and\ \bibinfo {author} {\bibfnamefont {M.}~\bibnamefont {Valli}},\ }\href@noop {} {\  (\bibinfo {year} {2024})},\ \Eprint {http://arxiv.org/abs/2403.18895} {arXiv:2403.18895 [hep-ph]} \BibitemShut {NoStop}%
\bibitem [{\citenamefont {Facchinetti}\ \emph {et~al.}(2023)\citenamefont {Facchinetti}, \citenamefont {Lucca},\ and\ \citenamefont {Clesse}}]{Facchinetti:2022kbg}%
  \BibitemOpen
  \bibfield  {author} {\bibinfo {author} {\bibfnamefont {G.}~\bibnamefont {Facchinetti}}, \bibinfo {author} {\bibfnamefont {M.}~\bibnamefont {Lucca}}, \ and\ \bibinfo {author} {\bibfnamefont {S.}~\bibnamefont {Clesse}},\ }\href {\doibase 10.1103/PhysRevD.107.043537} {\bibfield  {journal} {\bibinfo  {journal} {Phys. Rev. D}\ }\textbf {\bibinfo {volume} {107}},\ \bibinfo {pages} {043537} (\bibinfo {year} {2023})},\ \Eprint {http://arxiv.org/abs/2212.07969} {arXiv:2212.07969 [astro-ph.CO]} \BibitemShut {NoStop}%
\bibitem [{\citenamefont {Brandt}(2016)}]{Brandt:2016aco}%
  \BibitemOpen
  \bibfield  {author} {\bibinfo {author} {\bibfnamefont {T.~D.}\ \bibnamefont {Brandt}},\ }\href {\doibase 10.3847/2041-8205/824/2/L31} {\bibfield  {journal} {\bibinfo  {journal} {Astrophys. J. Lett.}\ }\textbf {\bibinfo {volume} {824}},\ \bibinfo {pages} {L31} (\bibinfo {year} {2016})},\ \Eprint {http://arxiv.org/abs/1605.03665} {arXiv:1605.03665 [astro-ph.GA]} \BibitemShut {NoStop}%
\bibitem [{\citenamefont {Monroy-Rodr\'\i{}guez}\ and\ \citenamefont {Allen}(2014)}]{Monroy-Rodriguez:2014ula}%
  \BibitemOpen
  \bibfield  {author} {\bibinfo {author} {\bibfnamefont {M.~A.}\ \bibnamefont {Monroy-Rodr\'\i{}guez}}\ and\ \bibinfo {author} {\bibfnamefont {C.}~\bibnamefont {Allen}},\ }\href {\doibase 10.1088/0004-637X/790/2/159} {\bibfield  {journal} {\bibinfo  {journal} {Astrophys. J.}\ }\textbf {\bibinfo {volume} {790}},\ \bibinfo {pages} {159} (\bibinfo {year} {2014})},\ \Eprint {http://arxiv.org/abs/1406.5169} {arXiv:1406.5169 [astro-ph.GA]} \BibitemShut {NoStop}%
\bibitem [{\citenamefont {Murgia}\ \emph {et~al.}(2019)\citenamefont {Murgia}, \citenamefont {Scelfo}, \citenamefont {Viel},\ and\ \citenamefont {Raccanelli}}]{Murgia:2019duy}%
  \BibitemOpen
  \bibfield  {author} {\bibinfo {author} {\bibfnamefont {R.}~\bibnamefont {Murgia}}, \bibinfo {author} {\bibfnamefont {G.}~\bibnamefont {Scelfo}}, \bibinfo {author} {\bibfnamefont {M.}~\bibnamefont {Viel}}, \ and\ \bibinfo {author} {\bibfnamefont {A.}~\bibnamefont {Raccanelli}},\ }\href {\doibase 10.1103/PhysRevLett.123.071102} {\bibfield  {journal} {\bibinfo  {journal} {Phys. Rev. Lett.}\ }\textbf {\bibinfo {volume} {123}},\ \bibinfo {pages} {071102} (\bibinfo {year} {2019})},\ \Eprint {http://arxiv.org/abs/1903.10509} {arXiv:1903.10509 [astro-ph.CO]} \BibitemShut {NoStop}%
\bibitem [{\citenamefont {Zumalacarregui}\ and\ \citenamefont {Seljak}(2018)}]{Zumalacarregui:2017qqd}%
  \BibitemOpen
  \bibfield  {author} {\bibinfo {author} {\bibfnamefont {M.}~\bibnamefont {Zumalacarregui}}\ and\ \bibinfo {author} {\bibfnamefont {U.}~\bibnamefont {Seljak}},\ }\href {\doibase 10.1103/PhysRevLett.121.141101} {\bibfield  {journal} {\bibinfo  {journal} {Phys. Rev. Lett.}\ }\textbf {\bibinfo {volume} {121}},\ \bibinfo {pages} {141101} (\bibinfo {year} {2018})},\ \Eprint {http://arxiv.org/abs/1712.02240} {arXiv:1712.02240 [astro-ph.CO]} \BibitemShut {NoStop}%
\bibitem [{\citenamefont {Tomita}(1975)}]{Tomita:1975kj}%
  \BibitemOpen
  \bibfield  {author} {\bibinfo {author} {\bibfnamefont {K.}~\bibnamefont {Tomita}},\ }\href {\doibase 10.1143/PTP.54.730} {\bibfield  {journal} {\bibinfo  {journal} {Prog. Theor. Phys.}\ }\textbf {\bibinfo {volume} {54}},\ \bibinfo {pages} {730} (\bibinfo {year} {1975})}\BibitemShut {NoStop}%
\bibitem [{\citenamefont {Matarrese}\ \emph {et~al.}(1994)\citenamefont {Matarrese}, \citenamefont {Pantano},\ and\ \citenamefont {Saez}}]{Matarrese:1993zf}%
  \BibitemOpen
  \bibfield  {author} {\bibinfo {author} {\bibfnamefont {S.}~\bibnamefont {Matarrese}}, \bibinfo {author} {\bibfnamefont {O.}~\bibnamefont {Pantano}}, \ and\ \bibinfo {author} {\bibfnamefont {D.}~\bibnamefont {Saez}},\ }\href {\doibase 10.1103/PhysRevLett.72.320} {\bibfield  {journal} {\bibinfo  {journal} {Phys. Rev. Lett.}\ }\textbf {\bibinfo {volume} {72}},\ \bibinfo {pages} {320} (\bibinfo {year} {1994})},\ \Eprint {http://arxiv.org/abs/astro-ph/9310036} {arXiv:astro-ph/9310036} \BibitemShut {NoStop}%
\bibitem [{\citenamefont {Acquaviva}\ \emph {et~al.}(2003)\citenamefont {Acquaviva}, \citenamefont {Bartolo}, \citenamefont {Matarrese},\ and\ \citenamefont {Riotto}}]{Acquaviva:2002ud}%
  \BibitemOpen
  \bibfield  {author} {\bibinfo {author} {\bibfnamefont {V.}~\bibnamefont {Acquaviva}}, \bibinfo {author} {\bibfnamefont {N.}~\bibnamefont {Bartolo}}, \bibinfo {author} {\bibfnamefont {S.}~\bibnamefont {Matarrese}}, \ and\ \bibinfo {author} {\bibfnamefont {A.}~\bibnamefont {Riotto}},\ }\href {\doibase 10.1016/S0550-3213(03)00550-9} {\bibfield  {journal} {\bibinfo  {journal} {Nucl. Phys. B}\ }\textbf {\bibinfo {volume} {667}},\ \bibinfo {pages} {119} (\bibinfo {year} {2003})},\ \Eprint {http://arxiv.org/abs/astro-ph/0209156} {arXiv:astro-ph/0209156} \BibitemShut {NoStop}%
\bibitem [{\citenamefont {Mollerach}\ \emph {et~al.}(2004)\citenamefont {Mollerach}, \citenamefont {Harari},\ and\ \citenamefont {Matarrese}}]{Mollerach:2003nq}%
  \BibitemOpen
  \bibfield  {author} {\bibinfo {author} {\bibfnamefont {S.}~\bibnamefont {Mollerach}}, \bibinfo {author} {\bibfnamefont {D.}~\bibnamefont {Harari}}, \ and\ \bibinfo {author} {\bibfnamefont {S.}~\bibnamefont {Matarrese}},\ }\href {\doibase 10.1103/PhysRevD.69.063002} {\bibfield  {journal} {\bibinfo  {journal} {Phys. Rev. D}\ }\textbf {\bibinfo {volume} {69}},\ \bibinfo {pages} {063002} (\bibinfo {year} {2004})},\ \Eprint {http://arxiv.org/abs/astro-ph/0310711} {arXiv:astro-ph/0310711} \BibitemShut {NoStop}%
\bibitem [{\citenamefont {Ananda}\ \emph {et~al.}(2007)\citenamefont {Ananda}, \citenamefont {Clarkson},\ and\ \citenamefont {Wands}}]{Ananda:2006af}%
  \BibitemOpen
  \bibfield  {author} {\bibinfo {author} {\bibfnamefont {K.~N.}\ \bibnamefont {Ananda}}, \bibinfo {author} {\bibfnamefont {C.}~\bibnamefont {Clarkson}}, \ and\ \bibinfo {author} {\bibfnamefont {D.}~\bibnamefont {Wands}},\ }\href {\doibase 10.1103/PhysRevD.75.123518} {\bibfield  {journal} {\bibinfo  {journal} {Phys. Rev. D}\ }\textbf {\bibinfo {volume} {75}},\ \bibinfo {pages} {123518} (\bibinfo {year} {2007})},\ \Eprint {http://arxiv.org/abs/gr-qc/0612013} {arXiv:gr-qc/0612013} \BibitemShut {NoStop}%
\bibitem [{\citenamefont {Baumann}\ \emph {et~al.}(2007)\citenamefont {Baumann}, \citenamefont {Steinhardt}, \citenamefont {Takahashi},\ and\ \citenamefont {Ichiki}}]{Baumann:2007zm}%
  \BibitemOpen
  \bibfield  {author} {\bibinfo {author} {\bibfnamefont {D.}~\bibnamefont {Baumann}}, \bibinfo {author} {\bibfnamefont {P.~J.}\ \bibnamefont {Steinhardt}}, \bibinfo {author} {\bibfnamefont {K.}~\bibnamefont {Takahashi}}, \ and\ \bibinfo {author} {\bibfnamefont {K.}~\bibnamefont {Ichiki}},\ }\href {\doibase 10.1103/PhysRevD.76.084019} {\bibfield  {journal} {\bibinfo  {journal} {Phys. Rev. D}\ }\textbf {\bibinfo {volume} {76}},\ \bibinfo {pages} {084019} (\bibinfo {year} {2007})},\ \Eprint {http://arxiv.org/abs/hep-th/0703290} {arXiv:hep-th/0703290} \BibitemShut {NoStop}%
\bibitem [{\citenamefont {Dom\`enech}(2021)}]{Domenech:2021ztg}%
  \BibitemOpen
  \bibfield  {author} {\bibinfo {author} {\bibfnamefont {G.}~\bibnamefont {Dom\`enech}},\ }\href {\doibase 10.3390/universe7110398} {\bibfield  {journal} {\bibinfo  {journal} {Universe}\ }\textbf {\bibinfo {volume} {7}},\ \bibinfo {pages} {398} (\bibinfo {year} {2021})},\ \Eprint {http://arxiv.org/abs/2109.01398} {arXiv:2109.01398 [gr-qc]} \BibitemShut {NoStop}%
\bibitem [{\citenamefont {Agazie}\ \emph {et~al.}(2023)\citenamefont {Agazie} \emph {et~al.}}]{NANOGrav:2023gor}%
  \BibitemOpen
  \bibfield  {author} {\bibinfo {author} {\bibfnamefont {G.}~\bibnamefont {Agazie}} \emph {et~al.} (\bibinfo {collaboration} {NANOGrav}),\ }\href {\doibase 10.3847/2041-8213/acdac6} {\bibfield  {journal} {\bibinfo  {journal} {Astrophys. J. Lett.}\ }\textbf {\bibinfo {volume} {951}},\ \bibinfo {pages} {L8} (\bibinfo {year} {2023})},\ \Eprint {http://arxiv.org/abs/2306.16213} {arXiv:2306.16213 [astro-ph.HE]} \BibitemShut {NoStop}%
\bibitem [{\citenamefont {Antoniadis}\ \emph {et~al.}(2023)\citenamefont {Antoniadis} \emph {et~al.}}]{EPTA:2023fyk}%
  \BibitemOpen
  \bibfield  {author} {\bibinfo {author} {\bibfnamefont {J.}~\bibnamefont {Antoniadis}} \emph {et~al.} (\bibinfo {collaboration} {EPTA}),\ }\href@noop {} {\  (\bibinfo {year} {2023})},\ \Eprint {http://arxiv.org/abs/2306.16214} {arXiv:2306.16214 [astro-ph.HE]} \BibitemShut {NoStop}%
\bibitem [{\citenamefont {Reardon}\ \emph {et~al.}(2023)\citenamefont {Reardon} \emph {et~al.}}]{Reardon:2023gzh}%
  \BibitemOpen
  \bibfield  {author} {\bibinfo {author} {\bibfnamefont {D.~J.}\ \bibnamefont {Reardon}} \emph {et~al.},\ }\href {\doibase 10.3847/2041-8213/acdd02} {\bibfield  {journal} {\bibinfo  {journal} {Astrophys. J. Lett.}\ }\textbf {\bibinfo {volume} {951}},\ \bibinfo {pages} {L6} (\bibinfo {year} {2023})},\ \Eprint {http://arxiv.org/abs/2306.16215} {arXiv:2306.16215 [astro-ph.HE]} \BibitemShut {NoStop}%
\bibitem [{\citenamefont {Xu}\ \emph {et~al.}(2023)\citenamefont {Xu} \emph {et~al.}}]{Xu:2023wog}%
  \BibitemOpen
  \bibfield  {author} {\bibinfo {author} {\bibfnamefont {H.}~\bibnamefont {Xu}} \emph {et~al.},\ }\href {\doibase 10.1088/1674-4527/acdfa5} {\bibfield  {journal} {\bibinfo  {journal} {Res. Astron. Astrophys.}\ }\textbf {\bibinfo {volume} {23}},\ \bibinfo {pages} {075024} (\bibinfo {year} {2023})},\ \Eprint {http://arxiv.org/abs/2306.16216} {arXiv:2306.16216 [astro-ph.HE]} \BibitemShut {NoStop}%
\bibitem [{\citenamefont {Franciolini}\ \emph {et~al.}(2023{\natexlab{b}})\citenamefont {Franciolini}, \citenamefont {Iovino}, \citenamefont {Vaskonen},\ and\ \citenamefont {Veermae}}]{Franciolini:2023pbf}%
  \BibitemOpen
  \bibfield  {author} {\bibinfo {author} {\bibfnamefont {G.}~\bibnamefont {Franciolini}}, \bibinfo {author} {\bibfnamefont {A.}~\bibnamefont {Iovino}, \bibfnamefont {Junior.}}, \bibinfo {author} {\bibfnamefont {V.}~\bibnamefont {Vaskonen}}, \ and\ \bibinfo {author} {\bibfnamefont {H.}~\bibnamefont {Veermae}},\ }\href {\doibase 10.1103/PhysRevLett.131.201401} {\bibfield  {journal} {\bibinfo  {journal} {Phys. Rev. Lett.}\ }\textbf {\bibinfo {volume} {131}},\ \bibinfo {pages} {201401} (\bibinfo {year} {2023}{\natexlab{b}})},\ \Eprint {http://arxiv.org/abs/2306.17149} {arXiv:2306.17149 [astro-ph.CO]} \BibitemShut {NoStop}%
\bibitem [{\citenamefont {Ellis}\ \emph {et~al.}(2024)\citenamefont {Ellis}, \citenamefont {Fairbairn}, \citenamefont {Franciolini}, \citenamefont {H\"utsi}, \citenamefont {Iovino}, \citenamefont {Lewicki}, \citenamefont {Raidal}, \citenamefont {Urrutia}, \citenamefont {Vaskonen},\ and\ \citenamefont {Veerm\"ae}}]{Ellis:2023oxs}%
  \BibitemOpen
  \bibfield  {author} {\bibinfo {author} {\bibfnamefont {J.}~\bibnamefont {Ellis}}, \bibinfo {author} {\bibfnamefont {M.}~\bibnamefont {Fairbairn}}, \bibinfo {author} {\bibfnamefont {G.}~\bibnamefont {Franciolini}}, \bibinfo {author} {\bibfnamefont {G.}~\bibnamefont {H\"utsi}}, \bibinfo {author} {\bibfnamefont {A.}~\bibnamefont {Iovino}}, \bibinfo {author} {\bibfnamefont {M.}~\bibnamefont {Lewicki}}, \bibinfo {author} {\bibfnamefont {M.}~\bibnamefont {Raidal}}, \bibinfo {author} {\bibfnamefont {J.}~\bibnamefont {Urrutia}}, \bibinfo {author} {\bibfnamefont {V.}~\bibnamefont {Vaskonen}}, \ and\ \bibinfo {author} {\bibfnamefont {H.}~\bibnamefont {Veerm\"ae}},\ }\href {\doibase 10.1103/PhysRevD.109.023522} {\bibfield  {journal} {\bibinfo  {journal} {Phys. Rev. D}\ }\textbf {\bibinfo {volume} {109}},\ \bibinfo {pages} {023522} (\bibinfo {year} {2024})},\ \Eprint {http://arxiv.org/abs/2308.08546} {arXiv:2308.08546 [astro-ph.CO]} \BibitemShut {NoStop}%
\bibitem [{\citenamefont {De~Luca}\ \emph {et~al.}(2023)\citenamefont {De~Luca}, \citenamefont {Kehagias},\ and\ \citenamefont {Riotto}}]{DeLuca:2023tun}%
  \BibitemOpen
  \bibfield  {author} {\bibinfo {author} {\bibfnamefont {V.}~\bibnamefont {De~Luca}}, \bibinfo {author} {\bibfnamefont {A.}~\bibnamefont {Kehagias}}, \ and\ \bibinfo {author} {\bibfnamefont {A.}~\bibnamefont {Riotto}},\ }\href {\doibase 10.1103/PhysRevD.108.063531} {\bibfield  {journal} {\bibinfo  {journal} {Phys. Rev. D}\ }\textbf {\bibinfo {volume} {108}},\ \bibinfo {pages} {063531} (\bibinfo {year} {2023})},\ \Eprint {http://arxiv.org/abs/2307.13633} {arXiv:2307.13633 [astro-ph.CO]} \BibitemShut {NoStop}%
\bibitem [{\citenamefont {Iovino}\ \emph {et~al.}(2024)\citenamefont {Iovino}, \citenamefont {Perna}, \citenamefont {Riotto},\ and\ \citenamefont {Veerm\"ae}}]{Iovino:2024uxp}%
  \BibitemOpen
  \bibfield  {author} {\bibinfo {author} {\bibfnamefont {A.~J.}\ \bibnamefont {Iovino}}, \bibinfo {author} {\bibfnamefont {G.}~\bibnamefont {Perna}}, \bibinfo {author} {\bibfnamefont {A.}~\bibnamefont {Riotto}}, \ and\ \bibinfo {author} {\bibfnamefont {H.}~\bibnamefont {Veerm\"ae}},\ }\href@noop {} {\  (\bibinfo {year} {2024})},\ \Eprint {http://arxiv.org/abs/2406.20089} {arXiv:2406.20089 [astro-ph.CO]} \BibitemShut {NoStop}%
\bibitem [{\citenamefont {Foreman-Mackey}(2016)}]{corner}%
  \BibitemOpen
  \bibfield  {author} {\bibinfo {author} {\bibfnamefont {D.}~\bibnamefont {Foreman-Mackey}},\ }\href {\doibase 10.21105/joss.00024} {\bibfield  {journal} {\bibinfo  {journal} {JOSS}\ }\textbf {\bibinfo {volume} {1}},\ \bibinfo {pages} {24} (\bibinfo {year} {2016})}\BibitemShut {NoStop}%
\end{thebibliography}%

\end{document}